\documentclass[11pt]{article}

\usepackage{dsfont}

\usepackage{amsthm}
\usepackage{amsfonts}
\usepackage{amsmath}
\usepackage{bbm}
\usepackage[
backend=biber,
maxbibnames=99,
style=alphabetic,url=false,doi=false
]{biblatex}
\usepackage{subcaption}

\usepackage[a4paper, total={6.5in, 9.5in}]{geometry}

\usepackage{amssymb,bm}

\usepackage{quantikz}
\usetikzlibrary{external}

\usepackage[noblocks]{authblk}

\usepackage{graphicx}
\usepackage{hyperref}

\usepackage[ruled,vlined]{algorithm2e}
\usepackage[nopatch]{microtype}
\usepackage{stmaryrd}

\usepackage[ruled,vlined]{algorithm2e}
\usepackage[nopatch]{microtype}
\usepackage{booktabs}
\usepackage{stfloats} 
\usepackage{caption}
\usepackage{siunitx}
\renewcommand{\mod}

\usepackage[normalem]{ulem}
\DeclareSIUnit\angstrom{\text {Å}}

\usepackage[mode=build]{standalone}

\newtheorem{theorem}{Theorem}
\numberwithin{theorem}{section}

\newtheorem{lemma}[theorem]{Lemma}
\newtheorem*{lemma*}{Lemma}

\usepackage{mathtools}

\newcommand{\displacement}{\vec{x}(t)}
\newcommand{\displacementY}{\vec{y}(t)}
\newcommand{\displacementj}{x_j(t)}
\newcommand{\displacementk}{x_k(t)}
\newcommand{\velocity}{\dot{\vec{x}}(t)}

\newcommand{\accleration}{\ddot{\vec{x}}(t)}
\newcommand{\acclerationY}{\ddot{\vec{y}}(t)}
\newcommand{\acclerationj}{\ddot{x}_j(t)}

\newcommand{\idisplacement}{\vec{x}(0)}
\newcommand{\ivelocity}{\dot{\vec{x}}(0)}

\newcommand{\kjj}{\kappa_{jj}}
\newcommand{\kjk}{\kappa_{jk}}
\newcommand{\MassM}{\mathbf{M}}

\newcommand{\LaplacianF}{\mathbf{F}}
\newcommand{\NtN}{N \times N}

\newcommand{\poly}{\operatorname{poly}}

\newtheorem{corollary}[theorem]{Corollary}

\definecolor{adiyellow}{RGB}{200,150,0}    

\definecolor{adigreen}{RGB}{50,150,50}

\definecolor{petrosblue}{RGB}{0,0,200}    
\definecolor{seanred}{RGB}{200,0,0}      
\definecolor{stuartgreen}{RGB}{0,150,0}  
\definecolor{ioanniscyan}{RGB}{0,150,150} 
\definecolor{juliencoral}{RGB}{255,127,80} 

\begin{document}

\title{Quantum Elastic Network Models and their Application to Graphene}

\author[1,$\dagger$]{Ioannis Kolotouros}

\author[2,$\dagger$, *]{Adithya Sireesh}

\author[2, $\dagger$]{Stuart Ferguson}

\author[2]{Sean Thrasher}

\author[2]{Petros Wallden}

\author[3]{Julien Michel}

\affil[1]{Q-CTRL, Sydney, NSW Australia}
\affil[2]{Quantum Software Lab, School of Informatics, University of Edinburgh, Scotland, United Kingdom}
\affil[3]{EaStCHEM School of Chemistry, University of Edinburgh, United Kingdom}

\affil[$\dagger$]{These authors contributed equally to this work.}
\affil[*]{Corresponding authors: \url{asireesh@ed.ac.uk},\url{s1616497@ed.ac.uk}}

\maketitle

\begin{abstract}

Molecular dynamics simulations are a central computational methodology in materials design for relating atomic composition to mechanical properties. However, simulating materials with atomic-level resolution on a macroscopic scale is infeasible on current classical hardware, even when using the simplest elastic network models (ENMs) that represent molecular vibrations as a network of coupled oscillators. To address this issue, we introduce Quantum Elastic Network Models (QENMs) and utilize the quantum algorithm of Babbush et al. (PRX, 2023), which offers an exponential advantage when simulating systems of coupled oscillators. Here, we extend their algorithm in 2D systems and demonstrate how our method enables the efficient simulation of planar materials. As an example, we apply our algorithm to the task of simulating a 2D graphene sheet. We analyze the complexity for initial-state preparation, Hamiltonian simulation, and measurement of this material, and provide two real-world applications: heat transfer and the out-of-plane rippling effect. We estimate that an atomistic simulation of a graphene sheet on the centimeter scale, classically requiring hundreds of petabytes of memory and prohibitive runtimes, could be encoded and simulated with as few as $\sim 160$ logical qubits.

\end{abstract}

\section{Introduction}

Computer simulations of materials \cite{frenkel2023understanding, monticelli2013biomolecular, hollingsworth2018molecular} are among the dominant applications of scientific high-performance computing. A major focus of computational materials science is relating atomic arrangements at the nanometre length scale to bulk material properties observed on centimeter scales and beyond. Bridging these roughly seven orders of magnitude in length is essential for understanding how impurities, mixtures, or manufacturing-related defects influence material behaviour on length scales relevant to everyday life.

Unfortunately, atomistic simulations are severely constrained in size by their computational demands. To put the challenge in perspective, a perfect graphene sheet of 1 cm² manufactured in a laboratory via chemical vapour deposition would contain approximately 3.8 quadrillion carbon atoms. Simulating classically the vibrational dynamics of such a sheet with atomic resolution would require storing six double-precision values per atom to track coordinates and velocities in 3D space—amounting to roughly 180 petabytes of memory. This figure is more than thirty times greater than the total system memory available on El Capitan, the most powerful supercomputer currently in operation. 

At the same time, quantum computing aims to revolutionize the computational landscape by providing efficient algorithms for previously intractable problems \cite{alam2025fermionic, robledo2025chemistry}. Quantum hardware has experienced tremendous growth over the past few years \cite{google2025quantum}, with fault-tolerant quantum computers anticipated by 2030. Although advances in hardware and quantum error correction are conspicuous, there is still a great need to understand the advantage that these devices can offer in practical and useful applications \cite{babbush2025grand}.

The quantum algorithm landscape can, so far, be divided into two main categories. The first category contains the near-term quantum algorithms (such as NISQ algorithms) that aim to exploit the power of the available quantum computers. This category includes algorithms for classical optimization \cite{vcepaite2025quantum}, such as VQAs \cite{cerezo2021variational}, imaginary-time evolution \cite{mcardle2019variational, kolotouros2025accelerating}, or sampling algorithms such as quantum-enhanced Markov chain Monte Carlo \cite{layden2023quantum, ferguson2025quantum}. However, these algorithms are mostly heuristics, so their actual performance can only be evaluated in practice in tandem with the available quantum hardware.

The latter category is fault-tolerant quantum algorithms. For these algorithms to be executed appropriately, large error-corrected quantum devices are necessary. Some of the most prominent examples of such algorithms are the HHL algorithm \cite{harrow2009quantum} used to solve linear systems of equations, Grover's algorithm for unstructured database search \cite{grover1996fast}, Shor's algorithm for prime factorization \cite{shor1999polynomial}, or the simulation of quantum mechanical systems \cite{low2019hamiltonian}, to name a few. In comparison to the former category, fault-tolerant quantum algorithms have proven theoretical guarantees about their performance with asymptotic advantage over their classical counterparts.

On top of the aforementioned algorithms, quantum computing offers an alternative path towards scalable molecular dynamics simulations. Recently, Babbush et al. \cite{babbush2023exponential} reported that a quantum computer can simulate a system of coupled oscillators exponentially faster than a classical computer, provided that certain assumptions (sparsity in the connectivity and limited non-zero initial conditions) are satisfied. The algorithm can be used to estimate global quantities such as the kinetic or potential energy of subsystems or the entire system.

However, the question of whether this algorithm can be used for any practical applications is still open. In this paper, we aim to provide an answer by introducing an algorithm that can simulate the dynamics of large molecular assemblies, utilizing the results of \cite{babbush2023exponential}. Specifically, we introduce a quantum version of the widely used Elastic Network Model (ENM) \cite{tirion1996large}, which we call a Quantum Elastic Network Model (QENM), and show how to apply our algorithm in a graphene molecule. We estimate that the algorithm would allow for simulations of a square graphene sheet of 1cm\textsuperscript{2} with around 150-200 logical qubits. 
When evaluating material properties, the computational speedup of this quantum algorithm is highly application-specific. Certain experimental setups may inherently require exponential runtime, or they might allow for the algorithm to be dequantized entirely \cite{sakamoto2025quantum}. Nevertheless, as discussed in Sec. \ref{sec:discussion}, our framework still provides significant resource advantages in those cases.

Elastic network models \cite{tirion1996large, eyal2006anisotropic, togashi2018coarse, lezon2009elastic} are widely used as a coarse-grained method to study the slow dynamics (and conformational changes) of large molecules and materials. Instead of describing the interactions between atoms by complex force fields, ENMs consider only harmonic interactions between atoms within a certain cutoff distance, effectively representing the molecular system as a network of nodes connected by elastic springs. By simplifying the potential energy surface into a set of harmonic oscillators, ENMs allow researchers to bypass the high-frequency ``noise'' of local atomic vibrations and focus on the low-frequency, large-scale collective motions that are often functionally relevant for proteins and molecular complexes. However, normal mode analyses of ENMs run into memory limitation for models that contain more than millions of atoms, whereas numerical integration may be tractable on classical devices for up to billions of atoms. This means that direct simulation of materials and molecular assemblies on a macroscopic scale at atomic resolution is beyond the reach of classical HPC, and compromises such as use of coarse-grain simulations or finite elements modelling are necessary to simulate materials on a macroscopic scale. 

Recent quantum algorithms for molecular dynamics \cite{simon2024improved, ollitrault2020nonadiabatic, o2022efficient, fedorov2021ab} primarily focus on the electronic structure problem. Because they rely on exact quantum mechanical descriptions, such as \textit{ab initio} calculations or the Born-Oppenheimer approximation, they are computationally restricted to microscopic systems, like the $\text{H}_2$ molecule. While highly accurate at the atomic level, these methods cannot scale to approximate macroscopic phenomena.

Quantum Elastic Network Models (QENMs) bypass these limitations by trading electronic exactness for scalability. By applying the simulation framework of Babbush et al. \cite{babbush2023exponential} to coarse-grained interactions, QENMs allow us to simulate materials at the macroscopic scale. In practice, materials scientists can use QENMs to rapidly test different dopant configurations and structural defects in centimeter-scale materials \textit{in silico}. Ultimately, this allows one to directly predict macroscopic behaviors without needing to physically synthesize and characterize the material in a laboratory first.

\subsection{Main results}
Our \emph{main contributions} are summarized below.
\begin{itemize}
    \item We introduce the quantum analogue of the elastic network models and discuss when these models can be efficiently implemented on a quantum device.

    \item We introduce a discretization method to load $2^n$ samples of the Maxwell-Boltzmann distribution onto the amplitudes of an $n$-qubit quantum state using $\mathcal{O}(n)$ resources.

    \item We extend the original algorithm of Babbush et al. \cite{babbush2023exponential} to $D>1$ \emph{coupled} dimensions.

    \item We show that molecules that exhibit a well-defined structure, all the oracles required for the algorithm introduced by Babbush et al. \cite{babbush2023exponential} can be executed efficiently. Furthermore, we introduce necessary oracles that need to be efficiently executable, when moving from 1D to larger dimensions.

    \item We introduce an efficient and practical construction of the \emph{connectivity oracle} of a graphene sheet and discuss an extension of this construction in the case where \emph{defects} are present in the lattice.

    \item We analyze the complexity of preparing various initial states for our algorithm, depending on the observables of interest, and show that the procedure remains efficient in the case of the graphene sheet.

    \item We introduce two practical applications for our model, \emph{heat transfer} and out-of-plane \emph{rippling effect}, and discuss the existing bottlenecks that need to be addressed in order to make our model more realistic.
\end{itemize}

\subsection{Structure}

The remainder of the paper is organized as follows. In Sec. \ref{sec:preliminaries}, we introduce the coupled harmonic oscillator problem, we give the essential background on elastic network models, and on the quantum algorithm for simulating coupled classical oscillators. We conclude by motivating the study of molecular dynamics through its connection to an abstraction called Elastic Network Models (ENMs). In Sec. \ref{sec:QENMs}, we introduce our framework for quantum elastic network models, we provide an algorithm to efficiently load the initial conditions onto the amplitudes of a quantum state, and discuss how to construct the necessary oracles efficiently. In Sec. \ref{section:application_to_2d}, we extend the algorithm of Babbush et al. to $D=2$ dimensions. In Sec. ~\ref{sec:graphene}, we apply our algorithm on a graphene sheet. We thoroughly explain how to prepare the initial state, how to simulate the material, and finally, we propose two practical applications for such a simulation. In Sec. \ref{sec:discussion}, we discuss the immediate next steps to make our model more realistic. Finally, in Sec. \ref{sec:conclusion}, we conclude our work by addressing some limitations and directions for future research.

\section{Preliminaries}
\label{sec:preliminaries}

As mentioned previously, systems of coupled classical oscillators serve as a ubiquitous mathematical framework for describing a wide variety of physical phenomena. Thus, overcoming the computational bottlenecks inherent in solving these systems unlocks better capabilities for modelling the complex physical reality that surrounds us. In this section, we first introduce the coupled harmonic oscillator problem and the quantum algorithm developed by Babbush et al.~\cite{babbush2023exponential} to solve it. We then shift our focus to physical domains where harmonic oscillators arise, and review classical Elastic Network Models (ENMs), ultimately leading to our formulation of Quantum Elastic Network Models (QENMs).

\subsection{Classical Harmonic Oscillator Systems}
\begin{figure}[!h]
    \centering
    \includegraphics[width=0.5\textwidth]{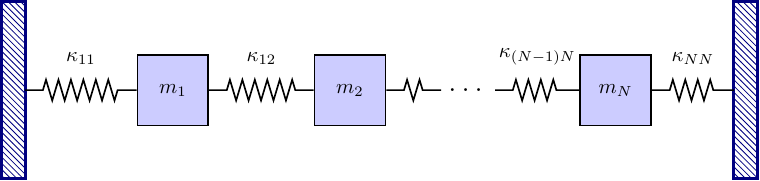}
    \vspace{0.5cm}
    \caption{An example of a $D=1$ system of oscillators with $N$ masses. Masses $m_1$ and $m_N$ are coupled to the left and right walls respectively. The only nonzero spring constants here are $\{\kappa_{i(i+1)}\}_{i=1}^{N-1}, \kappa_{11}$ and $\kappa_{NN}$.}
    \label{fig:1d-system}
\end{figure}

\begin{figure}[!h]
    \centering
    \includegraphics[width=0.4\textwidth]{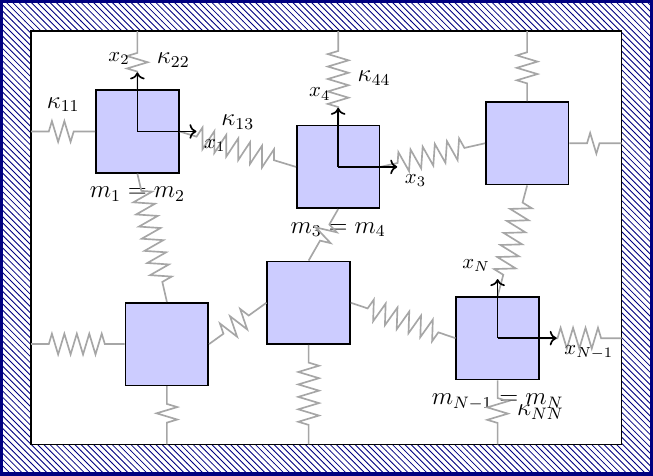}
    \vspace{0.5cm}
    \caption{An example of a $D=2$ system of oscillators with $N/2$ masses.}
    \label{fig:2d-system}
\end{figure}

In a very general setting (and as described in~\cite{babbush2023exponential} which we detail here for completeness), a system of coupled harmonic oscillators is defined by a set of $N$ masses $m_1, m_2,\cdots, m_N$ that are connected to each other by springs. At any time $t\geq 0$, the displacements (w.r.t the rest position) and velocities are given by $\vec{x}(t)=(x_1(t), x_2(t),\cdots, x_N(t))^T\in \mathbb{R}^N$ and $\dot{\vec{x}}(t)=(\dot{x}_1(t), \dot{x}_2(t),\cdots, \dot{x}_N(t))^T\in \mathbb{R}^N$, respectively where $\dot{a}(t)=\frac{d}{dt}a(t)$ and $\ddot{a}(t)=\frac{d^2}{dt^2}a(t)$ (see fig.~\ref{fig:1d-system}). Let $\kappa_{kj} = \kappa_{jk} \geq 0$, be the spring constants that couple the $j^{th}$ and $k^{th}$ oscillators, and $\kappa_{jj}\geq 0$ be the spring constant of the spring that connects the $j^{th}$ oscillator to a wall. The above setting would be for a single dimensional system i.e. $D=1$. Similarly, for a system of $D$ spatial dimensions, one would need $\displacement \in \mathbb{R}^{DN}$, and $\velocity \in \mathbb{R}^{DN}$, i.e. $D$ separate coordinates for each of the $N$ oscillators (see fig.~\ref{fig:2d-system} for an example of a $D=2$ system).

The harmonic approximation assumes that any deviation of a mass from its equilibrium position results in a restoring force proportional to and in the opposite direction of the displacement. Simulating such systems, in the harmonic approximation with initial configuration $\idisplacement$ and $\ivelocity$, is done by solving Newton's equations (for all $j\in \{1,\cdots,N\}$):
\begin{equation}
    m_j \acclerationj = \sum_{k\neq j}\kappa_{jk}(\displacementk - \displacementj) - \kjj\displacementj
\end{equation}

A more succinct representation in matrix form gives $\MassM\accleration = -\LaplacianF\displacement$, where $\MassM$ is an $\NtN$ diagonal matrix with entries $m_j > 0$ and $\LaplacianF$ is the $\NtN$ weighted graph laplacian of our spring constant matrix, with $f_{jj}=\sum_k \kjk$ and $f_{jk} = -\kjk$

Traditionally, in order to solve this system, one would apply either differential equations solvers to solve the problem numerically, or use normal mode analysis to get an analytical solution.
One would have to transform the second-order differential equation representing the dynamics of the system to mass-weighted coordinates $\displacementY=\MassM^{1/2}\displacement$ and $\mathbf{A} = \MassM^{-1/2}\LaplacianF\MassM^{-1/2}\succeq 0$, which yields
\begin{equation}
    \acclerationY = - \mathbf{A}\displacementY 
\end{equation}
Because the mass-weighted Hessian matrix $\mathbf{A}$ is real, symmetric, and positive semidefinite ($\mathbf{A} \succeq 0$), it can be diagonalized by an orthogonal matrix $\mathbf{U}$ whose columns are the eigenvectors of $\mathbf{A}$. This yields:
\begin{equation}
    \mathbf{U}^T \mathbf{A} \mathbf{U} = \mathbf{\Lambda}
\end{equation}
where $\mathbf{\Lambda}$ is a diagonal matrix containing the eigenvalues $\lambda_k$. The positive semidefinite nature of $\mathbf{A}$ is physically significant; it guarantees that all eigenvalues are non-negative ($\lambda_k \geq 0$). These eigenvalues correspond to the squared angular frequencies of the system's normal modes, $\lambda_k = \omega_k^2$, ensuring that the frequencies are strictly real.

To decouple the equations of motion, we further introduce a coordinate transformation into the normal mode space, defining the normal coordinates $\mathbf{q}$ as:
\begin{equation}
    \displacementY = \mathbf{U} \mathbf{q}
\end{equation}
Substituting this transformation into our mass-weighted differential equation and multiplying from the left by $\mathbf{U}^T$, we leverage the orthogonality condition ($\mathbf{U}^T \mathbf{U} = \mathbf{I}$) to obtain the decoupled equations of motion:
\begin{equation}
    \ddot{\mathbf{q}} = - \mathbf{\Lambda} \mathbf{q}
\end{equation}
This transformation reduces the complex, $N$-body problem into a set of independent, one-dimensional harmonic oscillators. The equation of motion for each individual normal mode $k$ is given by:
\begin{equation}
    \ddot{q}_k = - \omega_k^2 q_k
\end{equation}
which admits the familiar analytical solution:
\begin{equation}
    q_k(t) = c_k \cos(\omega_k t + \phi_k)
\end{equation}
where the amplitude $c_k$ and phase angle $\phi_k$ are determined by the initial configuration $\idisplacement$ and initial velocities $\ivelocity$ of the network. Finally, the physical displacements of the original masses can be reconstructed by applying the inverse transformations:
\begin{equation}
    \displacement = \MassM^{-1/2} \mathbf{U} \mathbf{q}(t)
\end{equation}

The cost of the diagonalization procedure is actually the primary bottleneck of classical Normal Mode Analysis with time complexity of the order of $\mathcal{O}(N^3)$ and space complexity $\mathcal{O}(N^2)$, i.e. $\poly(N)$ or equivalently $\exp{(n)}$, where $N=2^n$. The goal of~\cite{babbush2023exponential} (which we detail in the next section) was to develop a quantum algorithm to simulate the dynamics of such coupled oscillator systems in time $\poly(n)$.

\subsection{Simulating coupled classical oscillators exponentially faster}
\label{subsec:simulating_coupled_classical_oscillators}

\begin{figure}[!t]
    \centering
    \resizebox{1\textwidth}{!}{%
    \begin{tikzpicture}[
    font=\sffamily\large,
    >={Stealth[length=5pt]},
    node distance=1cm and 1.5cm,
    unitaryprocess/.style={
        rectangle, draw, fill=blue!10, text width=4cm, 
        align=center, minimum height=1cm, rounded corners=5pt
    },
    nonunitaryprocess/.style={
        rectangle, draw, fill=gray!10, text width=4cm, 
        align=center, minimum height=1cm, rounded corners=5pt
    },
    state/.style={
        rectangle, draw, fill=white, text width=4.2cm, 
        align=center, minimum height=0.8cm, rounded corners=5pt
    },
    detailnode/.style={
        rectangle, draw, text width=7.5cm, 
        align=center, minimum height=1cm, rounded corners=2pt, fill=orange!10,
        font=\sffamily\small
    },
    glossarynode/.style={
        rectangle, draw, text width=5cm, 
        align=left, minimum height=2cm, rounded corners=5pt, fill=green!5,
        font=\sffamily\small
    },
    group/.style={
        draw, dashed, inner sep=15pt, rounded corners=5pt, minimum width=6cm
    }
]

    
    \node [state] (s1) {$\mathbf{|0\rangle^{\otimes\mathcal{O}(log(N))}}$};
    \node [unitaryprocess, below=of s1] (p1) {$\mathcal{W}$};
    \node [state, below=of p1] (s2) {$|\psi(0)\rangle$};

    \node [unitaryprocess, below=4cm of s2] (p2) {$e^{-i\mathbf{H}t}$};
    \node [state, below=of p2] (s3) {$|\psi(t)\rangle$};

    \node [nonunitaryprocess, below=4cm of s3] (p3) {$\langle\psi(t)| P_V|\psi(t)\rangle$};
    \node [state, below=of p3] (s4) {$\hat{k}_v$, $\hat{u}_v$};

    \draw [->, thick] (s1) -- (p1);
    \draw [->, thick] (p1) -- (s2);
    \draw [->, thick] (s2) -- (p2);
    \draw [->, thick] (p2) -- (s3);
    \draw [->, thick] (s3) -- (p3);
    \draw [->, thick] (p3) -- (s4);

    \node [group, fit=(s1) (p1) (s2), label=left:\rotatebox{90}{\textbf{Initial state preparation}}] (g1) {};
    \node [group, fit=(p2) (s3), label=left:\rotatebox{90}{\textbf{Hamiltonian Simulation}}] (g2) {};
    \node [group, fit=(p3) (s4), label=left:\rotatebox{90}{\textbf{Measurement}}] (g3) {};

    \node [detailnode, below right=0.5cm and 3cm of g1.north east, anchor=north west, text width=200pt] (stepA1) {
        \textbf{Choose encoding}\\
        \hspace{-5pt} Energies, velocities  \hspace{5pt}   $\vert$   \hspace{5pt}   Displacements       
    };
    
    \node [detailnode, below left=0.5cm and 0.5cm of stepA1.south, text width=4cm] (stepA2) {
        \textbf{Initial state ($K_V, U_V, \dot{x}_V)$}\\
        $|\psi(0)\rangle=\frac{1}{\sqrt{2 E}}\binom{\sqrt{\mathrm{M}} \dot{\boldsymbol{x}}(0)}{i \boldsymbol{\mu}(0)}$
    };
    
    \node [detailnode, below=0.3cm of stepA2, fill=cyan!5, text width=3cm] (stepA3) {
        \textbf{Complexity}\\
        $\mathcal{O}\left(\sqrt{E_{\max } d / E}\right)$ calls to $\mathcal{U}, \mathcal{S}$
    };

    \node [detailnode, below right=0.5cm and 0.5cm of stepA1.south, text width=4cm] (stepAa2) {
        \textbf{Initial state ($x_V$)}\\
        $\ket{\psi(0)} = \frac{1}{\sqrt{2F}}\binom{\mathbf{P}\boldsymbol{y}(0)}{-i\mathbf{B}^+\mathbf{P}\boldsymbol{\dot{y}}(0)}$
    };
    \node [detailnode, below=0.3cm of stepAa2, fill=cyan!5, text width=3cm] (stepAa3) {
        \textbf{Complexity}\\
        $\tilde{\mathcal{O}}(\text{cond}(\mathbf{B})s\log N)$ quantum gates
        
    };

    \node [detailnode, above right=2.5cm and 3cm of g2.north east, anchor=north west] (stepB1) {
        \textbf{Classical equation of motion}\\
        $\mathbf{y}(t) = \sqrt{\mathbf{M}}\mathbf{x}(t)$, $\ddot{\mathbf{y}}(t) = -\mathbf{A}\mathbf{y}(t)$ \\
        where $\mathbf{A} = \sqrt{\mathbf{M}}^{-1} \mathbf{K} \sqrt{\mathbf{M}}^{-1}$
    };

    \node [detailnode, below=0.4cm of stepB1] (stepB2) {
        \textbf{Map to Schrödinger equation }\\
        $\ddot{\mathbf{y}} + i\sqrt{\mathbf{A}}\dot{\mathbf{y}} = i\sqrt{\mathbf{A}}(\dot{\mathbf{y}} + i\sqrt{\mathbf{A}}\mathbf{y}) \implies \frac{d}{dt}|\psi\rangle = -i\mathbf{H}|\psi\rangle$
    };

    \node [detailnode, below=0.4cm of stepB2] (stepB3) {
        \textbf{Hamiltonian via block encoding}\\
        $\mathbf{H} = -\begin{pmatrix} \mathbf{0} & \mathbf{B} \\ \mathbf{B}^\dagger & \mathbf{0} \end{pmatrix}$ s.t. $\mathbf{B}\mathbf{B}^\dagger = \mathbf{A}$
    };

    \node [detailnode, below=0.4cm of stepB3, fill=cyan!5, text width=3.5cm] (stepB4) {
        \textbf{Complexity}\\
        $\mathcal{O}\left(t\sqrt{d \frac{\kappa_{\text{max}}}{m_{\text{min}}}} + \log(1/\epsilon)\right)$ calls to $\mathbf{K}, \mathbf{M}$
    };

    \node [detailnode, above right = 1cm and 2cm of g3.east, text width = 120] (stepC1) {
        \textbf{Observable: $K_V$}\\
        Estimate $K_V$, of subset, $V$: \\
        $P_V=(\mathds{1}-\mathcal{V}) / 2$ \\
        $\langle \psi(t) | P_V | \psi(t) \rangle = \frac{\mathcal{K}_V(t)}{E}$
    };

    \node [detailnode, below=0.6cm of stepC1, text width = 100] (stepC2) {
        \textbf{Output: $\hat{k}_V(t)$}\\
        $|\hat{k}_V(t)-K_V(t) / E| \leq \epsilon$
    };

    \node [detailnode, right = 1cm of stepC1, text width = 120] (stepCc1) {
        \textbf{Observable: $U_V$}\\
        Estimate $U_V$ of subset $V$: \\
        $P_V=(\mathds{1}-\mathcal{V}) / 2$ \\
        $\langle \psi(t) | P_V | \psi(t) \rangle = \frac{\mathcal{U}_V(t)}{E}$
    };

    \node [detailnode, below=0.6cm of stepCc1, text width = 100] (stepCc2) {
        \textbf{Output: $\hat{u}_V(t)$}\\
        $|\hat{u}_V(t)-U_V(t) / E| \leq \epsilon$
    };

    \node [detailnode, fill=cyan!5, text width=150] 
        at ($(stepC2.south)!0.5!(stepCc2.south) + (0,-1.5cm)$) (stepShared) {
        \textbf{Complexity}\\
        High-confidence amplitude estimation\\
        $\mathcal{O}(\log (1 / \delta) / \epsilon)$ calls to $\mathcal{V}$
    };


    \draw [->] (stepC1) -- (stepC2);
    \draw [->] (stepCc1) -- (stepCc2);

    \draw [->] (stepB1) -- (stepB2);
    \draw [->] (stepB2) -- (stepB3);
    \draw [->] (stepB3) -- (stepB4);
    
    \draw [->] (stepC2.south) -- (stepShared.north);
    \draw [->] (stepCc2.south) -- (stepShared.north);

    \draw [->, dashed, gray] (g1.east) -- (stepA1.west);
    \draw [->] (stepA1) -- (stepA2);
    \draw [->] (stepA1) -- (stepAa2);
    \draw [->] (stepA2) -- (stepA3);
    \draw [->] (stepAa2) -- (stepAa3);
    
    \draw [->, dashed, gray] (g2.east) -- (stepB1.west);
    \draw [->, dashed, gray] (g3.east) -- (stepC1.west);

\end{tikzpicture}
    }
    \vspace{0.5cm}
    \caption{ Overview of the algorithm introduced in \cite{babbush2023exponential}. The algorithm is divided into three main subroutines (left), each of which is outlined in more detail (right).} 
\label{fig:overview_of_algorithm}   
\end{figure}
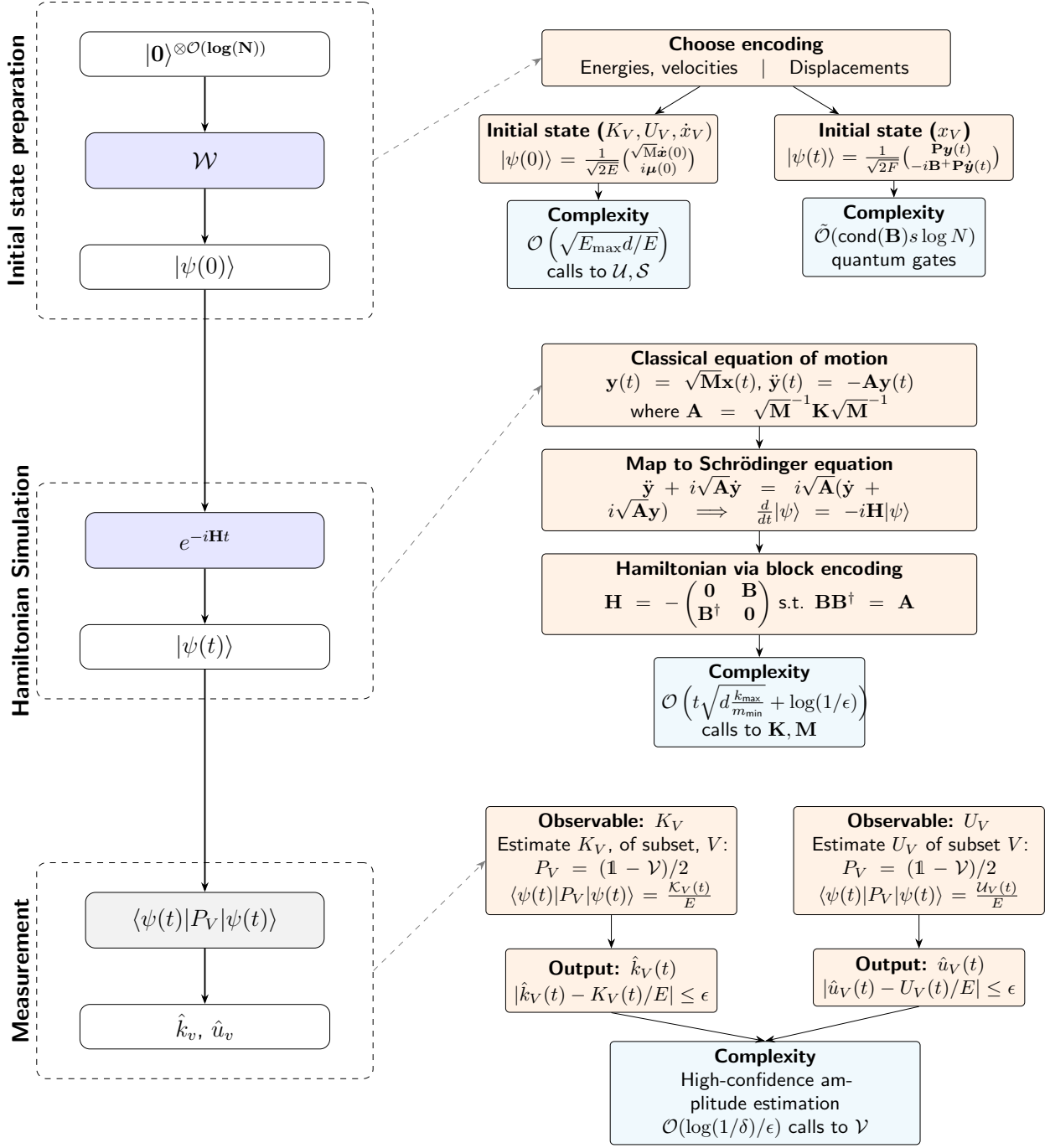

In this section, we provide a high-level overview of the recent quantum algorithm introduced by \cite{babbush2023exponential}. We do this for the example case of measuring kinetic or potential energies, but other observables are discussed in Sec. \ref{subsec:obs_quants}. The algorithm is divided into three steps: i) initial state preparation ii) Hamiltonian simulation and iii) measurement. It is then left to the following sections to analyze how one can efficiently execute these processes, including the required oracles, on digital quantum computers for given problems. The general framework for this algorithm is outlined in Fig. \ref{fig:overview_of_algorithm}.

\subsubsection{Problem formulation}

The end-goal of their algorithm is to simulate the dynamics of a system of $N$ coupled oscillators at a time $t\geq0$. Using the same problem setting as described in the previous section, the classical dynamics of any such system are governed by Newton's equation:
\begin{equation}
\textbf{M}\ddot{\boldsymbol{x}}(t) = -\textbf{F}\boldsymbol{x}(t)
\label{eq:Newtons_equation}
\end{equation}
  Solving this system of equations to obtain the positions and velocities of all atoms at different times can be done efficiently (polynomially) in $N$. Even to simply readout the different positions and momenta of the solution would take linear time in $N$. The choice is made to reformulate the problem as that of computing the time dynamics of $N = 2^n$ oscillators efficiently in $n$, or equivalently with complexity that is $polylog(N)$.

The main idea is to map the dynamics of a classical system of $N$ oscillators  into a quantum system whose dynamics are governed by the Schrödinger equation:
\begin{equation}
    \frac{\partial \ket{\psi(t)}}{\partial t} = -i\textbf{H}\ket{\psi(t)}
\label{eq:schrodinger_eq}
\end{equation}
where $\mathbf{H}$ denotes the Hamiltonian of the system. This system is described by $polylog(N)$ qubits.

\subsubsection{Encoding}
\label{sec:orig_encoding}
This exponential space saving relies on efficiently encoding the positions and velocities in the amplitudes of a quantum state. To achieve this reduction, one needs to carefully consider how the oscillator variables (and initial conditions) are mapped to qubits. Depending on what type of information one wants to extract from the classical evolution of this system, the initial state should be chosen accordingly. 

If we are interested in measuring either the kinetic energy,

\begin{equation}
\label{eq:kinetic_energy}
    K(t)=\frac{1}{2} \sum_j m_j \dot{x}_j(t)^2,
\end{equation}
or the potential energy,
\begin{equation}
\label{eq:potential_energy}
U(t)=\frac{1}{2}\left(\sum_j \kappa_{j j} x_j(t)^2+\sum_{k>j} \kappa_{j k}\left(x_j(t)-x_k(t)\right)^2\right)
\end{equation}
then we can initialize the system in the state:

\begin{equation}
    \ket{\psi(0)} = \frac{1}{\sqrt{2E}} \begin{pmatrix}
        \sqrt{\mathbf{M}} \dot{\boldsymbol{x}}(0)\\
        i\boldsymbol{\mu}(0)
    \end{pmatrix}
\label{eq:initial_state}
\end{equation}
where $\dot{\boldsymbol{x}}(0)$ is the velocity vector at time $t=0$, $\boldsymbol{\mu}(0)$ is a vector encoding the positions of the atoms and their pairwise differences, and $E$ is the total energy of the system (which remains constant throughout the time evolution). The vector $\vec{\boldsymbol{\mu}}(t) \in \mathbb{R}^M[M:=N(N+1) / 2]$ is a vector with $N$ entries $\sqrt{\kappa_{j j}} x_j(t)$ and $N(N-1) / 2$ entries $\sqrt{\kappa_{j k}}\left(x_j(t)-x_k(t)\right)$, with $k>j$. This immediately follows from the fact that the total probability (amplitude squared) of measuring a given subset of basis-states corresponds directly to the energy contribution of that basis state. 
In Sec. \ref{subsec:encodings} we analyse another encoding that is also a solution to the Schrödinger equation (Eq. \eqref{eq:schrodinger_eq}), and allows for the extraction of squared displacements. We then discuss some additional observable quantities, using either encoding, in Sec. \ref{subsec:obs_quants}.

In \cite{babbush2023exponential}, it is shown that $\ket{\psi(0)}$ can be efficiently prepared by an oracle, $\mathcal{W}$, which requires $\mathcal{O}\left(\sqrt{E_{\max } d / E}\right)$ calls to oracles $\mathcal{U}$ and $\mathcal{S}$, and its inverses, where $\mathcal{U}$ is an oracle that encodes the initial conditions, $\mathcal{S}$ is an oracle that carries information about the mass and coupling coefficients, $d$ is the sparsity of $\mathbf{K}$, and $E_{\max}$ is the energy of a system of $N$ uncoupled oscillators with masses $m_{\max}$ and individual coupling coefficients $\kappa_{\max}$. \footnote{Here, $m_{\text{max}}$ and $k_{\text{max}}$ are simply the known extrema of the mass and coupling coefficients respectively.}

\subsubsection{Hamiltonian simulation}

Given the initial state, $|\psi(0)\rangle$, the job of Hamiltonian simulation is to evolve the physical system to time, $t$, to a state that is $\varepsilon$-close to $|\psi(t)\rangle$ in Euclidean norm. This corresponds to the map $|\psi(0)\rangle \mapsto |\psi(t)\rangle$. Here, the dynamics of the classical system (Eq. \ref{eq:Newtons_equation}) are mapped to the dynamics of a quantum mechanical system (Eq. \ref{eq:schrodinger_eq}) by considering the set of variables:
\begin{equation}
    \boldsymbol{y}(t) = \sqrt{\textbf{M}}\boldsymbol{x}(t) \implies \ddot{\boldsymbol{x}}(t) = \sqrt{\textbf{M}}^{-1} \ddot{\boldsymbol{y}}(t)
\end{equation}
Thus, by substituting $\boldsymbol{y}(t)$ in Newton's Equation in Eq. \eqref{eq:Newtons_equation} we get:
\begin{equation}
    \ddot{\boldsymbol{y}}(t) = -\sqrt{\textbf{M}}^{-1} \textbf{F}\sqrt{\textbf{M}}^{-1} \boldsymbol{y}(t) \implies  \ddot{\boldsymbol{y}}(t) = -\textbf{A}\boldsymbol{y}(t)
\label{eq:newtons_equation_new_variables}
\end{equation}
where $\textbf{A} = -\sqrt{\textbf{M}}^{-1} \textbf{F}\sqrt{\textbf{M}}^{-1} \succcurlyeq 0$. By adding $i\sqrt{\textbf{A}}
\dot{\boldsymbol{y}}(t)$ in both sides we can rewrite Eq. \eqref{eq:newtons_equation_new_variables} as:
\begin{equation}
    \ddot{\boldsymbol{y}}(t) + i\sqrt{\textbf{A}}
\dot{\boldsymbol{y}}(t) = i\sqrt{\textbf{A}}[\dot{\boldsymbol{y}}(t) + i\sqrt{\textbf{A}}\boldsymbol{y}(t)]
\label{eq:Schrodinger_equation_transformation}
\end{equation}
which is exactly Schrödinger equation (Eq. \eqref{eq:schrodinger_eq}) for Hamiltonian $-\sqrt{\textbf{A}}$. Thus:
\begin{equation}
    \dot{\boldsymbol{y}}(t) + i\sqrt{\textbf{A}}
\boldsymbol{y}(t) = e^{i\sqrt{\textbf{A}}t}(\dot{\boldsymbol{y}}(0) + i\sqrt{\textbf{A}}
\boldsymbol{y}(0))
\end{equation}
The question that naturally arises, is how one could simulate a quantum mechanical system that evolves under the hamiltonian $-\sqrt{\textbf{A}}$? It turns out that simulating $-\sqrt{\mathbf{A}}$, given access to only $\mathbf{A}$ is computationally suboptimal.  Instead, Babbush et al. \cite{babbush2023exponential} demonstrated that we can construct a block Hamiltonian $\mathbf{H}$, whose evolution naturally encodes the classical system's state:
\begin{equation}
    \mathbf{H} = -\begin{pmatrix}
        \mathbf{0} & \mathbf{B}\\
        \mathbf{B}^\dagger & \mathbf{0} 
    \end{pmatrix} \in \mathbb{C}^{(N+M)\times (N+M)}
\label{eq:problem_hamiltonian}
\end{equation}
where $\mathbf{B}\in \mathbb{C}^{N\times M}$ is related to the \emph{incidence matrix} of the graph and satisfies $\mathbf{B}\mathbf{B}^{\dagger} =\mathbf{A}$. For a given graph $G(V,E)$ with vertices $V=\{v_1, v_2, \ldots, v_{|V|}\}$, edges $E=\{e_1, e_2, \ldots, e_{|E|}\}$, and weights $W=\{w_1, w_2, \ldots, w_{|E|}\}$, $\mathbf{B}$ is an $|V|\times|E|$ matrix with rows indexed by vertices and columns indexed by edges. The way to construct it is to start by arbitrarily assigning orientations to the edges of the graph (this affects the form of $\mathbf{B}$ but does not affect $\mathbf{B}\mathbf{B}^{\dagger}$). Then, the matrix elements $\mathbf{B}_{ij}$ (corresponding to node $v_i$ and edge $e_j$) are defined as:
\begin{equation}
    \mathbf{B}_{ij} = \begin{cases}
        -\sqrt{w_j} \text{ if edge $e_j$ leaves vertex $v_i$} \\
        \sqrt{w_j} \text{ if edge $e_j$ is a self-loop of vertex $v_i$} \\
        \sqrt{w_j} \text{ if edge $e_j$ enters vertex $v_i$} \\
        0 \text{ otherwise}
    \end{cases}
\end{equation}

The operator $\mathbf{0}$ is the matrix with all elements equal to zero. It can be easily shown that both states in Eq. \eqref{eq:initial_state} and Eq. \eqref{eq:initial_state_alternate} satisfy Schrödinger equation (Eq. \eqref{eq:schrodinger_eq}) for the Hamiltonian in Eq. \eqref{eq:problem_hamiltonian}. Since $\mathbf{B}$ acts on an $N\times M$ space, we need to pad additional $0$ and let $\mathbf{B}$ act on the space with basis $\{\ket{j,k}: \; j\leq k \in [N]\}$. Intuitively, we consider that the graph is all-to-all connected but with most of the weights equal to zero. The (nonunitary) operator $\mathbf{B}$ then acts as:
\begin{equation}
    \mathbf{B} \ket{j} \ket{k} = \begin{cases}
        \sqrt{\frac{\kappa_{jj}}{m_j}} \ket{j} \text{ if $j=k$}\\
        \sqrt{\frac{\kappa_{jk}}{m_j}}\ket{j}- \sqrt{\frac{\kappa_{jk}}{m_k}}\ket{k} \text{ if $j<k$}
    \end{cases} 
    \label{eq:incidence_matrix_def}
\end{equation}
With this definition, it is easy to check that $\sqrt{\mathbf{M}}\mathbf{B} (\sqrt{\mathbf{M}}\mathbf{B})^{\dagger} = \mathbf{F}$ and as such, $\mathbf{B}\mathbf{B}^{\dagger} = \mathbf{A}$. In \cite{babbush2023exponential}, it is shown that by using this block encoding, the Hamiltonian simulation requires $\mathcal{O}\Big(t\sqrt{d \frac{\kappa_{\text{max}}}{{m_{\text{min}}}}}+\log (1 / \epsilon)\Big)$ calls to oracles describing $\mathbf{K}$ and $\mathbf{M}$.

\subsubsection{Measurement}

Assuming that we can efficiently prepare the initial encoding ($\mathcal{W}|0\rangle \mapsto|\psi(0)\rangle$, Eq. \ref{eq:initial_state}), and efficiently perform Hamiltonian simulation ($|\psi(0)\rangle \mapsto|\psi(t)\rangle$), then we need to consider what observables can be extracted. The kinetic and potential energy of a subset of oscillators can be efficiently determined with error, $\epsilon$ and success probability $\delta$ by making $\mathcal{O}(\log (1 / \delta) / \epsilon)$ calls to a simple oracle $\mathcal{V}$. This adds a phase to the particular basis states which encode the (kinetic or potential) energy of the subset of oscillators, $V$. 
For the example case of kinetic energy, $\epsilon$ is defined  as

\begin{equation}
\label{eqn:K_epsilon}
|\hat{k}_V(t)-K_V(t) / E| \leq \epsilon
\end{equation}
where $\hat{k}_V(t)$ is the estimated quantity, $K_V(t)$ is the exact kinetic energy and $E$ is the total energy. This can be done by measuring the projector onto $V$, $P_V=(\mathbbm{1}-\mathcal{V}) / 2$, due to the fact that $K_V(t) / E= \langle\psi(t)| P_V|\psi(t)\rangle$. In fact the problem is essentially that of estimating the expectation value of $\mathcal{V}$, which can be done by measuring many copies of $|\psi(t)\rangle$. The optimal approach is to use high-confidence amplitude estimation \cite{knill2007optimal}. Note that the exact same process is used for the potential energy, but the oracle $\mathcal{V}$ is constructed to mark basis states corresponding to the potential energy contribution. See Sec. \ref{subsec:obs_quants} for more details on extracting other observables of interest.

Now that we have introduced the methods for tackling coupled harmonic oscillator systems, we shift our focus to the physical domains where these mathematical abstractions have proven indispensable. One such area, and the central focus of this paper, is the study of molecular dynamics through Elastic Network Models (ENMs).

The original quantum algorithm for coupled oscillators was designed to demonstrate an exponential quantum advantage, not to necessarily to simulate real physical systems. As a result, it focused on a single spatial dimension, assuming the model could be extended to higher dimensions ($D>1$) simply by scaling to $D$ independent nodes for each node. However, this simplification breaks down in realistic dynamics simulations because motion along different spatial axes is inherently coupled. To capture this physical behavior and reflect the true geometry of the system, we introduce additional quantum oracles (detailed in Sec.~\ref{section:application_to_2d}) specifically designed to handle these multidimensional interdependencies.
\subsection{Elastic network models}\label{subsec:elastic_network_models}

ENMs are particularly useful in fields such as biochemistry, where the structural dynamics of large proteins can be understood \cite{togashi2018coarse, hayward2008normal}. In these models, each atom is represented as a point-like mass while the inter-atomic forces are modeled by spring forces. The advantage of this simplified picture is that the computational cost is significantly lower than a full MD simulation. More involved simulations require prohibitively large computational resources to accurately simulate molecular assemblies due to the need for accurate force fields, and the practical time complexity of state-of-the-art classical algorithms \cite{durrant2011molecular,kim2014vibrational, zhang2016optimized}. 

Despite their simplistic nature, ENMs can provide detailed insight into the slow conformational motions of a wide variety of materials. Each oscillator is placed according to the given reference structure of the molecule, while the connections are determined by the local molecular topology. In the original ENM model proposed by Tirion \cite{tirion1996large}, each atom was replaced by a mass while later, coarse-grained ENMs \cite{bahar1997direct, haliloglu1997gaussian} were introduced. In the latter, instead of mapping every atom into a mass, atoms are grouped (for example, it is common to label a carbon atom bonded to a functional group as $\alpha$-carbon ($C^{\alpha}$)), and each group is treated as an inextricable unit. Connections (springs) between masses are defined given a maximum cutoff distance: masses are only allowed to be connected when their distance is lower than a predefined cutoff.

The fundamental principle of ENMs
is that the potential energy is assumed to be a quadratic function around the optimal energy conformation:
\begin{equation}
    U = \frac{\gamma}{2} \sum_{\substack{i,j \\ r_{ij}<R_c}}(r_{ij} - r_{ij}^0)^2
\end{equation}
where $\gamma$ is the spring constant, $r_{ij} = \Vert\boldsymbol{r}_i - \boldsymbol{r}_j\Vert$ is the distance between masses $i$ and $j$, $r_{ij}^0$ is the distance between atoms in the reference structure,
and $R_c$ is the cut-off distance. This harmonic approximation is the exact starting point of the quantum algorithm described in \cite{babbush2023exponential}, and this similarity drives the motivation to adapt their algorithm to simulate ENMs.

Using this formalism, one can now solve for and extract various properties related to the molecules we wish to study, at a lower computational cost than a full MD simulation. ENMs can be used as a predictive tool to describe collective global motions, and can be used to model near-equilibrium dynamics. As noted in \cite{lezon2009elastic}, the partition function of a system of coupled oscillators is $\mathcal{Z}\sim [\text{det}(\mathbf{F}^{-1})]^{1/2}$, where $\mathbf{F}$ is the Laplacian matrix. As a result, deriving the low-frequency, or else \emph{slow} modes, determine the most probable global fluctuations of a system of oscillators.

A valuable property of ENMs is that it allows to compare several quantities with experimental measurements. For example, B-factors provide a measure of the mean-square fluctuations of individual atoms. These can be calculated using the ENM framework as:
\begin{equation}
    \langle \Delta r_i^2\rangle = \frac{3k_B T}{\gamma}[\mathbf{F}^{-1}]_{ii}
\end{equation}
 and can be compared with real B-Factors obtained through X-ray crystallography. Moreover, while B-Factors give information about how an atom moves, the cross-correlations describe how much atoms move in relation to one another. This is represented by the covariance matrix $C_{ij}$, and its matrix elements can be calculated as:
 \begin{equation}
     \langle \Delta r_i \Delta r_j\rangle = \frac{3k_B T}{\gamma} (\mathbf{F}^{-1})_{ij}.
 \end{equation}
 
Beyond fluctuation properties, ENMs serve as a versatile predictive tool for a range of biophysical phenomena. They are used to identify hinge sites \cite{khade2020characterizing}, i.e. regions where the molecule preferentially bends during conformational transitions, which are directly encoded in the low-frequency normal modes of the Laplacian. ENMs also provide a natural framework for studying allosteric signaling \cite{feher2014computational}; by applying a small perturbation to the force constants at a given network node and analyzing the resulting shift in the eigenvalue spectrum, one can assess which residues most strongly influence the global dynamics of the system. Those nodes whose perturbation produces the largest spectral response are identified as the most likely allosteric mediators \cite{feher2014computational}. Furthermore, ENMs have been successfully applied beyond single proteins, to systems such as membrane proteins and viral capsids, where the collective mechanical properties of large assemblies are of primary interest.

 As such, ENMs provide a very useful tool, compared to computationally expensive (but more accurate) force fields. These models can capture the essential low-frequency dynamics of large molecules at a fraction of the cost. However, when scaling ENMs to model these large molecules, the computational cost of $\mathcal{O}(N^3)$ to diagonalize the Laplacian matrix can become prohibitively expensive, as the number of atoms $N$ grows. One proposed remedy is further coarse-graining by bundling groups of nodes into single effective nodes, but this risks washing out the very structural detail that makes ENMs informative. In the next section, we discuss how the algorithm of Babbush et al. \cite{babbush2023exponential} (see Sec. \ref{subsec:simulating_coupled_classical_oscillators}) can be adapted to the ENM setting, enabling the simulation of very large system of oscillators with a significant reduction in computational cost.

\section{Quantum Elastic Network Models}
\label{sec:QENMs}
As we discussed in Section \ref{subsec:elastic_network_models}, ENMs provide a very useful tool that is used to study the slow structural dynamics of molecules. Instead of mapping each atom in a point-like mass, ENMs usually consider only the $\alpha$-carbon atoms in amino acids. Two atoms are considered connected if their distance is smaller than a predefined cutoff.

In this paper, we show that, if certain conditions (that later are discussed) are satisfied, we can map the classical elastic network model to a quantum mechanical system that can simulate the dynamics of an ENM with speedup over a classical computer. Our results are based on the recent paper \cite{babbush2023exponential} (see \ref{subsec:simulating_coupled_classical_oscillators} for details). In order for this result to hold, certain assumptions are made. We discuss how these assumptions are satisfied in a QENM. The particular quantum advantage depends on the computational experiment that one considers \cite{sakamoto2025quantum}; therefore, in this section we focus on the efficient construction of the oracles required by the quantum algorithm, deferring the analysis of speedups relative to classical methods to the discussion of specific applications outlined in Sec. \ref{sec:graphene}.

Our algorithm is divided into three main steps, all of which are analyzed in the following sections. In Sec. \ref{subsec:encodings} we discuss the different initial states that be chosen as a starting point to the algorithm. Then, we show, similar to an ENM simulation, how the initial positions and velocities can be encoded efficiently in the initial quantum state. Next, in Sec. \ref{subsec:constructing_efficient_oracles}, we explain how to construct efficient oracles for QENM simulation.

\subsection{Encodings}
\label{subsec:encodings}

The algorithm in \cite{babbush2023exponential}, starts by preparing an initial state that is a solution to the Schrödinger equation in Eq. \eqref{eq:schrodinger_eq}. The choice of the initial state depends on the initial conditions but also on what observables we are interested in measuring at the end of the algorithm. The initial quantum state described previously (Eq. \ref{eq:initial_state}) allows for measurement of kinetic or potential energy. A different encoding is required if the user is interested in a subset of the displacements at time $t$. An example of this would be if we want to estimate $B$-factors, Mean-Squared Deviation (MSD), or root mean-squared fluctuations (RMSF). In this case, we can initialize the system in the state:
\begin{equation}
    \ket{\psi(0)} = \frac{1}{\sqrt{2F}}\begin{pmatrix}
        \mathbf{P}\sqrt{\mathbf{M}}\boldsymbol{x}(0)\\
        -i\mathbf{B}^+\mathbf{P}\sqrt{\mathbf{M}}\dot{\boldsymbol{x}}(0)
    \end{pmatrix}
\label{eq:initial_state_alternate}
\end{equation}
where $F$ is a time-conserved quantity ($\dot{F}=0$), $\mathbf{P}$ is the projector on the subspace orthogonal to the null space of $\mathbf{A}$ (or $\mathbf{B}^{\dagger}$).

The authors in \cite{babbush2023exponential} argued that only a polynomial number of (arbitrary) nonzero initial velocities and positions can be chosen, otherwise the initial state preparation would require exponentially large time. One would have to query over all possible values to encode them in a quantum state, eliminating any exponential advantage, but also one would have to calculate the initial energy, which would also take $\mathcal{O}(N)$ time. In Sec. \ref{subsec:loading_intial_velocities}, we argue that the user can still choose an exponentially large number of non-zero velocities, as long as these velocities come from the Maxwell-Boltzmann distribution.

\subsubsection{Loading initial velocities}
\label{subsec:loading_intial_velocities}

\begin{figure}[!ht]
    \centering
    \includegraphics[width=0.45\textwidth]{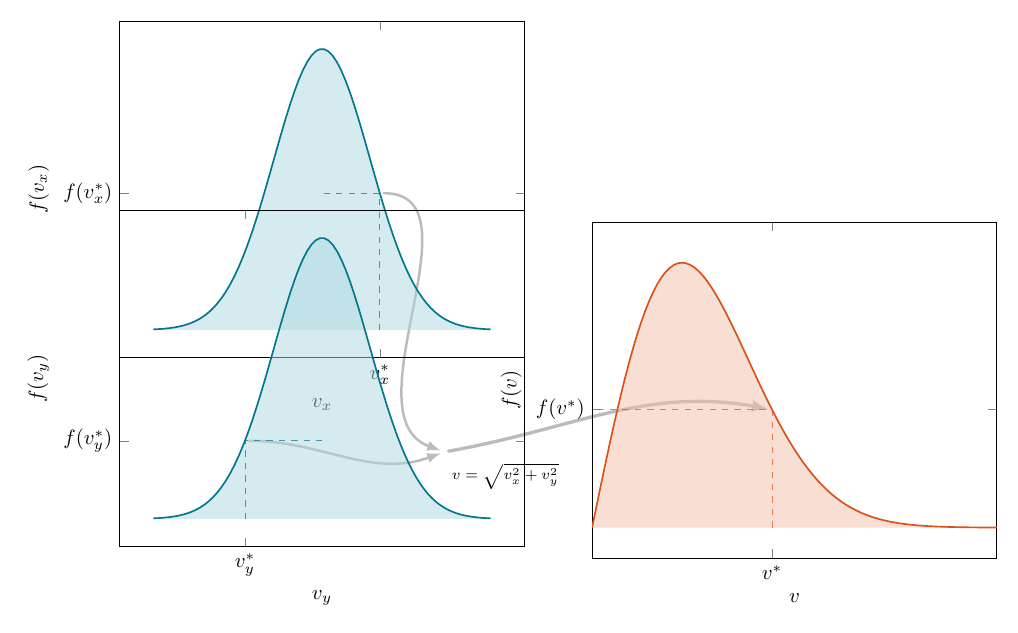}
    \caption{Illustration of sampling from Maxwell-Boltzmann distribution by sampling from $D=2$ independent Gaussians }
    \label{fig:boltzmann-from-independant-gaussians}
\end{figure}

In MD simulations, achieving the target thermodynamic temperature (T) is a fundamental requirement for accurate sampling. Traditionally, there are two main methodologies. The first is to add a thermostat, which acts as a heat bath to exchange energy with the system of interest. This is not ideal in this case, as it requires the addition of virtual atoms and complicates the required connectivity oracle. The second approach is to simply initialize the system in the desired temperature by sampling the initial velocities from the \emph{Maxwell-Boltzmann distribution}. The Maxwell-Boltzmann (MB) probability distribution describes the distribution of velocities for a system of atoms (or molecules) that have reached thermodynamic equilibrium. The probability density function of the velocity (in a 3-dimensional space) for a given mass $m$ is given by

\begin{equation}
f(v) = \Bigg[\frac{m}{2\pi k_B T} \Bigg]^{3/2} 4\pi v^2 e^{-\frac{mv^2}{2k_B T}}
\end{equation}
where $k_B$ is the Boltzmann constant. For each velocity coordinate ($v_x,v_y,v_z$), the MB distribution is a \emph{normal} distribution with $\mu =0$ and standard deviation $\sigma=\sqrt{k_BT/m}$ (see Fig.~\ref{fig:boltzmann-from-independant-gaussians}). Our goal is to be able to prepare efficiently a quantum state
\begin{equation}
  \ket{\psi}_{\mathcal{B}}= \frac{1}{\Big[\sum_{i=0}^{N-1}v_i^2\Big]^{1/2} }\sum_{i=0}^{N-1}v_i \ket{i} 
\label{eq:boltzmann_quantum_state}
\end{equation}

Intuitively, each amplitude $v_i$ of every computational basis state $\ket{i}$ indicates the velocity of atom $i$ (in 1D) that is sampled from the Maxwell-Boltzmann probability distribution. Naively preparing a dense array of floats in the amplitude of the input state would take exponential time. As mentioned, even classically sampling such values would require exponential resources in $n$. We argue that this challenge can be overcome by discretizing the input distribution. For simplicity, we assume uniform mass $m$ across all $N$ nodes (which is true in an ENM simulation if one considers only the $C^{\alpha}$ atoms, or in the graphene simulation that we analyze later in the paper), though this method can accommodate up to $\mathcal{O}(\mathrm{polylog}(N))$ distinct mass values without major changes. Then for each node $j$, the initial velocity $\dot{x}_j(0)$ is sampled independently as:
\begin{equation}    
\dot{x}_j(0) \sim \mathcal{B}_D(m, T) = \mathcal{N}(0, k_BT/m),
\end{equation}
where $\mathcal{B}(m, T)$ is the Maxwell-Boltzmann distribution for a mass $m$ at temperature $T$, and $\mathcal{N}(0, k_BT/m)$ is the normal distribution with mean $\mu=0$ and variance $\sigma^2 = k_BT/m$.

Classically sampling $N = 2^n$ independent velocities and encoding them into amplitudes of a quantum state would require $\mathcal{O}(2^n)$ time and memory, violating our aim of $polylog(N)$ resource scaling. Instead, we propose a discretized strategy: rather than sampling velocities directly, we assign each node to one of $k$ coarse-grained velocity buckets, chosen such that the bucket distribution approximates a Maxwell-Boltzmann distribution. 

In our case, we denote $\mathcal{B}_D^k(m,T)$ the discrete approximation of the Maxwell-Boltzmann distribution into $k$ distinct velocity buckets, where each mass can be assigned to each of the $k$ distinct buckets with a given probability $P_i$. To construct this distribution, we first divide the velocity interval $[-v_{\text{max}}, v_{\text{max}}]$  into $k$ (not necessarily equally spaced) intervals\footnote{You can think of $v_{\max}>0$ as a user's choice for the maximum available velocity for the simulation. In practice, it can be chosen so that $\int_{v_{\max}}^\infty f(v)dv = \int_{-\infty}^{-v_{\max}}f(v)dv \approx 0$.}. The probability that the mass $m$ has a velocity within a given interval $[v_i, v_{i+1}]$ is fixed, and is given as $P_i = \int_{v_i}^{v_{i+1}}f(v)dv$. Clearly, the total probability sums to 1 since:
\begin{equation*}
    \sum_{i=1}^k P_i = \sum_{i=1}^k\int_{v_i}^{v_{i+1}}f(v)dv = \int_{-v_{\text{max}}}^{v_{\max}}f(v)dv \approx \int_{-\infty}^{\infty}f(v)dv = 1
\end{equation*}

\begin{figure}[!ht]
    \centering
    \includegraphics[width=0.45\textwidth]{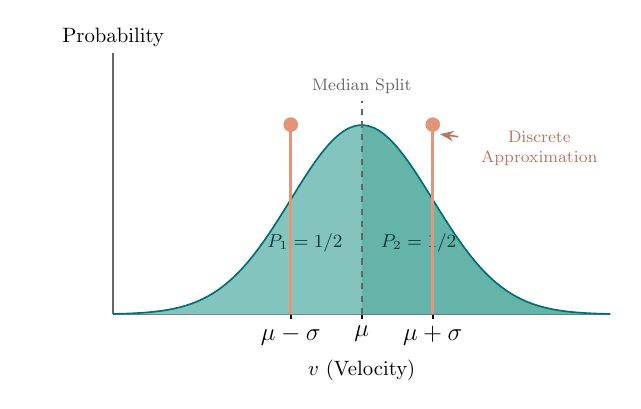}
\caption{\textbf{Two-point approximation of the Maxwell-Boltzmann distribution.} The probability mass is partitioned exactly at the median velocity to ensure equiprobable buckets ($P_0=P_1=0.5$). The representative velocities $\tilde{v}_0$ and $\tilde{v}_1$ are chosen symmetrically around the mean ($\mu \pm \sigma$) to strictly conserve the system's kinetic energy and momentum moments.}
\label{fig:discretized-boltzmann}
\end{figure}

A natural question that arises is how one can choose a representative velocity $\tilde{v}_i$ for each bucket $i$ and how this is connected to the approximation error of the discrete probability distribution. To answer this question, we note that the error between the continuous and the discrete probability distribution can be quantified by the error in their respective moments \footnote{The $n$-th moment is defined as $\mu_n = \int_{-\infty}^{\infty} x^n f(x) \, dx$,  when we assume zero mean.}. In this paper, we only consider up to the second moment. We start by defining the discrete probability distribution function as:
\begin{equation}
    f_D^k(v) = \sum_{i=1}^k P_i \delta(v-\tilde{v}_i)
\end{equation}
where $\tilde{v}_i$ is the velocity of the bucket $i$, and $\delta$ is the Dirac delta function. It is easy to see that:
\begin{equation}
    \int_{-\infty}^{\infty}f_D^k(v)dv = \int_{-\infty}^{\infty}\sum_{i=1}^kP_i \delta(v-\tilde{v}_i) dv = \sum_{i=1}^k P_i\int_{-\infty}^\infty \delta(v-\tilde{v}_i)dv = 1
\end{equation}
where we used the delta function property that $\int_{-\infty}^\infty \delta(v-\tilde{v}_i)dv = 1$. As such, the 0-moments of the distribution are equal. The first moment of the distribution can provide guidance on how to choose the velocities of each bucket. If we want the first moments to match, we can calculate the corresponding velocities $\tilde{v}_i$ so that the following is satisfied:
\begin{equation}
    \int_{v_i}^{v_{i+1}}vf(v)dv = \int_{v_i}^{v_{i+1}}v'f_D^k(v')dv' \text{   for all intervals $[v_i, v_{i+1}]$}
\end{equation}
The right hand side can be rewritten as:
\begin{equation}
\int_{v_i}^{v_{i+1}}v'f_D^k(v')dv' = \int_{v_i}^{v_{i+1}}v'\sum_{j=1}^k P_j\delta(v'-\tilde{v}_j)dv' = \sum_{j=1}^k P_j \int_{v_i}^{v_{i+1}}v'\delta(v'-\tilde{v}_j)dv' = P_i\tilde{v}_i
\end{equation}
Thus, in order for the first moments to match, the velocities can be chosen according to the rule:
\begin{equation}
    \tilde{v}_i = \frac{\int_{v_i}^{v_{i+1}}vf(v)dv}{P_i}
\end{equation}
The analysis can be slightly more complicated when we focus on second, or higher order moments. In this paper, we are only interested to match the distribution up to the second moment, which is related to the kinetic energy. In that case, we need to solve the following non-linear system to derive the corresponding velocities and their respective probabilities. The system is derived by demanding that all moments (up to the second order) match, i.e: 
\begin{equation}
\begin{gathered}
    P_1 + P_2 + \ldots P_k = \int_{-\infty}^{\infty}v^0f(v)dv = 1 \\
    P_1\tilde{v}_0 + P_2\tilde{v}_1 + \ldots P_k \tilde{v}_k = \int_{-\infty}^{\infty}vf(v)dv = 0 \\
    P_1 \tilde{v}_0^2 + P_2 \tilde{v}_1^2 + \ldots + P_k \tilde{v}_k^2 = \int_{-\infty}^{\infty}v^2f(v)dv = \frac{k_B T}{m} 
\end{gathered}
\end{equation}
It turns out that if we are interested in matching only up to the second moment, a discretization into $k=2$ buckets is sufficient. A choice of the corresponding parameters can easily be verified to be $P_1=P_2 = 1/2$, $\tilde{v}_1=\mu+\sigma$, and $\tilde{v}_2 = \mu-\sigma$, i.e. if we perform a median split (see Fig. \ref{fig:discretized-boltzmann}). On the other hand, if we are interested in matching higher-order moments, larger number of discretization buckets is necessary. However, due to the symmetry of the normal distribution, our approximation matches the continuous Boltzmann distribution up to the third moment. \footnote{To be exact, our approximation matches the continuous Maxwell-Boltzmann distribution for all odd moments, since all of them are zero as $f(v)$ is an even function.} 

Note that the above discretization choice is not unique. The resulting errors (for each choice) can be quantified by measuring the discrepancy between each discretization and the higher-order moments of the original distribution. But in the case where we are interested in matching only the first two moments, the two different choices are indistinguishable. In this paper, we analyze how to discretize the continuous Maxwell-Boltzmann to $k$ \emph{equiprobable} velocity buckets and provide an algorithm to encode the $2^n$ samples from the $\mathcal{B}^k_D(m,T)$ distribution onto a quantum state (see Eq. \eqref{eq:boltzmann_quantum_state}) efficiently.

We now detail our method to load an exponentially large velocity vector into our quantum state. This is required to prepare the initial state for the QENM simulation. Rather than storing a random distribution in classical memory (which scales linearly with system size), we generate the velocities procedurally using quantum arithmetic. Here, we restrict our analysis to sampling from 2 buckets, for reasons detailed earlier in this section. We classify nodes into two buckets $v_0$ and $v_1$ respectively. The state preparation proceeds as follows:

\begin{figure}[!ht]
\centering
    \includegraphics[width=0.75\textwidth]{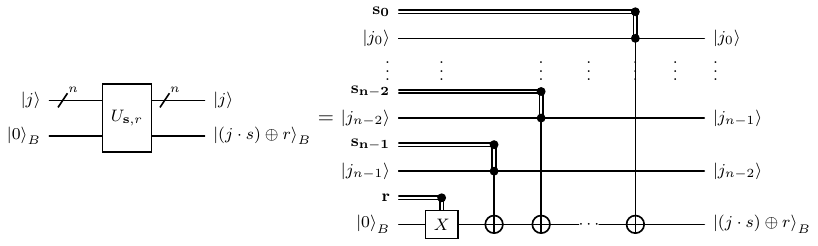}
    \vspace{0.25cm}
\caption{\emph{Circuit for randomized bucket assigment}: Given a randomly sampled $\theta=(s,r) \in \{0,1\}^{n+1}$, and an $n$ qubit quantum register $\ket{j}$, the above construction computes $(j\cdot s) \oplus r$, where the dot product is in $\mathbb{F}_2$}
\label{fig:randomized-bucket-assignment}
\end{figure}

\paragraph{1. Randomized Bucket Assignment}
We assign each node index $j$ to a bucket $b_j \in \{0, 1\}$ using a randomized parity check. We select a random key $\theta=(s,r)$ consisting of an $n$-bit string $s$ and a single bit $r$. The bucket index is computed as the dot product of the node index and the string $s$:
\begin{equation}
    b_j = (j \cdot s) \oplus r \pmod 2
\end{equation}
This is implemented efficiently using a sequence of CNOT gates classically controlled on the bits of $s$:
\begin{align}
    \ket{j}\ket{0}_B \xrightarrow{X^r} \ket{j}\ket{r}_B \xrightarrow{\text{CNOTs}} \ket{j}\ket{r \oplus \bigoplus_{k: s_k=1} j_k}_B = \ket{j}\ket{b_j}_B
\end{align}
If the key is chosen uniformly randomly, the assignment is uniformly random because for any nonzero input $i$, the parity of a uniformly random subset of its bits (i.e. the result of the dot product) is itself uniformly random, and the inclusion of a global random offset $r$ ensures exact uniformity even for $j=0$. Refer to Fig.~\ref{fig:randomized-bucket-assignment} for the circuit to randomly assign velocity buckets.

\begin{figure}[!ht]
\centering
    \includegraphics[width=0.75\textwidth]{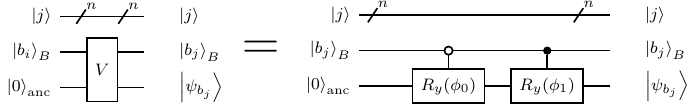}
    \vspace{0.25cm}
\caption{\textbf{Velocity Amplitude Encoding Circuit.} The abstract operation $V$ (left) is implemented via controlled rotations (right). The flag qubit $\ket{b_j}$ selects between rotation angles $\phi_0$ and $\phi_1$, mapping the normalized velocity $\tilde{v}$ into the ancilla amplitudes such that $\ket{\psi_{b_j}} = \sqrt{1 - \tilde{v}_{b_j}^2} \ket{0}_{\text{anc}} + \tilde{v}_{b_j} \ket{1}_{\text{anc}}$.}\label{fig:velocity-amplitude-encoding}
\end{figure}

\paragraph{2. Velocity Amplitude Encoding}
To load the velocities into the amplitudes, we bypass the arithmetic complexity of full inequality testing. Since we only have two buckets, the velocity is strictly determined by the single flag qubit $\ket{b_j}$. We apply a controlled rotation $R_y(\phi)$ on an auxiliary qubit, where the angles $\phi_0$ and $\phi_1$ are chosen such that $\sin(\phi_x) \propto v_x$ (see Fig.~\ref{fig:velocity-amplitude-encoding}):
\begin{equation}
    \ket{j}\ket{b_j}\ket{0}_{\text{anc}} \xrightarrow{CR_y} \ket{j}\ket{b_j} \left( \sqrt{1 - \tilde{v}_{b_j}^2} \ket{0}_{\text{anc}} + \tilde{v}_{b_j} \ket{1}_{\text{anc}} \right)
\end{equation}
where $\tilde{v}$ are the normalized velocities. Measuring the ancilla in state $\ket{1}$ projects the system onto the target state $\sum v_{b_j} \ket{j}$ with probability proportional to the kinetic energy. This step is preceded by Amplitude Amplification.

\paragraph{3. Uncomputation and Complexity}
To finalize the state preparation, we must disentangle the auxiliary registers. We apply the inverse of the \textbf{Randomized Bucket Assignment} (Step 1). This returns the bucket registers to $\ket{0}$, leaving the system in the required state $\ket{\psi(0)} \propto \sum_j v_j \ket{j}$ (see Fig.~\ref{fig:randomized-bucket-assignment}).

\emph{Note: The above steps would have to be repeated for each spatial dimension in our system. The steps are controlled on the qubits representing the spatial dimensions.}

\paragraph{4. Generalized Loading}
The previously described steps can be generalised to $n_b=2^k$ equiprobable buckets. We generate a $k$-bit bucket index $\ket{B_i}$ by executing Step 1 $k$ times independently using independent random keys $\{\theta_0, \dots, \theta_{k-1}\}$. Controlled on this index, we load the representative velocity into an ancilla register $\ket{V}$ using a logic-based quantum lookup table:
\begin{equation}
    \ket{j}\ket{B_j}\ket{0}_V \xrightarrow{\text{Lookup}} \ket{j}\ket{B_j} \ket{v_{B_j}}_V
\end{equation}
where $v_{B_i} \in \{v_0, \dots, v_{n_b-1} \}$. Using this value register, we can now perform standard inequality testing as detailed in~\cite{babbush2023exponential} to encode the amplitudes. Finally, we uncompute $B_i$ to reset the bucket index register.

\subsubsection{Loading initial displacements}
\label{subsection:loading_initial_displacements}

The type of initial displacements depends on the practitioner's preferences \cite{frenkel2023understanding, rapaport2004art}. Our goal is to prepare a quantum state $\ket{\psi(0)} \sim \sum_{i=0}^{N-1}x_i(0)\ket{i}$, where $x_i(0)$ are the displacements (in 1D) of each oscillator at time $t=0$. The choice of the initial displacements will determine the complexity of the initial state preparation.

First of all, the simplest choice is to initialize all atoms in their rest positions (i.e. the reference structure) \cite{bacstuug2012molecular}. This corresponds to a starting displacement $x_i(0) =0$ for all masses. As we later show, this choice is the simplest and requires the least amount of gates in order to prepare the initial state in Eq. \eqref{eq:initial_state}.

A second type of initial condition that can be chosen, is to add small random perturbations on the initial positions of at most $\mathcal{O}(\text{poly}\log(N))$ atoms. These perturbations are usually chosen to be $ \leq 1\si{\angstrom}$ \cite{gur2013global}. In this case, the quantum state can be efficiently prepared through the use of lookup tables. We outline the procedure below:
\begin{enumerate}
    \item We initialize the system in the uniform superposition state $\frac{1}{\sqrt{N}}\sum_{i=0}^{N-1}\ket{i}$.

    \item We employ an ancilla register and use a predefined lookup table to encode displacements as:
    \begin{equation}
        \frac{1}{\sqrt{N}}\sum_{i=0}^{N-1}\ket{i} \ket{0} \xrightarrow{\text{Lookup Table}} \frac{1}{\sqrt{N}}\sum_{i=0}^{N-1}\ket{i}\ket{x_i(0)}
    \end{equation}

    \item Finally, we perform inequality testing \cite{sanders2019black} to get:
    \begin{equation}
        \frac{1}{\sqrt{N}}\sum_{i=0}^{N-1}\ket{i}\ket{x_i(0)} \xrightarrow{\text{Inequality Test}} \frac{1}{\sqrt{\sum_{i=0}^{N-1}x_i^2(0)}}\sum_{i=0}^{N-1}x_i(0)\ket{i}
    \end{equation}
    
\end{enumerate}

Other types of initialization can be also be used. For example, in more expensive ENM simulations \cite{costa2015exploring}, the lowest normal modes are used to perturb the atoms from the initial positions. However, such an initialization would require exponential classical resources to obtain the lowest normal modes of the system.

\subsection{Constructing efficient oracles}
\label{subsec:constructing_efficient_oracles}

In the original \cite{babbush2023exponential} paper, the authors assumed that several operations can be performed efficiently on a quantum computer. Some of these procedures are, for example, the encoding of the initial positions and velocities in a quantum state. We thoroughly analyzed how this can be done efficiently in the case of a QENM simulation in the previous subsections.

In order to prepare the quantum state in Eq. \eqref{eq:initial_state} and then simulate the system under the unitary induced by the Hamiltonian in Eq. \eqref{eq:problem_hamiltonian}, several quantum operators (\emph{oracles}) should be efficiently implementable. Examples of such oracles are, for example, those that encode the mass or connectivity of the problem. For highly unstructured organic molecules, efficiently realizing these oracles is non-trivial, if even possible. In this paper, we restrict ourselves to chemical problems with underlying structure that can be exploited.

The efficiency (and practicality) of a quantum algorithm depends on the resources required to implement each quantum oracle -- an oracle that requires an exponential number of resources would eliminate any quantum advantage\footnote{Or more precisely, it would eliminate any hope for exponential quantum advantage, while it may still be possible to offer polynomial quantum advantage.}. At this point, we will discuss the efficiency of the aforementioned oracles.\\

\noindent \textbf{Mass oracle:}\\

The first type of oracle that is used in this work is the mass oracle that performs the transformation:
\begin{equation*}   \ket{j,z}\xrightarrow{\mathcal{O}_m}\ket{j,z\oplus \Bar{m}_j}
\label{eq:mass_oracle}
\end{equation*}
where $\Bar{m}_j$ is the bitstring representation of the mass $m_j$.  In order for such a circuit to be efficient, the function $\bar{m}_j : [N] \rightarrow \{0,1\}^n$ must be a function that can be implemented with \emph{at most} $\mathcal{O}(\text{poly}\log(N))$ number of gates. In the case of ENMs, and in general for molecular dynamics, the number of different masses is fixed. More precisely, in ENMs such as Gaussian Network Models (GNMs) \cite{bahar1997direct} or Anisotropic Network Models (ANMs) \cite{eyal2006anisotropic}, we assume that all masses are equal and correspond to $C^{\alpha}$ atoms with ($m_{C} = 12\si{amu}$). This also holds for the graphene simulation that we analyze in Sec. \ref{sec:graphene}.

In this case, constructing the circuit that implements the mass oracle in Eq. \eqref{eq:mass_oracle} is rather easy. Specifically, the circuit can be executed by directly applying $X$ gates on the second register:
\begin{equation}
    \ket{j}\ket{z} \xrightarrow{X^{\bar{m}_{C^{\alpha}}}} \ket{j}\ket{z\oplus \bar{m}_{C^{\alpha}}}
\end{equation}
with the number of $X$ gates specified by the bitsring $\bar{m}_{C^{\alpha}}$.\\

\noindent \textbf{Connectivity Oracle:}\\

Implementing the connectivity oracle is not a trivial task and requires some form of structure of the underlying molecule. The connectivity oracle is defined as
\begin{equation}
    \ket{j, \ell} \xrightarrow{\mathcal{O}_l} \ket{j, a(j,\ell)}
\label{eq:connectivity_oracle}
\end{equation}
where $a(j,\ell)$ is the column index of the $\ell$-th nonzero entry in the $j$-th row of the coupling coefficient matrix $\mathbf{K}$. In the case where atoms are structured in a grid or a chain, constructing this oracle requires simple quantum arithmetic and a small overhead in qubits, while in other cases it can be more complicated. In the next section, we choose graphene due to it exhibiting a periodic structure and detail how to construct the connectivity oracle in Sec. \ref{sec:connectivity_oracle}.\\

\noindent \textbf{Further assumptions:}\\

As it was noted in \cite{babbush2023exponential}, preparing the initial state of Eq. \eqref{eq:initial_state} for a system of different masses, or for a system with many different coupling coefficients requires the user (at some point of the algorithm) to be able to efficiently perform the transformation:
\begin{equation}
    \ket{0}_1 \rightarrow \frac{1}{\sqrt{m_{\max}\alpha^2 + 2\kappa_{\max} d \beta^2}}\Big(\sqrt{m_{\max}}\alpha\ket{0}_1 + i\sqrt{2\kappa_{\max} d}\beta \ket{1}_1\Big)
\label{eq:single_qubit_transformation}
\end{equation}
where $\alpha^2 = \sum_{j=1}^N\dot{x}_j(0)^2$, $\beta^2 = \sum_{i=1}^N x_j(0)^2$, $m_{\max}$ is the maximum mass of the system, and $\kappa_{\max}$ is the maximum coupling strength. Although in certain type of simulations, such as ENMs (GNNs and ANMs), all masses and coupling coefficients are assumed to be equal, in more complex ENMs models the coupling strength may differ between different interactions. For example, in \cite{yang2009protein}, the authors introduced distance-dependent coupling strengths that give better predictions to crystallographic $B$-factors

As such, this can impose a practical constraint, if one cannot estimate $\alpha, \beta$ (or their approximations) efficiently. As we discussed in Sec. \ref{subsection:loading_initial_displacements}, because the number of non-zero displacements are chosen to be at most $\mathcal{O}(\text{poly}\log(N))$, then $\beta$ can be estimated efficiently. However, in the case of Maxwell-Boltzmann velocities initialization, there are $\mathcal{O}(\exp(n))$ non-zero initial velocities, thus calculating $\alpha$ by querying each velocity is inefficient. Lemma \ref{lemma:relative_fluctuations} aims to address this concern.

\begin{lemma}
    Consider a system of $N$ atoms of mass $m$ in $D$ dimensions, whose velocities are sampled from the Maxwell-Boltzmann distribution. Then, the relative fluctuations of the kinetic energy is:
    \begin{equation}
        \frac{\sigma_K}{\langle K\rangle} = \sqrt{\frac{2}{DN}}
    \end{equation}
    where $\sigma_K$, $\langle K\rangle$ is the standard deviation and mean of the kinetic energy $K$ respectively.
\label{lemma:relative_fluctuations}
\end{lemma}

\begin{proof}
     We start by calculating the average kinetic energy:
    \begin{equation}
        \label{eqn:avg_K_boltzmann}
        \langle K \rangle = \frac{m}{2}\sum_{j=0} ^{N-1} \langle v_j^2\rangle = \frac{Nm}{2} \int v^2f(v)dv = \frac{NDk_BT}{2}
    \end{equation}
    where we used the fact that $\langle v^2 \rangle = \int v^2f(v)dv = \frac{Dk_B T}{m}$, for a mass $m$ in $D$ dimensions.  If we then divide the kinetic energy by the variance of the random velocities, we observe that the (standardized) kinetic energy is the sum of $N$ random normal variables and thus can be described by the $\chi^2_{DN}$ random distribution whose mean and variance are known. Thus:
    \begin{equation}
        \text{Var}\Bigg(\frac{K}{k_B T}\Bigg)= 2DN \implies \text{Var}(K) =\frac{DN}{2}(k_BT)^2
    \end{equation}
    which implies that the standard deviation is $\sigma_K = \sqrt{\frac{DN}{2}} k_B T$. As such, we have proven our result.
\end{proof}

The result in Lemma \ref{lemma:relative_fluctuations} implies that for systems of large size (also known as thermodynamic limit), the relative fluctuations are practically zero, and we can replace the actual kinetic energy by its mean $\langle K\rangle$. Thus, in our case, $\sqrt{m} \alpha = \sqrt{2K} \approx \sqrt{2\langle K \rangle}$ for large $N$. This is in agreement with the law of large numbers, since increasing $N$, increases the number of independent samples from the Maxwell-Boltzmann distribution, thus converging to the true average.

However, in our case, we do not sample from the continuous Maxwell-Boltzmann distribution, but from $\mathcal{B}^2_D(m,T)$ which approximates the Boltzmann distribution up to the second moment. On top of that, due to its construction, all atoms will have velocity either $\mu-\sigma$ or $\mu+\sigma$ with equal probability. This implies the following Corollary.

\begin{corollary}
    For the probability distribution $\mathcal{B}^2_D(m,T)$, that approximates the Maxwell-Boltzmann distribution up to second moment, we have that $\alpha = \sqrt{2\langle K \rangle } = \sqrt{NDk_BT}$.
\end{corollary}
We can then conclude that executing the transformation in Eq. \eqref{eq:single_qubit_transformation}, can be done by applying an $R_y(\theta)$ rotation, with $\theta = \text{arc} \cos \Big(\frac{\sqrt{2\langle K \rangle }}{\sqrt{2 \langle K \rangle+ 2\kappa d\beta^2}}\Big)$, and then by applying an $S$ gate.

\section{Application to 2D materials}
\label{section:application_to_2d}

In \cite{babbush2023exponential}, the authors demonstrate the mapping of 1D coupled oscillator dynamics onto the Schrödinger equation. As noted earlier, extending this framework to higher dimensions, however, introduces cross-coordinate coupling. In a 2D lattice, for instance, the longitudinal and transverse degrees of freedom are no longer independent; a displacement in the $y$-direction inherently alters the restorative forces and subsequent motion in the $x$-direction due to the geometric constraints of the lattice. In this section, we analyze how the algorithm of \cite{babbush2023exponential} can be extended in $D>1$ dimensions, in which the motions in each coordinate are \emph{coupled}, and we explain how the QENM algorithm can be applied to model 2D materials. We focus on graphene for simplicity, but our procedure can easily be extended to represent other classes of two-dimensional materials, such as carbides or nitrides. 

As we discussed earlier, the procedure to load the initial velocities and positions is similar for different molecules, and depends on the practitioner's preferences. However, the methodology to apply the connectivity oracle is different on every molecule and depends on its structure. In the worst case, where the target molecule exhibits no structure, constructing the connectivity oracle requires $\Omega(N)$ time. If, on the other hand, the molecule exhibits periodicity or symmetries (e.g. crystalline molecules), then it is easier to construct these circuits by using a unit cell framework and quantum arithmetic (see our analysis in Sec. \ref{sec:connectivity_oracle}).

We are interested in simulating the dynamics of a large \emph{graphene} sheet. We assume $N=2^n$ point-like carbon atoms of mass $m_C$ that are coupled with each other by spring-like forces. At any $t\geq 0$, the position of a mass $i$ is described by the vector $\mathbf{r}_i(t) = (x_i(t), y_i(t))$, since the graphene sheet is a $2D$ material in a $3$-dimensional space. The two coordinates describe the displacements in the $x$ and $y$ coordinates respectively. Our analysis below can easily be generalized in the case of $D=3$ dimensions.

For a system of $N$ masses in 2D with equal coupling strengths $\kappa$, the total harmonic potential can be written as:
\begin{equation}
    U = \sum_{i,j} U_{ij} =  \frac{\kappa}{2} \sum_{\substack{i,j \\ \text{i,j connected}}}(r_{ij} - r_{ij}^0)^2
\end{equation}
where $r_{ij} = |\boldsymbol{r}_i - \boldsymbol{r}_j| = \sqrt{(x_i - x_j)^2 + (y_i-y_j)^2}$ is the distance between atoms $i$ and $j$, $r_{ij}^0$ is the distance between atoms in the reference structure, or else the \emph{bond length}, and $\boldsymbol{r}_i$ is the absolute coordinate of the mass $i$. The (absolute) equilibrium coordinates of any mass $i$ in the reference structure are denoted as $(x_i^0, y_i^0)$. The bond length is then defined as
\begin{equation}
   r_{ij}^0 = |\boldsymbol{r}_i^0 - \boldsymbol{r}_j^0| = \sqrt{(x_i^0 - x_j^0)^2 + (y_i^0-y_j^0)^2}.
\end{equation}
In the case of graphene, we can assume that, without loss of generality, $r_{ij}^0 = a$ for all connected $i,j$. The algorithm can be easily extended in the case where different atom types have different bond lengths. Thus, the total potential can be written as:
\begin{equation}
    U =\frac{\kappa}{2} \sum_{i=0}^{N-1}\sum_{j\in \text{neighbors($i$)}}(r_{ij} - a)^2
\end{equation}
The dynamics of each carbon atom $i$ are described by Newton's equation:
\begin{equation}
    m_i \ddot{\boldsymbol{u}}_i = \boldsymbol{F}_i
\label{eq:Newtons_equation_single_mass_2d}
\end{equation}
where $\boldsymbol{u}_i = (u_{i,x}, u_{i,y})$ are the relative displacements of the carbon atom $i$ in the $x$ and $y$ direction respectively:
\begin{equation}
    u_{x,i} = x_i - x_i^0, \; u_{y,i} = y_i - y_i^0
\end{equation}
Since we know the total potential, we can calculate the force $\boldsymbol{F}_i$ as:
\begin{equation}
    \boldsymbol{F}_i = -\nabla_i U = -\Bigg(\frac{\partial U}{\partial x_i}\hat{\boldsymbol{x}} + \frac{\partial U}{\partial y_i} \hat{\boldsymbol{y}}\Bigg) = -\kappa \sum_{j\in \text{neighbors($i$)}} (r_{ij}-a)\hat{\boldsymbol{n}}_{ij}
\end{equation}
where $\hat{\boldsymbol{n}}_{ij} = \frac{\boldsymbol{r}_i - \boldsymbol{r}_j}{r_{ij}}$ is the unit vector pointing in the direction that connects masses $i$ and $j$. The force $\boldsymbol{F}_i$ can be decomposed onto its $x$ and $y$ components as:
\begin{equation}
\begin{gathered}
    F_{i,x} = -\kappa \sum_{j\in \text{neighbors($i$)}} (r_{ij}-a)\frac{x_i-x_j}{r_{ij}} \\
    F_{i,y} = -\kappa \sum_{j\in \text{neighbors($i$)}} (r_{ij}-a)\frac{y_i-y_j}{r_{ij}}
\end{gathered}
\label{eq:force_components}
\end{equation}
The forces can be rewritten in a more convenient form if we assume that the displacements of the oscillators are much smaller than the bond length $a$. In that case, we have that:
\begin{gather}
    x_i^0 - x_j^0 = a\cos \theta_{ij} \\
    y_i^0 - y_j^0 = a \sin \theta_{ij}
\end{gather}
where $\theta_{ij} = \arccos [(x_i^0-x_j^0)/a]$. Under a small perturbation from the reference structure, we can write:
\begin{gather}
    x_i - x_j = (x_i^0 - x_j^0) + \Delta x = a\cos \theta_{ij} + \Delta x \\
    y_i - y_j = (y_i^0 -y_j^0) + \Delta y = a\sin \theta_{ij} + \Delta y
\end{gather}
where $\Delta x = u_{x,i} - u_{x,j}$ and $\Delta y = u_{y,i} - u_{y,j}$. Thus, the distance $r_{ij}$ can be rewritten as:
\begin{gather*}
    r_{ij} = |\boldsymbol{r}_i - \boldsymbol{r}_j| = \sqrt{(x_i - x_j)^2 + (y_i-y_j)^2} 
    = \sqrt{(a\cos \theta_{ij} + \Delta x)^2 + (a\sin \theta_{ij} + \Delta y)^2} \\
    = \sqrt{a^2 \cos^2 \theta_{ij} + 2a\cos\theta_{ij} \Delta x + \Delta x^2 + a^2 \sin^2 \theta_{ij} + 2a\sin\theta_{ij} \Delta y + \Delta y^2} \\ = \sqrt{a^2 + 2a(\Delta x \cos \theta_{ij} + \Delta y \sin \theta_{ij}) + \Delta x^2 + \Delta y^2}
\end{gather*}
Since $\Delta x, \Delta y\ll a$, we can neglect the second order terms, and if we use the binomial approximation, we get:
\begin{equation}
    r_{ij} \approx a\Bigg(1 + \frac{2a(\Delta x \cos \theta_{ij} + \Delta y \sin\theta_{ij})}{2a^2} \Bigg) = a + (\Delta x \cos \theta_{ij} + \Delta y \sin \theta_{ij})
\end{equation}
and thus \emph{the stretch} of the spring is:
\begin{equation}
    r_{ij} - a \approx \Delta x \cos \theta_{ij} + \Delta y \sin \theta_{ij}
\end{equation}
If we look at the unit vector components $\frac{x_i-x_j}{r_{ij}}$ and $\frac{y_i-y_j}{r_{ij}}$ in Eq. \eqref{eq:force_components}, the change in direction is negligible for the force calculation and thus:
\begin{equation}
\begin{gathered}
    \frac{x_i - x_j}{r_{ij}} = \frac{a\cos \theta_{ij} + \Delta x}{a} \approx \cos \theta_{ij} \\
    \frac{y_i - y_j}{r_{ij}} = \frac{a\sin \theta_{ij} + \Delta y}{a} \approx \sin \theta_{ij}
\end{gathered}
\end{equation}
Putting it all together, the forces in Eq. \eqref{eq:force_components} are written as:
\begin{equation}
\begin{gathered}
    F_{i,x} = -\kappa \sum_{j\in \text{neighbors($i$)}} \Big[(u_{x,i} - u_{x,j})\cos^2 \theta_{ij} + (u_{y,i} - u_{y,j}) \cos\theta_{ij} \sin \theta_{ij} \Big] \\
    F_{i,y} = -\kappa \sum_{j\in \text{neighbors($i$)}} \Big[(u_{x,i} - u_{x,j})\cos \theta_{ij} \sin \theta_{ij} + (u_{y,i} - u_{y,j}) \sin^2 \theta_{ij} \Big]
\end{gathered}
\label{eq:forces_graphene}
\end{equation}
As we can see in Eq. \eqref{eq:force_components} the movements in the $x$ and $y$ axes cannot be analyzed independently since forces on the $x$ direction depend on the relative displacements on the $y$ axis and vice versa (i.e. the system cannot be decoupled). Moreover, in our analysis we can assume that the angles $\theta_{ij}$ remain constant throughout the evolution. This is true, for example, when the displacements are much smaller compared to the bond length.

 Hence, we can rewrite the Newton's equation (Eq. \eqref{eq:Newtons_equation_single_mass_2d}) for a given mass $i$ in matrix form as:
\begin{equation}
    m_i \ddot{\boldsymbol{u}}_i = -\sum_{j\in \text{neighbors($i$)}}\mathbf{k}_{ij} (\boldsymbol{u_i}-\boldsymbol{u_j})
\end{equation}
where $\mathbf{k}_{ij}$ is defined as:
\begin{equation}
    \mathbf{k}_{ij} = \kappa \begin{pmatrix}
        \cos^2 \theta_{ij} & \cos \theta_{ij}\sin \theta_{ij}\\
        \sin \theta_{ij} \cos \theta_{ij} & \sin^2 \theta_{ij}
    \end{pmatrix}
\end{equation}
We can now proceed and write Newton's equation for the total system as:
\begin{equation}
    \mathbf{M} \ddot{\boldsymbol{u}} = -\mathbf{F} \mathbf{u}
\end{equation}
where $\boldsymbol{u} = (u_{x,1}, u_{y,1}, u_{x,2}, u_{y,2}, \ldots, u_{x,N}, u_{y,N})$ is the displacement vector for the whole system, $\mathbf{M} = \text{diag}(m_0, m_0, m_1, m_1, \ldots)$ is the diagonal mass matrix with double the number of masses, and $\mathbf{F}$ is the Laplacian matrix with elements comprised of $2\times 2$ blocks:
\begin{equation}
    \mathbf{F}_{ij} = \begin{cases}
        \sum_{j\in \text{neighbors($i$)}}\mathbf{k}_{ij} \text{ if $i=j$} \\
        -\mathbf{k}_{ij} \text{ if $i\neq j$ and $j\in \text{neighbors($i$)}$}\\
        \mathbf{0} \text{ otherwise}
    \end{cases}
\end{equation}

One interesting thing about graphene is that it exhibits a well-defined structure. As we later discuss in Sec. \ref{sec:connectivity_oracle}, the graphene molecule can be divided into unit cells comprised of $A$ and $B$ type atoms. In both cases, the angles are known and are $\theta_{ij} \in \{\pi/6, 5\pi/6, 3\pi/2\}$ or $\theta_{ij} \in \{\pi/2, 7\pi/6, 11\pi/6\}$, whether atom $i$ corresponds to an $A$ or $B$ atom respectively. The same applies for the atoms in the boundaries, but are connected to one less carbon atom.

Next, we need to find a matrix $\mathbf{B}$, that satisfies $\mathbf{B}\mathbf{B^\dagger} = \mathbf{A}$. However, in contrast to the 1D case, there are certain coupling coefficients that are negative. For this reason, we cannot use the same methodology as in \cite{babbush2023exponential}, but we still need to find a matrix $\mathbf{B}$ such that $\mathbf{B}\mathbf{B}^\dagger = \mathbf{A}$. The first thing that we notice is that matrices $\boldsymbol{k}_{ij}$ are rank-1 matrices, since $\boldsymbol{k}_{ij} = \boldsymbol{v}\boldsymbol{v}^T$ for $\boldsymbol{v} = \sqrt{\kappa}\begin{pmatrix}
    \cos \theta_{ij} \\
    \sin \theta_{ij}
\end{pmatrix}.$ This rank-1 decomposition provides the recipe for constructing the operator $\mathbf{B}$. Consider the block-operator $\mathbf{B}$, of size $2N\times M$, with $M=\frac{N(N+1)}{2}$. Intuitively, the operator $\mathbf{B}$, similar to the 1D case, can be thought as a generalization of the incidence matrix in 2D, where it acts on the space $\ket{j,k}: \; j \leq k \in [N]$. The block operator $\mathbf{B}$ can be written as:
\begin{equation}
    \mathbf{B} = 
    \begin{pmatrix}
        \mathbf{B}_{11} & \mathbf{B}_{12} & \ldots  &\mathbf{B}_{1M} \\
        \mathbf{B}_{21} & \mathbf{B}_{22} & \ldots &\mathbf{B}_{2M} \\
        \vdots & \vdots & \vdots & \vdots \\
        \mathbf{B}_{N1} & \mathbf{B}_{N2} & \vdots & \mathbf{B}_{NM}
    \end{pmatrix}
\end{equation}
where $\mathbf{B}_{il}$ is a $2\times 1$ block matrix, which is nonzero when the edge $l$ corresponds to an edge that connects node $i$ with node $j$, with $i\leq j$ (similar to the 1D incidence matrix) and is defined as:
\begin{equation}
 \mathbf{B}_{il} = \sqrt{\frac{\kappa}{m}} \begin{pmatrix}
     \cos \theta_{ij} \\
     \sin \theta_{ij}
 \end{pmatrix} \; \text{ if $i\leq j$}
\end{equation}
Similarly, the block $\mathbf{B}_{jl}$ (with $j>l$) is defined with a negative sign\footnote{Recall that in the 1D case, the matrix elements of the incidence matrix are assigned with a random orientation that preserves the product $\mathbf{B}\mathbf{B}^\dagger$.}:
\begin{equation}
 \mathbf{B}_{jl} = -\sqrt{\frac{\kappa}{m}} \begin{pmatrix}
     \cos \theta_{ij} \\
     \sin \theta_{ij}
 \end{pmatrix} \; \text{ if $i> j$}
\end{equation}
Since the matrix elements of $\mathbf{B}$ are real, the conjugate transpose matrix $\mathbf{B}^{\dagger}$ is an $M\times 2N$ operator with elements:
\begin{equation}
    \mathbf{B}^\dagger = 
    \begin{pmatrix}
        \mathbf{B}_{11}^T & \mathbf{B}_{21}^T & \ldots  &\mathbf{B}_{N1}^T \\
        \mathbf{B}_{12}^T & \mathbf{B}_{22}^T & \ldots &\mathbf{B}_{N2}^T \\
        \vdots & \vdots & \vdots & \vdots \\
        \mathbf{B}_{1M}^T & \mathbf{B}_{2M}^T & \vdots & \mathbf{B}_{NM}^T
    \end{pmatrix}
\end{equation}
By multiplying $\mathbf{B}\mathbf{B}^\dagger = \mathbf{A}$, we get that the block elements are:
\begin{equation}
    \mathbf{A}_{ij} = \sum_{l=1}^M \mathbf{B}_{il}\mathbf{B}_{jl}^T = \begin{cases}
        -\frac{1}{m}\boldsymbol{k}_{ij} \text{ if $i\neq j$ and  $j\in \text{neighbors($i$)}$}\\
        \frac{1}{m}\sum_{j \in \text{neighbors($i$)}} \boldsymbol{k}_{ij} \text{ if $i=j$}
    \end{cases}
\end{equation}
or else:
\begin{equation}
    \mathbf{B}\mathbf{B}^\dagger = \mathbf{M}^{-1/2} \mathbf{F} \mathbf{M}^{-1/2}
\end{equation}

Before we explain how to construct a block encoding operator $\mathcal{U}_\mathbf{B}$ for the operator $\mathbf{B}$ in Sec. \ref{sec:hamiltonian_simulation}, it will be very useful to introduce some primitives that will be useful in our construction. We start by defining a tensor $\mathbf{K}$ of dimension $2\times N\times N$ with elements $\kappa_{jk}^p$ with $j,k\in [N]$ and $p\in \{0,1\}$:
    \begin{equation}
        \kappa_{jk}^p = \begin{cases}
            \sqrt{\kappa_{jk}}\cos \theta_{jk} \text{ if $p=0$}\\
            \sqrt{\kappa_{jk}}\sin\theta_{jk} \text{ if $p=1$}
        \end{cases}
    \end{equation}
where $\kappa_{jk}$ corresponds to the coupling strength between nodes $j$ and $k$. Then, the action of $\mathbf{B}^{\dagger}$ can be written more formally as:
\begin{equation}
    \mathbf{B}^{\dagger}\ket{l}_{n+1} =  \sum_{k\geq j} \frac{\kappa_{jk}^p}{\sqrt{m}}\ket{j}_n\ket{k}_n - \sum_{k<j} \frac{\kappa_{jk}^p}{\sqrt{m}} \ket{k}_n\ket{j}_n
\label{eq:action_of_B_transpose2}
\end{equation}
where $l=2j + p$. Or else, if we write the matrix elements $\ket{l}_{n+1} \rightarrow \ket{p}_1\ket{j}_n$ where $p\in \{0,1\}$ denotes the $x$ and $y$ coordinates respectively, we can write the action of $\mathbf{B}$ and $\mathbf{B}^{\dagger}$ as:
\begin{equation}
\begin{gathered}
    \mathbf{B}^{\dagger}\ket{p}_1\ket{j}_n =  \sum_{k\geq j} \frac{\kappa_{jk}^p}{\sqrt{m}}\ket{j}_n\ket{k}_n - \sum_{k<j} \frac{\kappa_{jk}^p}{\sqrt{m}} \ket{k}_n\ket{j}_n \\
    \mathbf{B}\ket{j}_n\ket{k}_n = \sum_{p\in\{0,1\}} \frac{\kappa_{jk}^p}{\sqrt{m}}\Big(\ket{p}_1\ket{j}_n - \ket{p}_1\ket{k}_n\Big)
\end{gathered}
\label{eq:action_of_B_and_B_transpose}
\end{equation}
for all $k\in \text{neighbors($j$)}$.

Our goal is to apply the QENM algorithm described in Sec. \ref{sec:QENMs}. As a first step, we introduce additional oracles that are needed for a $D=2$ system and explain the methodology to construct an efficient connectivity oracle for a graphene sheet. Then, we analyze how to prepare the quantum state in Eq. \eqref{eq:initial_state} or the alternative initial state in Eq. \eqref{eq:initial_state_alternate}, and analyze the total resources needed. Following that, we explain how to simulate our system by constructing the unitary induced by the Hamiltonian in Eq. \eqref{eq:problem_hamiltonian}. Finally, we discuss realistic and practical applications for this type of simulation.

\section{QENMs: Application to Graphene}
\label{sec:graphene}

\subsection{Angle and connectivity-strength oracles}

In contrast to the $D=1$ case analyzed in Babbush et al. \cite{babbush2023exponential}, applying the algorithm in a $D>1$ case requires additional quantum oracles. For $D=2$ systems, such as the one analyzed in this paper, we need to introduce the following oracles.\\

\noindent \textbf{Angle Oracle:}\\

As we already discussed, it is important to encode the angle between any two connected masses $j,k$. Recall that in the case of graphene, the angles are multiples of $\pi/6$, i.e.
\begin{equation}
    \theta_{jk} \in \Bigg\{ \frac{\pi}{6}, \frac{3\pi}{6}, \frac{5\pi}{6}, \frac{7\pi}{6}, \frac{9\pi}{6}, \frac{11\pi}{6}\Bigg\}
\end{equation}
for a total of 6 different angles. We can thus encode in a register of $3$ qubits, the odd multiples of $\frac{\pi}{6}$ as:
\begin{equation}
    \ket{000} \rightarrow \ket{\overline{\pi/6}}, \ket{001} \rightarrow\ket{\overline{3\pi/6}}, \ldots
\end{equation}
By taking advantage of this encoding, we can introduce the \emph{angle oracle}:
\begin{equation}
    \ket{j}_n \ket{k}_n \ket{0}_{r_{\theta}} \xrightarrow{O_{\theta}}\ket{j}_n \ket{k}_n \ket{\bar{\theta}_{jk}}_{r_{\theta}}
\label{eq:angle_oracle}
\end{equation}
where $r_{\theta}$ is the number of qubits of the last register that determines the accuracy of the representation, and is equal to 3 for the case of graphene.\\

\noindent \textbf{Trigonometric Oracle:}\\

Next, we define the \emph{trigonometric oracle} that is defined as:
\begin{equation}
    \ket{p}_1 \ket{\theta_{jk}}_3\ket{0}_{r_\text{trig}} \xrightarrow{O_{\text{trig}}} \begin{cases}
        \ket{0}_1 \ket{\theta_{jk}}_3\ket{\overline{\cos \theta_{jk}}}_{r_\text{trig}} \text{ if $p=0$}\\
        \ket{1}_1 \ket{\theta_{jk}}_3\ket{\overline{\sin \theta_{jk}}}_{r_\text{trig}} \text{ if $p=1$}
    \end{cases}
\end{equation}
Constructing this oracle can be done easily with the use of a \emph{look-up table}. Since the number of different angles is 6, the size of the lookup table will be $2^{\lceil \log 6 \rceil+1}$, as we also have to look at the value of $p$.\\

\noindent \textbf{Connectivity-strength Oracle:}\\

When moving from $D=1$ to $D=2$, the coupling strength between two atoms is determined by the orientation of the spring that connects the two masses. As a result, the total force between two masses $j$ and $k$, at each component ($x$ or $y$), depends on their angle $\theta_{jk}$ (measured with respect to the $x$ axis). For this reason, we introduce the \emph{connectivity-strength oracle}, defined as:
\begin{equation}
    \ket{p}_1\ket{j}_n\ket{k}_n \ket{0}_{r_{\kappa}} \xrightarrow{O_{\mathbf{K}}} \ket{p}_1\ket{j}_n\ket{k}_n \ket{\bar{\kappa}_{jk}^p}_{r_{\kappa}} =\begin{cases}
        \ket{0}_1\ket{j}_n\ket{k}_n \ket{\overline{\sqrt{\kappa_{jk}}\cos\theta_{jk}}}_{r_{\kappa}} \text{ if $p=0$} \\
        \ket{1}_1\ket{j}_n\ket{k}_n \ket{\overline{\sqrt{\kappa_{jk}}\sin\theta_{jk}}}_{r_{\kappa}} \text{ if $p=1$}
    \end{cases}
\label{eq:connectivity_strength_oracle}
\end{equation}
where the number of qubits $r_{\kappa}$ depends entirely on the precision to represent the values $\kappa_{jk}^p$, and $\bar{\kappa}_{jk}^p$ is the binary representation of $\kappa_{jk}^p$.\\

\noindent \textbf{Phase Oracle}:\\

Finally, the last oracle that we need to introduce is a \emph{phase oracle} that will be used both at the initial state preparation and the Hamiltonian simulation part. This oracle is defined as:
\begin{equation}
    \ket{p}\ket{j}\ket{k} \xrightarrow{O_{\text{phase}
    }} \text{sgn}(\kappa_{jk}^p) \ket{p}\ket{j}\ket{k}
\label{eq:phase_oracle}
\end{equation}
This oracle can also be constructed using a look-up table, as we have only $2 \times 6$ different values of $\kappa_{jk}^p$ for the graphene molecule.

\subsection{The Connectivity Oracle for Graphene}
\label{sec:connectivity_oracle}

In this section, we will outline how one can implement the connectivity oracle defined in Eq. \eqref{eq:connectivity_oracle}, for a graphene molecule of arbitrary size. In this case, the oracle performs the following map:
\begin{equation}
S_a: \ket{j}\ket{\ell} \mapsto \ket{j} \ket{a(j, \ell)}
\label{eq:graphene_connectivity_oracle}
\end{equation}
where $j \in \{0, \dots, N-1\}$ is the integer index of a carbon atom, $\ell \in \{0,\cdots,d-1\}$ with $d$ being the maximum sparsity of any node in the system, and $a(j ,\ell)$ is the index of the $\ell^{th}$ neighbor of node $j$.

For graphene we exploit the material's bounded sparsity ($d=3$) and crystalline symmetry to replace memory-intensive QRAM lookups with an efficient  algorithm. While graphene possess non-uniform sparsity at the boundaries, we construct the oracle to treat the lattice uniformly as a 3-regular graph. For boundary nodes, the third neighbor index generates a ``ghost'' coordinate which is subsequently flagged as invalid. This sets the effective spring constant to zero. This padding strategy allows for easier arithmetic and detection of boundaries.

We adopt a \emph{unit cell} framework to model the graphene lattice. A unit cell is a fundamental building block of a crystal, which consists of the smallest arrangement of atoms that when tiled in all directions, generates the entire lattice. This framework allows us to replace memory heavy $\mathsf{QRAM}$  lookups with basic quantum arithmetic. We are able to transform a non square lattice of nodes into a coordinate grid of unit cells, which are easier to reason about and index. Using this approach, we can map a global node index $j$ directly to the quantum coordinate registers $\ket{r}\ket{c}\ket{s}$, where the coordinate $(r,c)$ represents the coordinates of the unit cell in which the node is present, and $s$ identifies a specific node within that unit cell. The main idea then is to characterize neighbor shift vectors into memory which when added with the current node's coordinates gives the neighbors associated with the node.

\subsection*{Decoding}

\begin{figure}[!ht]
    \centering
    \begin{subfigure}{0.35\linewidth}
        \centering
        \includegraphics[width=\textwidth]{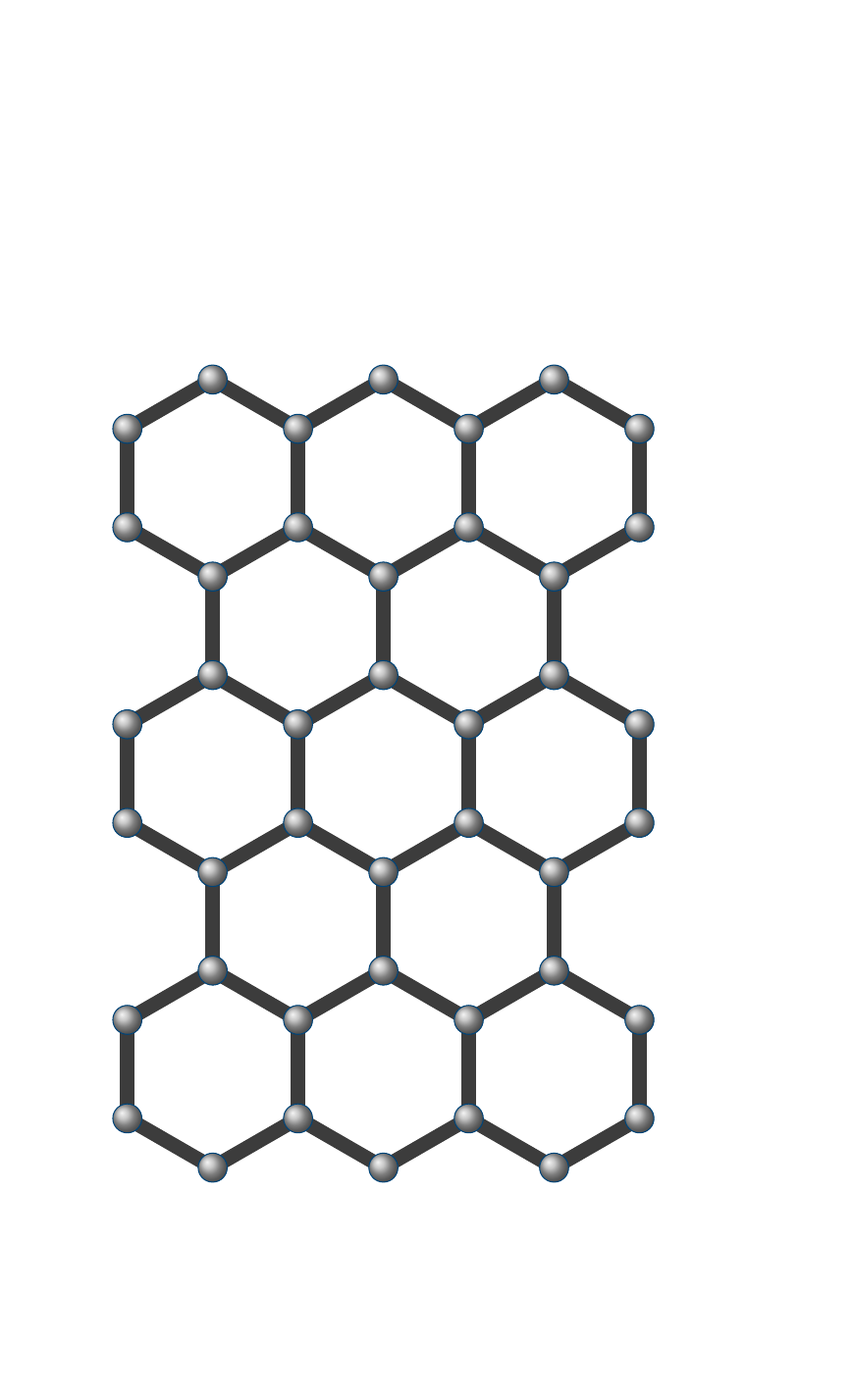}
        \caption{}
        \label{fig:graphene-unitcell-a}
    \end{subfigure}
    \begin{subfigure}{0.35\linewidth}
        \centering
        \includegraphics[width=\textwidth]{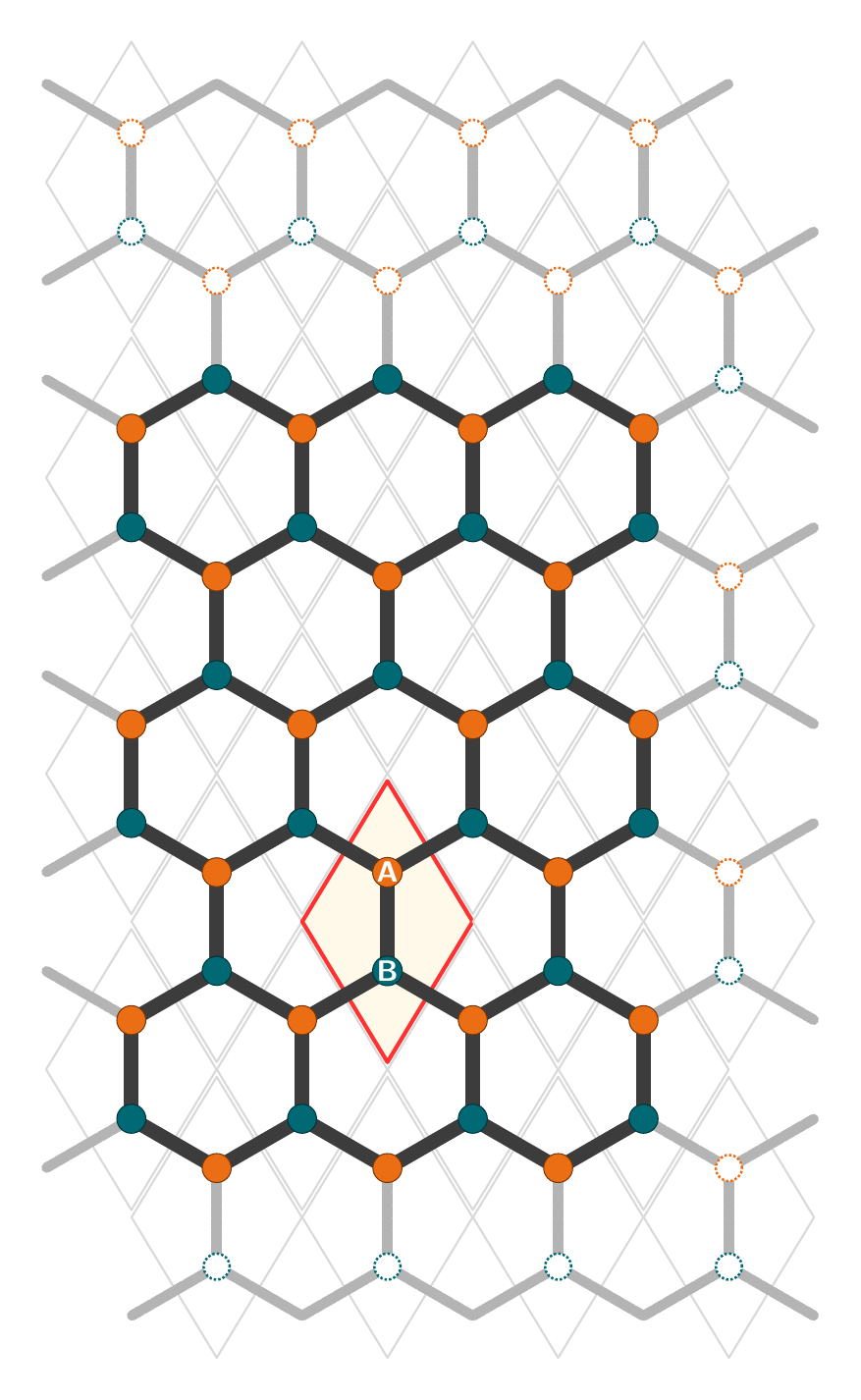}
        \caption{}
        \label{fig:graphene-unitcell-b}
    \end{subfigure}
    \vspace{0.5cm}
    \caption{\emph{Graphene unit cell geometry and the padding strategy used in this work}. (a) shows a 
     graphene sheet. (b) shows the graphene lattice with unit cells and sublattices (orange and blue nodes) explicitly labeled. Solid nodes and dark lines correspond to physical carbon atoms and bonds, while empty circles and gray lines denote dummy nodes and ghost bonds introduced by the padding and neighbor/edge calculation scheme. The illustration depicts an $8\times8$ padded lattice, with padding applied along the final row and column.}\label{fig:unit-cell-graphene}
\end{figure}

Our oracle for graphene is a mapping from a single integer index $j$ to a coordinate triplet $(r, c, s)$ that uniquely identifies each atom. To do so, as mentioned previously, we decompose the graphene sheet into unit cells. Each unit cell, contains two atoms, belonging to the `A' and `B' sublattices, hence the index $s\in\{0,1\}$. We can visualize all unit cells in Fig. \ref{fig:unit-cell-graphene}a, and then show they can be decomposed in sublattices `A' and `B' in Fig. \ref{fig:unit-cell-graphene}b, illustrated with orange and blue dots respectively.

In our model of the graphene sheet, there is an odd number of rows and columns filled with carbon atoms. We assume the number of filled rows and columns to be $1$ less than a power of two for simplicity. We add an extra row and column to make the number of rows even (and a power of 2) and to aid with boundary detection as detailed later in this section. We define a canonical mapping from the coordinate triplet $(r, c, s)$ to the single index $j$ as:
$$
j = 2^{n_c+1}r+2c + s = (r,c,s)
$$
where:
\begin{itemize}
    \item $r \in \{0, \dots, 2^{n_r}-1\}$ indicates the row index of the unit cell.
    \item $c \in \{0, \dots, 2^{n_c}-1\}$ indicates the column index of the unit cell.
    \item $s \in \{0, 1\}$: the sublattice index. We adopt the convention that $s=0$ corresponds to a `B' atom (blue) and $s=1$ corresponds to an `A' atom (orange) as illustrated in Fig.~\ref{fig:graphene-unitcell-b}.
\end{itemize}

Since the lattice dimensions are a power of 2, we can directly read off $r, c, s$ from $j$ as follows:
\begin{enumerate}
    \item $s$ is the least significant bit of $j$
    \item $c$ corresponds to the next $n_c$ bits of $j$
    \item finally, $r$ is the final $n_r$ bits of $j$. i.e. the $n_r$ most significant bits of $j$.
\end{enumerate}

We now detail how to compute the neighbors of a node with index $r,c,s$.

\subsubsection*{Step-by-Step State Evolution}

We define the oracle operation $S_a$ as a sequence of three unitary transformations: \textbf{Relative Shift Initialization} ($U_{\overrightarrow{\delta}}$), \textbf{Absolute Coordinate Calculation} ($U_{\text{calc}}$), and \textbf{Bond Validation} ($U_{\text{val}}$). The system state is defined by the source coordinates $\ket{j}_{RCS}$. We entangle this with a neighbor register $\ket{k}_{R'C'S'}$ and a validity flag $\ket{f}_F$. For a specific neighbor index $\ell \in \{0, 1, 2\}$, the state evolves as follows:

\begin{align*}
\ket{j}_{RCS}\ket{\ell}\ket{0}_{R'C'S'}\ket{0}_F 
\xmapsto{CC\text{-}U_{\overrightarrow{\delta}}} & 
\ket{j}_{RCS} \ket{\delta r, \delta c, 1}_{R'C'S'} \ket{0}_F 
\\
\xmapsto{U_{\text{calc}}} & 
\ket{j}_{RCS} \underbrace{\ket{r + \delta r, c + \delta c, s \oplus 1}_{R'C'S'}}_{\ket{k}} \ket{0}_F 
\\
\xmapsto{U_{\text{val}}} & 
\ket{j}_{RCS} \ket{k}_{R'C'S'} \ket{\mathcal{D}(j) \lor \mathcal{D}(k)}_F
\end{align*}

Here, $\delta r$ and $\delta c$ represent the relative coordinates of the $\ell^{\text{th}}$ neighbor of $j$. The resulting state $\ket{k}$ contains the absolute coordinates of that neighbor, and $\mathcal{D}(\cdot)$ represents the boolean condition for identifying a ``dummy'' (non-physical) atom. In the following sections, we will describe the construction of each of these unitaries.

\paragraph{1. Relative Shift Initialization ($U_{\overrightarrow{\delta}}$)}

\begin{figure}[!ht]
    \centering
    \begin{subfigure}[b]{0.3\textwidth}
        \centering
        \includegraphics[width=\textwidth]{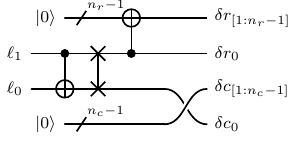}
        \caption{$r_0 = 0, s=0$}
        \label{fig:case0}
    \end{subfigure}
    \hspace{0.5cm}
    \begin{subfigure}[b]{0.3\textwidth}
        \centering
        \includegraphics[width=\textwidth]{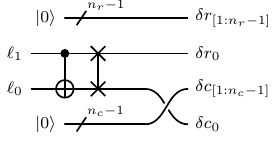}
        \caption{$r_0 = 0, s=1$}
        \label{fig:case1}
    \end{subfigure}
    \par\vspace{0.3cm}
    \begin{subfigure}[b]{0.3\textwidth}
        \centering
        \includegraphics[width=\textwidth]{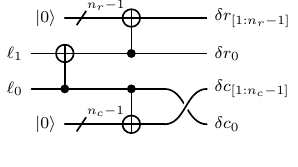}
        \caption{$r_0 = 1, s=0$}
        \label{fig:case2}
    \end{subfigure}
    \hspace{0.5cm}
    \begin{subfigure}[b]{0.3\textwidth}
        \centering
        \includegraphics[width=\textwidth]{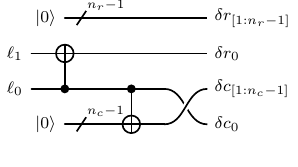}
        \caption{$r_0 = 1, s=1$}
        \label{fig:case3}
    \end{subfigure}
    \vspace{0.5cm}
    \caption{\emph{Quantum circuit implementation of the Shift Initialization unitary ($U_{\vec{\delta}}$)}. The circuits correspond to the four cases defined in Table~\ref{tab:shifts}. The neighbor index register $\ell$ controls the target registers to load the specific relative coordinates $\delta r$ and $\delta c$. The register $\ell = \ell_1\ell_0$ is a 2 qubit register initialized in the state $\ket{0} + \ket{1}+\ket{2}$. We then manipulate this register based on the condition on the $r$ and $s$ registers. Also, it is to be noted that these controlled operations can be simplified; there is clearly a factorizable structure in these circuits based on the cases when $s=0$ and $s=1$ (dictating whether the final fanout $\mathsf{CNOT}$ on $\delta r$ is to be applied) and similarly when $r_0=0$ and when $r_0=1$ (dictating the first two sequence of gates and the final fanout on $\delta c$ in the circuits).}
    \label{fig:all_cases}
\end{figure}

The unitary $U_{\overrightarrow{\delta}}$ prepares the shift vectors $\ket{\delta r, \delta c}$ controlled on the source node's row parity $r_0$ and sublattice $s$. In the unit cell framework we employ for graphene, any valid neighbor must belong to the opposite sublattice, thus the sublattice shift is always $\delta s = 1$. The mapping for the relative shifts is summarized in Table~\ref{tab:shifts}, and the corresponding circuit implementations for each case are illustrated in Fig.~\ref{fig:all_cases}.

\begin{table}[!ht]
\centering
\renewcommand{\arraystretch}{1} 
\setlength{\tabcolsep}{10pt}      
\begin{tabular}{|cc|c|c|c|}
\hline
\multicolumn{2}{|c|}{\textbf{Control Bits}} & \multicolumn{3}{c|}{\textbf{Relative Shift Vector} $(\delta r, \delta c)$} \\ \hline
$r_0$ & $s$ & $\ell=0$ & $\ell=1$ & $\ell=2$ \\ \hline
$0$ & $0$ & $(0, \phantom{+}0)$ & $(-1, \phantom{+}0)$ & $(-1, +1)$ \\ \hline
$0$ & $1$ & $(0, \phantom{+}0)$ & $(+1, \phantom{+}0)$ & $(+1, +1)$ \\ \hline
$1$ & $0$ & $(0, \phantom{+}0)$ & $(-1, -1)$ & $(-1, \phantom{+}0)$ \\ \hline
$1$ & $1$ & $(0, \phantom{+}0)$ & $(+1, -1)$ & $(+1, \phantom{+}0)$ \\ \hline
\end{tabular}
\vspace{0.5cm}
\caption{Mapping of neighbor index $\ell$ to relative shift vectors. The shifts $(\delta r, \delta c)$ i.e. relative coordinates are determined by the value $\ell$, row parity ($r_0$ in the case when we have $2^{n_r}$ rows) and sublattice $s$ of the source node $j$.}
\label{tab:shifts}
\end{table}

\begin{figure}[!ht]
\centering
    \includegraphics[width=0.65\textwidth]{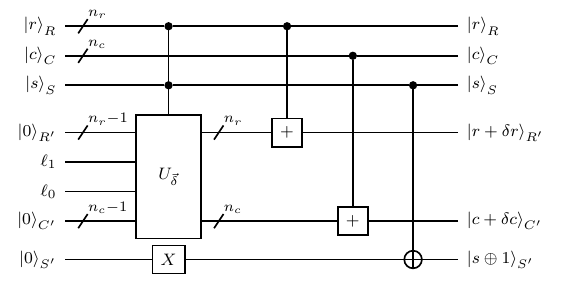}
    \vspace{0.25cm}
\caption{\emph{Circuit for computing neighbors by adding the relative shift vectors} First, the unitary $U_{\vec{\delta}}$ loads the relative shift vectors $(\delta r, \delta c)$ inplace into the $\ell$ register, controlled by the source node parameters $(r, s)$ and the neighbor index $\ell$. Second, quantum adders perform the operation $\ket{r'} \leftarrow \ket{r + \delta r}$ and $\ket{c'} \leftarrow \ket{c + \delta c}$ (modulo the lattice dimensions) to obtain the absolute coordinates of the neighbor. }\label{fig:shift-vectors-addition}
\end{figure}

\paragraph{2. Absolute Coordinate Calculation ($U_{\text{calc}}$)}
We perform the addition $\ket{r' \leftarrow r + \delta r}$ and $\ket{c' \leftarrow c + \delta c}$ using a quantum-quantum adder~\cite{cuccaro2004new}. This transforms the relative shift vectors in the neighbor register into absolute coordinates $\ket{k} = \ket{r', c', s'}$. The source register $\ket{j}$ is treated as the control and remains unchanged, creating the entangled edge state $\ket{j}\ket{k}$. Additions on the row coordinate are modulo $2^{n_r}$, additions on the column coordinate are modulo $2^{n_c}$ and additions on the sublattice coordinate $s$ are modulo $2$ i.e. we perform these modular additions using a non-modular adder and do not compute the final carry bit. This means boundary unit cells wrap around. Since we have padded the boundaries, and know which row/column indices the padded (i.e. dummy) nodes are associated with, we can easily flag any non-existent edges by detecting these dummy nodes. See Fig.~\ref{fig:shift-vectors-addition} for the circuit that computes the neighbor coordinates.

\paragraph{3. Bond Validation ($U_{\text{val}}$)}
The final task is to verify that the neighbor is a physical atom and not a ``ghost'' node in the padding area. We check both nodes against four boundary rules $\mathcal{C}_m$:
\begin{align*}
\mathcal{C}_1 &: (s = 0) \land (r = 0) & \text{(Bottom Edge)} \\
\mathcal{C}_2 &: (r = 2^{n_r} - 1) & \text{(Top Buffer)} \\
\mathcal{C}_3 &: (s = 1) \land (r = 2^{n_r} - 2) & \text{(Second-Last Row)} \\
\mathcal{C}_4 &: (c = 2^{n_c} - 1) \land (r_0 = 0) & \text{(Right Edge Buffer)}
\end{align*}
We compute $\mathcal{D}(\cdot)=\bigoplus_{i=1}^4\mathcal{C}_i$ by performing a logical OR of all computed conditions into the validity flag $\ket{f}$. If any condition is true for either $j$ or $k$ i.e. if either $\mathcal{D}(j)$ or $\mathcal{D}(k)$ are true, the edge is marked invalid ($f=1$). If $f=1$, the bond is marked as invalid, and its spring constant is to be set to zero. For this, we can use $f$ and apply amplitude amplification so that we are left with a state corresponding to only valid oscillators.

\emph{Note: While we have focused on the two-atom basis of graphene ($s \in \{0, 1\}$), this approach is natively extensible to any material with translational symmetry by adjusting the basis register size and the neighbor shift vectors.}

Until now, we analyzed the case of a pure graphene sheet. However, in many practical applications it may be of interest to simulate \emph{doped} graphene, where a certain percentage of the carbon atoms are replaced by a different element (e.g., Nitrogen or Silicon). Here, we explain how to extend our model to include some of these dopants without increasing the memory requirements.

\noindent In certain doped sheets, most atoms have the carbon mass $m_C$, but a small fraction of atoms have a different impurity mass $m_{\text{imp}}$ (for instance in the case of graphitic or pyridinic nitrogen doped graphene). We define the doping rate $p_d$ as the fraction of atoms that are replaced. For simplicity, let us assume $p_d \approx 1/2^f$ for some integer $f$. A naive approach would be to store a list of all impurity locations, but this would require $O(N)$ memory. Instead, we use a randomized approach to assign defects. This is similar to the \textbf{Randomized Bucket Assignment} approach we used for initializing velocities in Sec.~\ref{subsec:loading_intial_velocities}. In both cases, we use a simple arithmetic circuit to generate random features instead of storing them. To decide if an atom at index $j$ is an impurity, we generate $f$ random keys. Each key consists of a binary vector $\vec{h}_m$ and a single random bit $r_m$. For each key, we compute a bit $\sigma_m$ as follows:
\begin{equation}
    \sigma_m = (j \cdot \vec{h}_m) \oplus r_m \pmod 2
\end{equation}
This gives us a $k$-bit string $\sigma = \sigma_1 \sigma_2 \dots \sigma_f$. The probability that this string is exactly $00\dots0$ is $1/2^f$, which matches our doping rate $p$. The steps are as follows:
\begin{enumerate}
    \item Compute the $f$-bit string $\ket{\sigma}$ using CNOT gates (for the dot product) and X gates (for the $\oplus r_m$ shift).
    \item Check if $\ket{\sigma} = \ket{0}$.
    \item If it is $\ket{0}$, load the dopant mass $m_{\text{imp}}$. Otherwise, load $m_C$.
\end{enumerate}

This allows us to simulate a doped sheet with correct statistics using only minimal resources. The output of this oracle would be used as a control to load the altered masses and spring constants at the doped sites.

Now, using the flag for the doped atoms, the spring constant $\kappa_{jk}$ between any two nodes $j$ and $k$ can be loaded using a simple 2-bit lookup based on their respective doping flags. This 2-bit state directly indicates whether the bond connects two standard carbon atoms, a carbon and an impurity, or two impurities.

\subsection{Initial state}
\label{sec:initial_state_preparation}

As we already discussed, the dynamics of a planar graphene sheet can be described in a $D=2$ dimensional space. Since all the building blocks were properly defined in the previous sections, we can now explain the exact steps to prepare the initial state in Eq. \eqref{eq:initial_state}. We explain how the initial state can be prepared in the following two scenarios: i) when the carbon atoms are initialized in their reference structure (with zero relative displacements) and ii) when the atoms are initialized with small displacements $x_i \ll 1.42\si{\angstrom}$\footnote{The carbon-carbon bond length in graphene is approximately $1.42\si{\angstrom}$}.

 Recall that the quantum circuits that prepare the states encoding the Maxwell-Boltzmann velocities and the initial positions were defined in Sec. \ref{sec:QENMs}. From now on, we will refer to these unitaries as $U_{\text{vel}}$ and $U_{\text{pos}}$ respectively. We will also use a lower index (when needed) to indicate the number of qubits in each register. We will show how the initial state in Eq. \ref{eq:initial_state} can be encoded in a quantum state of $2n+3$ qubits as:
\begin{equation}
\begin{gathered}
\ket{\psi(0)} =  \frac{1}{\sqrt{m\alpha^2 +2\kappa d\beta^2}}\Bigg(\ket{0}_1\sum_{p,j} \sqrt{m}\dot{u}_{p,j}(0)\ket{p}_1\ket{0}_{1} \ket{j}_n\ket{0}_{n} + \\i \ket{1}_1 \sum_{j<k} \sqrt{\kappa_{jk}}\Big[\cos \theta_{jk} (u_{0,j}(0) -u_{0,k}(0)) +\sin \theta_{jk}  (u_{1,j}(0) -u_{1,k}(0))\Big] \ket{0}_1\ket{0}_{1}\ket{j}_{n}\ket{k}_{n}\Bigg)
\label{eq:initial_state_default_encoding}
\end{gathered}
\end{equation}

 We start by employing a register of $n+3$ qubits in the state $\ket{0}_1\ket{0}_1\ket{0}_1\ket{0}_{n}$, and define the operator $U_{\text{enc}}= \ket{0}\bra{0} \otimes U_{\text{vel}}  + \ket{1}\bra{1}\otimes U_{\text{pos}}$ acting as:
\begin{equation}
    \begin{gathered}
        \ket{0}_1 \ket{0}_1\ket{0}_1\ket{0}_{n}  \xrightarrow{U_{\text{enc}}} \frac{1}{\alpha}\ket{0}_1\sum_{p,j}\dot{u}_{p,j}(0)\ket{p}_{1} \ket{0}_{1}\ket{j}_{n} \\
       \ket{1}_1 \ket{0}_1\ket{0}_1\ket{0}_{n}  \xrightarrow{U_{\text{enc}}}  \frac{1}{\beta} \ket{1}_1  \sum_{p,j}u_{p,j}(0)\ket{0}_{1}\ket{p}_1\ket{j}_{n} 
    \end{gathered}
\end{equation}
where $\alpha^2 = \sum_{p,j}\dot{u}_{p,j}(0)^2,  \beta^2 = \sum_{p,j} u_{p,j}(0)^2$. The unitary $U_{\text{enc}}$ can be implemented by noting that the operator can be rewritten as:
\begin{equation}
    U_{\text{enc}} = \Big(\ket{0}\bra{0}\otimes \mathds{1}+ \ket{1}\bra{1}\otimes U_{\text{pos}}\Big)\Big(X\otimes \mathds{1}\Big)\\\Big(\ket{0}\bra{0}\otimes \mathds{1} + \ket{1}\bra{1}\otimes U_{\text{vel}}\Big)\Big(X\otimes \mathds{1}\Big)
\label{eq:encoding_position_and_velocity_unitary}
\end{equation}
where the first and third parentheses correspond to controlled $U_{\text{pos}}$ and $U_{\text{vel}}$ (controlled on the first register) that encode the initial positions and velocities respectively.

During the initial state preparation, several additional registers are required throughout the computation. These can be uncomputed and used further throughout the algorithm. We explain in detail all the resources required throughout the rest of this section.

In \cite{babbush2023exponential}, the authors gave a generic algorithm that prepares the initial state in Eq. \eqref{eq:initial_state} for $D=1$, when the system is comprised by $N$ different masses, and the coupling strength may differ for each interaction. In our case, however, the system is less complicated; all atoms have the same mass $m\equiv m_C$, and the number of different coupling strengths is constant and depends on the type of carbon atom as well as the angle in which it is connected to a neighbor carbon atom (see Sec. \ref{sec:connectivity_oracle}). Thus, several steps can be simplified, making the initial state preparation more efficient.

As we discussed, we start with three registers in the state $\ket{0}_1\ket{0}_{1}\ket{0}_1\ket{0}_{n}$.  We start by applying an $R_y(\theta)$ rotation, followed by an $S$ gate on the first qubit, with $\theta = 2\arccos \Big(\frac{\sqrt{2\langle K \rangle }}{\sqrt{2 \langle K \rangle+ 4\kappa d\beta^2}}\Big) = 2\arccos\Big(\frac{\sqrt{m}\alpha}{\sqrt{m\alpha^2 +4\kappa d\beta^2}}\Big) $: 
\begin{equation}
    \begin{aligned}
    \ket{0}_1\ket{0}_{1}\ket{0}_1\ket{0}_{n}
    \xrightarrow{SRy(\theta)} \frac{1}{\sqrt{m \alpha^2 + 4\kappa d\beta^2}}\Big(\sqrt{m}\alpha\ket{0}_1\ket{0}_{1}\ket{0}_1\ket{0}_{n}+i2\sqrt{\kappa d}\beta \ket{1}_1\ket{0}_{1}\ket{0}_1\ket{0}_{n}  \Big)
    \end{aligned}
\label{eq:controlled_rotation_init_state_prep}
\end{equation}
 Then, we apply the operator $U_{\text{enc}}$ that encodes the initial positions and velocities (see Eq. \eqref{eq:encoding_position_and_velocity_unitary}) on the whole system to get:
\begin{equation}
\begin{aligned}
    \xrightarrow{U_{\text{enc}}} \frac{1}{\sqrt{m \alpha^2 + 2\kappa d\beta^2}}\Big(\sqrt{m}\sum_{p,j}\dot{u}_{p,j}(0)\ket{0}_1\ket{p}_1\ket{0}_1\ket{j}_{n}+ i2\sqrt{\kappa d} \sum_{p,j}u_{p,j}(0) \ket{1}_1\ket{0}_{1}\ket{p}_1\ket{j}_{n}\Big)
\label{eq:state_preparation_first_stage}
\end{aligned}
\end{equation}
 Our end goal is to apply a phase $\mathbf{B}^{\dagger}\sqrt{\mathbf{M}}$ on the relative displacements $\ket{\boldsymbol{u}(0)}_{n+1}$ part and $\sqrt{\mathbf{M}}$ on the velocity $\ket{\dot{\boldsymbol{u}}(0)}_{n+1}$ part. In other words, we want to prepare the initial state:
 \begin{equation}
     \ket{\psi(0)} = \frac{1}{\sqrt{2E}}\begin{pmatrix}
         \sqrt{\mathbf{M}} \dot{\boldsymbol{u}}(0)\\
         i\boldsymbol{\mu}(0)
     \end{pmatrix}
 \end{equation}
where $\dot{\mathbf{u}}(0)^T\mathbf{M}\dot{\mathbf{u}}(0) = 2K_{\text{init}}$ and $-i\boldsymbol{\mu}(0)^Ti\boldsymbol{\mu}(0) = \boldsymbol{u}(0)^T\sqrt{\mathbf{M}}\mathbf{B}\mathbf{B}^{\dagger}\sqrt{\mathbf{M}}\boldsymbol{u}(0) = \boldsymbol{u}(0)^T\mathbf{F}\boldsymbol{u}(0) =2 U_{\text{init}}$.

\subsubsection{Maxwell-Boltzmann velocities}

Applying $\sqrt{\mathbf{M}}$ on the velocity part is straightforward. Since all masses are equal, this will induce global phase $\sqrt{m}$ in each basis state that corresponds to the velocities in either $x$ or $y$ axis.\footnote{When the masses are not equal, we need to apply inequality testing to apply a factor $\sqrt{\frac{m_j}{m_{\max}}}$ in each basis state.} But this has already been applied during the initial transformation of the second qubit in Eq. \eqref{eq:controlled_rotation_init_state_prep}, and as such, no other transformations are needed. In the case of more complex molecular systems with different masses (or impurities), the state preparation can become slightly more complex and will require an inequality test step.

\subsubsection{Non-zero displacements}
\label{sec:non_zero_displacements_preparation}

We first consider the case where a $\mathcal{O}(\text{poly}(\log N))$ number of displacements are chosen so that $x_i \ll 1.42\si{\angstrom}$, while the rest are set to zero. Note that our goal is to apply $\mathbf{B}^{\dagger}\sqrt{\mathbf{M}}$ only on the displacement part of Eq. \eqref{eq:state_preparation_first_stage}, while we want to leave the rest of the state unaffected. To do so, we need to control all the following operations on the first qubit. 

As such, for our analysis below, we will focus only on the components of Eq. \eqref{eq:state_preparation_first_stage} that contain the displacements $\boldsymbol{u}_j(0)$ for $j\in [N]$, as the rest of the components remain unaffected. Let $E$ be the initial energy of the system and
\begin{equation}
E_{\max} = K_{\max} + U_{\max} = \frac{m_{\max}}{2}\sum_{p,j}\dot{u}_{p,j}(0)^2 + \frac{\kappa_{\max}}{2}\sum_{p,j}u_{p,j}(0)^2
\end{equation}
is the maximum energy of the system when all masses are equal to $m_{\max} = \max_jm_j=m$ and uncoupled, with coupling strengths $\kappa_{\max} = \max_{jk}\kappa_{jk} = \kappa$.

Considering only the displacements part of \eqref{eq:state_preparation_first_stage}, and neglecting the normalization factors for now, we have:
\begin{equation}
    \ket{1}_1\ket{0}_{1}\sum_{p,j}u_{p,j}(0)\ket{p}_{1}\ket{j}_n
\end{equation}
Next, we employ an ancilla register of $n$ qubits in the $\ket{0}_{n}$ state, and use the connectivity oracle (see Sec. \ref{sec:connectivity_oracle}), to prepare the state:
\begin{equation}
\xrightarrow{\text{Connectivity Oracle}} \frac{1}{\sqrt{d}} \ket{1}_1\ket{0}_{1}\sum_{p,j}u_{p,j}(0)\ket{p}_1\ket{j}_{n} \sum_{k \in \text{neighbors($j$)}} \ket{k}_n
\end{equation}
Moreover, we prepare an additional register of $r_{\kappa}$ qubits in the state $\ket{0}_{r_{\kappa}}$, that will be used to store the (binary representation) of the coupling coefficients $\kappa_{jk}^p$. Thus, by employing this register and by using the connectivity strength oracle defined in Eq. \eqref{eq:connectivity_strength_oracle}, we have:
\begin{equation}
    \xrightarrow{O_{\mathbf{K}}} \frac{1}{\sqrt{d}}\ket{1}_1\ket{0}_{1}\sum_{p,j}u_{p,j}(0)\ket{p}_1\ket{j}_{n} \sum_{k \in \text{neighbors($j$)}} \ket{k}_n \ket{\bar{\kappa}_{jk}^p}_{r_{\kappa}}
\end{equation}
Once the coupling strengths are encoded in the additional register we need to encoded them into the amplitudes of the corresponding computational basis states. To do this, we employ additional $r_{\kappa} +1$ ancilla qubits in the state $\ket{0}_{r_{\kappa}+1}$ and apply \emph{inequality testing}. This gives:
\begin{equation}
    \xrightarrow{\text{Inequality Test}} \frac{1}{\sqrt{\kappa d}}\ket{1}_1\ket{0}_{1}\sum_{p,j,k}u_{p,j}(0) |\kappa_{jk}^p|\ket{p}_1\ket{j}_{n} \ket{k}_n \ket{0}_{r_{\kappa}+1} + \ket{\omega}
\end{equation}
where $\ket{\omega}$ is the state orthogonal to $\ket{0}_{r_{\kappa}+1}$ component. Next, we use the phase oracle (see Eq. \ref{eq:phase_oracle}), to get:
\begin{equation}
    \frac{1}{\sqrt{\kappa d}}\ket{1}_1\ket{0}_{1}\sum_{p,j,k}u_{p,j}(0) \text{sgn}(\kappa_{j,k}^p)\left|\kappa_{jk}^p\right|\ket{p}_1\ket{j}_{n} \ket{k}_n \ket{0}_{r_{\kappa}+1} + \ket{\omega}
\end{equation}
As such, we are left with the state
\begin{equation}
    \frac{1}{\sqrt{\kappa d}}\ket{1}_1\ket{0}_{1}\sum_{p,j,k}u_{p,j}(0) \kappa_{jk}^p\ket{p}_1\ket{j}_{n} \ket{k}_n \ket{0}_{r_{\kappa}+1} + \ket{\omega}
\end{equation}
Next, we employ an additional qubit and apply a quantum comparator between the two node index registers (flagging the result in the extra qubit) and then perform an $X$ gate on the additional qubit to get:
\begin{equation}
\begin{aligned}
    \frac{1}{\sqrt{\kappa d}}\ket{1}_1\ket{0}_{1}\Bigg(\sum_{k>j} \sqrt{\kappa_{jk}}\cos \theta_{jk}u_{0,j}(0) \ket{0}_1\ket{j}_{n}\ket{k}_{n}\ket{0}_1 + \sum_{k<j} \sqrt{\kappa_{jk}}\cos \theta_{jk}u_{0,j}(0) \ket{0}_1\ket{j}_{n}\ket{k}_{n}\ket{1}_1 \\
    + \sum_{k>j} \sqrt{\kappa_{jk}}\sin \theta_{jk}u_{1,j}(0) \ket{1}_1\ket{j}_{n}\ket{k}_{n}\ket{0}_1 + \sum_{k<j} \sqrt{\kappa_{jk}}\sin \theta_{jk}u_{1,j}(0) \ket{1}_1\ket{j}_{n}\ket{k}_{n}\ket{1}_1 \Bigg) + \ket{\omega'}
\end{aligned}
\end{equation}
where the last (single-qubit) register corresponds to the qubit flagging the result of the quantum comparator. If we then perform a $Z$ gate on the last qubit, relabel the third sum and perform a controlled-SWAP operation with the flag qubit as the control and the two node index registers as the target we get:
\begin{equation}
\begin{aligned}
    \frac{1}{\sqrt{\kappa d}}\ket{1}_1\ket{0}_{1}\Bigg(\sum_{j<k} \sqrt{\kappa_{jk}}\cos \theta_{jk} \ket{0}_1\ket{j}_{n}\ket{k}_{n}\Big(u_{0,j}(0)\ket{0}_1 -u_{0,k}(0)\ket{1}_1\Big)  + \\ \sum_{j<k}\sqrt{\kappa_{jk}}\sin \theta_{jk} \ket{1}_1\ket{j}_{n}\ket{k}_{n}\Big(u_{1,j}(0)\ket{0}_1 -u_{1,k}(0)\ket{1}_1\Big)\Bigg)+ \ket{\omega'}
\end{aligned}
\end{equation}
Next, we project the last register on the $\ket{+}$ state (by applying a Hadamard gate and then post-selecting the $\ket{0}$ state) to get:
\begin{equation}
\begin{gathered}
    \frac{1}{\sqrt{2\kappa d}}\ket{1}_1\ket{0}_{1}\Bigg(\sum_{j<k} \sqrt{\kappa_{jk}}\cos \theta_{jk} \Big(u_{0,j}(0) -u_{0,k}(0)\Big)\ket{0}_1\ket{j}_{n}\ket{k}_{n}  + \\ \sum_{j<k}\sqrt{\kappa_{jk}}\sin \theta_{jk}  \Big(u_{1,j}(0) -u_{1,k}(0)\Big) \ket{1}_1\ket{j}_{n}\ket{k}_{n}\Bigg)
\end{gathered}
\end{equation}
Then, we project the third qubit register onto the $\ket{+}$ state to get:
\begin{equation}
    \frac{1}{\sqrt{4\kappa d}}\ket{1}_1\ket{0}_{1}\Bigg(\sum_{j<k} \sqrt{\kappa_{jk}}\Big[\cos \theta_{jk} (u_{0,j}(0) -u_{0,k}(0)) +\sin \theta_{jk}  (u_{1,j}(0) -u_{1,k}(0))\Big] \ket{j}_{n}\ket{k}_{n}\Bigg)
\end{equation}
Finally, we can conclude that the action of all the aforementioned transformations will yield the overall state:
\begin{equation}
\begin{gathered}
  \frac{1}{\sqrt{m\alpha^2 +4\kappa d\beta^2}}\Bigg(\ket{0}_1\sum_{p,j} \sqrt{m}\dot{u}_{p,j}(0)\ket{p}_1\ket{0}_{1} \ket{j}_{n} \ket{0}_{n} + \\i \ket{1}_1 \sum_{j<k} \sqrt{\kappa_{jk}}\Big[\cos \theta_{jk} \big(u_{0,j}(0) -u_{0,k}(0)\big) +\sin \theta_{jk}  \big(u_{1,j}(0) -u_{1,k}(0)\big)\Big] \ket{0}_1\ket{0}_{1}\ket{j}_{n}\ket{k}_{n}\Bigg)
\end{gathered}
\end{equation}
which is subnormalized, since the state is not produced deterministically. To get the desired state, we perform rounds of amplitude amplification where the number of rounds scales as the inverse of the amplitude of the state in Eq. \eqref{eq:initial_state_default_encoding}
\begin{equation}
\mathcal{O}\Bigg(\sqrt{\frac{m_{\max}\alpha^2+4\kappa_{\max}d\beta^2}{2E}} \Bigg) = \mathcal{O}\Bigg(\sqrt{\frac{E_{\max}d}{E}} \Bigg)
\label{eq:amplitude_amplification_rounds}
\end{equation}

\subsubsection{Zero displacements}

The second case corresponds to the scenario where all displacements start at zero. If the user chooses that option, then the preparation of the initial state is significantly simplified. The aforementioned procedure in Sec. \ref{sec:non_zero_displacements_preparation} is neglected, and the amplitudes in the two last terms of Eq. \eqref{eq:initial_state_default_encoding} are zero. Moreover, no amplitude amplification is needed (since all masses are equal), and the initial state is prepared by only using the unitary that encodes the Maxwell-Boltzmann velocities.

\subsection{Alternative initial state}
\label{subsec:alternative_initial_state}

The initial state in Eq. \eqref{eq:initial_state} is not the only solution of the Schrödinger equation induced by the Hamiltonian in Eq. \eqref{eq:problem_hamiltonian}. As noted in \cite{babbush2023exponential} (see Appendix F), one can design different quantum states that are solutions of the Schrödinger equation. One such solution is the quantum state:
\begin{equation}
    \ket{\psi(t)} = \frac{1}{\sqrt{2F}}\begin{pmatrix}
    \mathbf{P}\boldsymbol{y}(t)\\
        -i\mathbf{B}^+\mathbf{P}\boldsymbol{\dot{y}}(t)
    \end{pmatrix}
\label{eq:alternative_quantum_state}
\end{equation}
where $\mathbf{B}^+$ is the Moore-Penrose pseudo-inverse of $\mathbf{B}$\footnote{The \emph{pseudo-inverse} of $\mathbf{B}$ is the inverse of $\mathbf{B}$, defined on the subspace orthogonal to its null space.} and $\mathbf{P}$ is the projector on the subspace orthogonal to the null space of $\boldsymbol{A}$\footnote{Since $\mathbf{A}=\mathbf{B}\mathbf{B^\dagger}$, then $\mathbf{A}$ shares the same null space as $\mathbf{B}^\dagger$.}. The pseudoinverse of the incidence matrix $\mathbf{B}$ satisfies:
\begin{equation}
    \begin{gathered}
        \mathbf{B}\mathbf{B}^+\mathbf{B} = \mathbf{B} \\
        \mathbf{B}^{\dagger}(\mathbf{B}\mathbf{B}^{\dagger})^+ = \mathbf{B}^\dagger \mathbf{A}^+=\mathbf{B}^+
    \end{gathered}
\end{equation}
The quantity $F>0$ ensures normalization, and is conserved throughout the evolution (i.e. $\dot{F} = 0$). Specifically, $F$ is defined as:
\begin{equation}
    F = \frac{1}{2}\boldsymbol{y}(t)^T\mathbf{P}\boldsymbol{y}(t) + \frac{1}{2} \boldsymbol{\dot{y}}(t)^T \mathbf{P}(\mathbf{B}^+)^{\dagger}\mathbf{B}^+\mathbf{P}\boldsymbol{\dot{y}}(t)
\label{eq:F_conserved_quantity}
\end{equation}

This alternative quantum state was first introduced in \cite{costa2019quantum}, where the authors designed a quantum algorithm to solve the wave equation given Dirichlet or Neumann boundary conditions. They argued that the most time-consuming part of their algorithm was the initial state preparation (of Eq. \eqref{eq:alternative_quantum_state}), as the time complexity depends on the condition number $\text{cond}(\mathbf{B})$ of the incidence matrix $\mathbf{B}$ and proving theoretical bounds on this number can be a very difficult task.

The main advantage of preparing this initial state is that it gives (almost) direct access to the displacements $\boldsymbol{y}(t)$. Specifically, with this encoding, we have access to $\mathbf{P}\boldsymbol{y}(t)$ at every time $t$, while using the other encoding in Eq. \eqref{eq:initial_state} can only have direct access to a displacement $y_j(t)$ whenever $\kappa_{jj}\neq 0$.

For a graphene molecule of $N$ atoms, the null space of the Laplacian $\mathbf{A}$ is 1-dimensional, spanned by the vector $v_1 = \frac{1}{\sqrt{N}}(1,1,\ldots, 1)$. To see this, recall that $\mathbf{A}$ is the graph Laplacian, and it is known \cite{chung1997spectral} that for a connected graph (and with equal weights in our case), the second eigenvalue is $\lambda_2>0$. As such $\mathbf{P}\boldsymbol{y}(t) = (\mathds{1}-v_1v_1^T)\boldsymbol{y}(t) \approx \boldsymbol{y}(t)$ with error $\mathcal{O}(1/N)$. Thus, in order to extract each displacement at a time $t$, we need to know the value of $F$, and this can be done once at the beginning of the algorithm. This step is done entirely classically, and its complexity depends on the initial conditions.

\begin{figure}[!ht]
\centering
\includegraphics[width=0.5\textwidth]{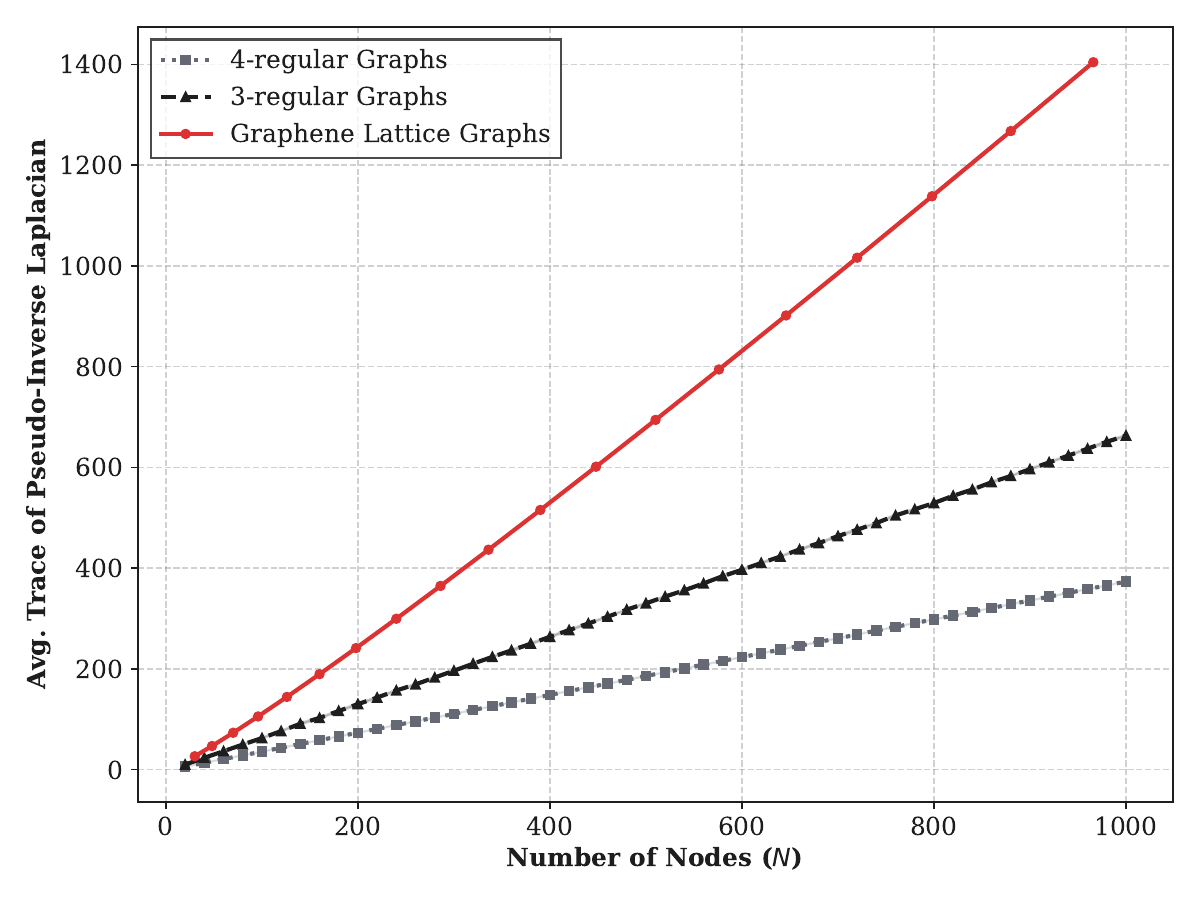}
    \caption{Scaling of the trace of the pseudo-inverse Laplacian $\text{Tr}(\mathbf{A}^+)$ with system size $N$. We compare the scaling behavior for three classes of graphs: random 4-regular graphs (dotted grey), random 3-regular graphs (dashed black), and hexagonal graphene lattice graphs (solid red). For the random regular graphs, data points represent the mean of 10 random instances per node count, with shaded regions indicating the standard deviation.}
\label{fig:pseudoinverse-trace}
\end{figure}

The first term of $F$, based on the initialization that we proposed in Sec. \ref{subsec:encodings}, can always be calculated in time $\mathcal{O}(\text{poly}\log N)$. For the second term we have:
\begin{equation}
\begin{gathered}
\boldsymbol{\dot{y}}(t)^T \mathbf{P}(\mathbf{B}^+)^{\dagger}\mathbf{B}^+\mathbf{P}\boldsymbol{\dot{y}}(t) = \boldsymbol{\dot{y}}(t)^T \mathbf{P}(\mathbf{B}^{\dagger}\mathbf{A}^+)^{\dagger}\mathbf{B}^+\mathbf{P}\boldsymbol{\dot{y}}(t)  \\
= \boldsymbol{\dot{y}}(t)^T \mathbf{P}(\mathbf{A}^+)^{\dagger}\mathbf{B}\mathbf{B}^+\mathbf{P}\boldsymbol{\dot{y}}(t) = \boldsymbol{\dot{y}}(t)^T \mathbf{P}\mathbf{A}^+\mathbf{P}\boldsymbol{\dot{y}}(t) = \boldsymbol{\dot{y}}(t)^T \mathbf{A}^+\boldsymbol{\dot{y}}(t)
\end{gathered}
\end{equation}
where we first used the fact that $\mathbf{B}\mathbf{B}^+ \mathbf{P} = \mathbf{P}$, and then that $(\mathbf{A}^+)^{\dagger} = \mathbf{A}^+$ for any real symmetric matrix $\mathbf{A}$. We now need to estimate $\mathbb{E}[\boldsymbol{\dot{y}}(t)^T \mathbf{A}^+\boldsymbol{\dot{y}}(t)]$ when $\dot{y}_i(t) \sim \mathcal{B}(m,T)$. We have:
\begin{equation}
\begin{gathered}
\mathbb{E}\Big[\boldsymbol{\dot{y}}(t)^T \mathbf{A}^+\boldsymbol{\dot{y}}(t)\Big] = \mathbb{E}\Bigg[\sum_{i,j=1}^N \mathbf{A}^+_{ij} \dot{y}_i(t)\dot{y}_j(t)\Bigg] = \sum_{i,j=1}^N \mathbf{A}^+_{ij} \mathbb{E}\Big[\dot{y}_i(t)\dot{y}_j(t)\Big] \\
= \sum_{i}^N \mathbf{A}^+_{ii}\mathbb{E}\Big[\dot{y}_i^2(t)\Big]  + \sum_{i\neq j}\mathbf{A}^+_{ij} \mathbb{E}\Big[\dot{y}_i(t)\Big]\mathbb{E}\Big[\dot{y}_j(t)\Big] \\
= \frac{k_B T}{m}\text{Tr}(\mathbf{A}^+)
\end{gathered}
\end{equation}
where we used the fact $\dot{y}_i(t), \dot{y}_j(t)$ are statistically independent and that $\mathbb{E}[\dot{y}_i(t)] = 0$ since it comes from the Maxwell-Boltzmann distribution. We can now use the fact that since the eigenvalues of $\mathbf{A}$ are $0=\lambda_1 <\lambda_2 \leq \lambda_3, \ldots \leq \lambda_N$, then the eigenvalues of the pseudo-inverse $\mathbf{A}^+$ are $\frac{1}{\lambda_2} \geq \frac{1}{\lambda_3} \ldots \geq \frac{1}{\lambda_N}$. Putting everything together, $F$ can be rewritten in average (and up to error $O(1/N)$) as:
\begin{equation}
    F = \frac{1}{2} \boldsymbol{y}(t)^T\boldsymbol{y}(t) + \frac{k_B T}{2m} \sum_{k=2}^N \frac{1}{\lambda_k}
\label{eq:average_F}
\end{equation}
It is interesting to understand how the second term of Eq. \eqref{eq:average_F} scales for different (well-defined) types of graphs, such as $d$-regular graphs (e.g. 3-regular graphs corresponding to the graphene molecule) or lattice-type graphs. As it is illustrated in Fig. \ref{fig:pseudoinverse-trace}, it turns out that the $\text{Tr}(\mathbf{A}^+)$ follows a linear correlation with the number of nodes of the graph, meaning that the value of $\text{Tr}(\mathbf{A}^+)$ can be inferred even for very large graphs. However, we argue that for unstructured graphs, or more complicated graphs than the ones above, one would need significantly more resources to estimate this quantity.

We need to stress that this analysis can be neglected in the case where we choose the initial velocities to be zero and allow only the initial positions to be nonzero. If we do so, then one does not need to calculate the second term of Eq. \eqref{eq:F_conserved_quantity}, but they would need to simulate the system for larger times, as the system should be left to thermalize \cite{berinskii2020equilibration}. This will result in larger simulation times $t$ to extract meaningful results; analyzing the time complexity in such a scenario is out-of-scope of this paper and is left for future work.

As proposed in \cite{costa2019quantum}, one way to prepare the initial state in Eq. \eqref{eq:alternative_quantum_state}, for $t=0$, is to employ the quantum algorithm proposed in \cite{childs2017quantum}. In that case, we can perform the transformation $\ket{\boldsymbol{\dot{y}}(t)}\rightarrow -i\mathbf{B}^+\mathbf{P}\ket{\boldsymbol{\dot{y}}(t)}$ using $\tilde{\mathcal{O}}(\text{cond}(\mathbf{B})s\log N)$ number of gates, where $s$ is the sparsity of $\mathbf{B}$. It is clear, that the condition number $\text{cond}(\mathbf{B})$ has a direct effect on the time needed to perform the transformation, and is highly dependent on the structure of the underlying graph. However, for graphs such as the graphene (with uniform weights), one can numerically estimate how this condition number scales with the number of nodes (atoms).

\begin{figure}[!ht]
\centering
\includegraphics[width=0.5\textwidth]{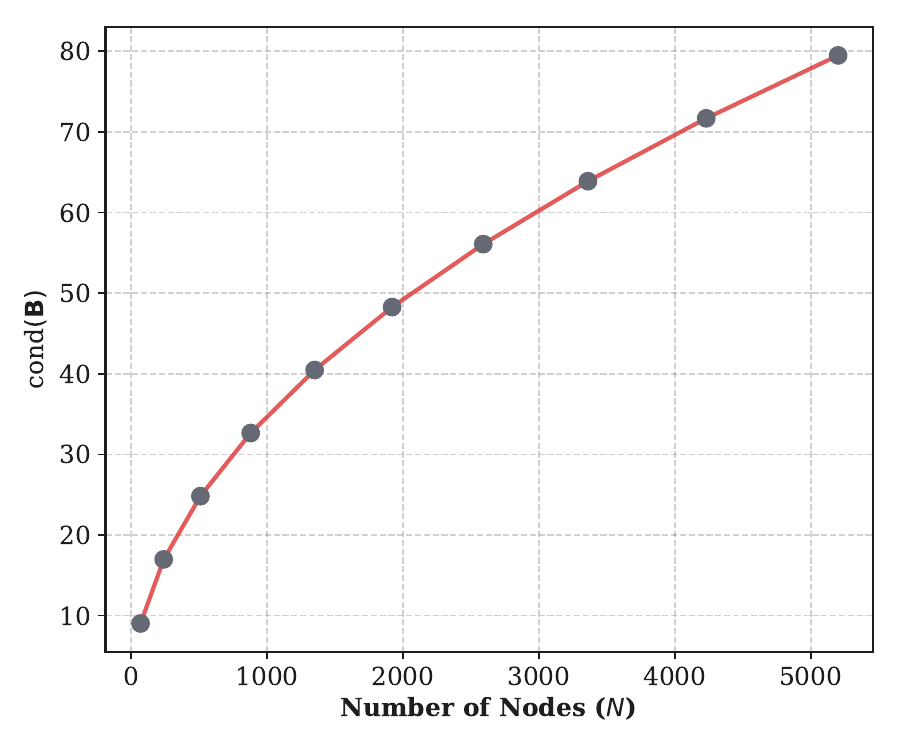}
\caption{Condition number of the incidence matrix $\mathbf{B}$, as a function of the number of nodes (carbon atoms). We can see that $\text{cond}(\mathbf{B}) \sim \mathcal{O}(\sqrt{N})$.}
\label{fig:condition_number_B}
\end{figure}

The results are illustrated in Fig.~\ref{fig:condition_number_B}. As we can clearly see, the condition number grows as the square root of the number of nodes in the graphene lattice. This indicates that the total gate complexity to prepare the state in Eq. \eqref{eq:alternative_quantum_state} is $\tilde{\mathcal{O}}(\sqrt{N}\log N)$. As such, having direct access to the displacements of the carbon masses comes at a higher cost, indicating a polynomial advantage over classical methods. On the other hand, having direct access on the energies of the masses (using the encoding in Eq. \eqref{eq:initial_state}) provides an exponential advantage over the classical counterparts.

\subsection{Hamiltonian Simulation}
\label{sec:hamiltonian_simulation}

In previous sections, we discussed the complexity of preparing different initial states for the graphene simulation, depending on the observable of interest. The initial state (independent of the choice) must then be evolved under the Hamiltonian in Eq. \eqref{eq:problem_hamiltonian}.

This type of Hamiltonian evolution falls into the more general category of \emph{sparse Hamiltonian simulation} \cite{berry2007efficient, berry2014exponential}. In this case, the user has access to a $d$-sparse Hamiltonian $\textbf{H}$, acting on $n$ qubits, through a quantum oracle $O_{\textbf{H}}$ that acts as:
\begin{equation}
    \ket{j} \ket{k} \xrightarrow{O_{\textbf{H}} } \ket{j}\ket{k}\ket{\textbf{H}_{jk}}
\end{equation}
Then, the goal is to construct a series of operations (i.e. a quantum circuit $\mathcal{Q}$), that will approximate the unitary evolution $e^{-i\textbf{H}t}$ (within error $\epsilon$)
\begin{equation}
    \lVert \mathcal{Q}- e^{-i\textbf{H}t}\rVert\leq \epsilon
\end{equation}

As explained in \cite{babbush2023exponential}, the key idea lies in implementing a block encoding of the incidence matrix $\mathbf{B}$, which will later be used for constructing a block-encoding for the Hamiltonian $\mathbf{H}$. It will be very useful if we pad the incidence matrix $\mathbf{B}$ with 0 in order to make its dimension $2N \times N^2$. Intuitively, we extend the action of $\mathbf{B}^{\dagger}$ on the space with basis $\{\ket{j,k}: \; j,k\in [N]\}$ \footnote{Previously, $\mathbf{B}$ acted on the space with basis $\{\ket{j,k}: \; j\leq k \in [N]\}$, of dimension $M\times N$.}. This does not affect its sparsity, but only increases the dimension of the matrix by adding extra zeroes.

We can now proceed and construct the block encoding of $\mathbf{B}^\dagger$ and use it to provide a block encoding of $\mathbf{B}$ by simply taking the conjugate transpose. The action of $\mathbf{B}^\dagger$ is given in Eq. \eqref{eq:action_of_B_transpose2}. We will now describe all steps to construct the unitary $\mathcal{U}_{\mathbf{B}}^\dagger$, that is a block encoding of $\mathbf{B}^{\dagger}$:
\begin{enumerate}
    \item Consider a basis element
    \begin{equation}
        \ket{p}_1\ket{j}_n
    \end{equation}
    Note that $j$ indicates the mass $j\in [N]$, while $p$ denotes the axis $x$ or $y$.

    \item We employ an additional register of $n$ qubits in the state $\ket{0}_n$ and we apply the connectivity oracle (see Sec. \ref{sec:connectivity_oracle}) between the last two registers to get:
        \begin{equation}
        \ket{p}_1\ket{j}_n \frac{1}{\sqrt{d}}\sum_{m=1}^d\ket{a(j,m)}_n \equiv  \ket{p}_1\ket{j}_n \frac{1}{\sqrt{d}}\sum_{{\substack{k=0 \\ \kappa_{jk} \neq 0}}}^{N-1}\ket{k}_n
        \end{equation}

    \item Next, we employ an additional register of $r_\kappa+1$ qubits in the state $\ket{0}_{r_{\kappa}}\ket{0}_1$  and apply the connectivity strength oracle (see Eq. \eqref{eq:connectivity_strength_oracle}) to get:
    \begin{equation}
        \ket{p}_1\ket{j}_n \frac{1}{\sqrt{d}}\sum_{{\substack{k=0 \\ \kappa_{jk} \neq 0}}}^{N-1}\ket{k}_n \ket{\bar{\kappa}_{jk}^p}_{r_{\kappa}}\ket{0}_1
    \end{equation}

    \item Once the connectivity strengths are encoded into computational basis states, we need to extract them as amplitudes in the corresponding states. To do so, we employ an additional $r_{\kappa}$ qubits, use inequality testing, and uncompute the $\ket{\bar{\kappa}^p_{jk}}$ register to get:
    \begin{equation}
        \ket{p}_1\ket{j}_n \frac{1}{\sqrt{\aleph d}}\sum_{k=0}^{N-1} \frac{\left|\kappa_{jk}^p\right|}{\sqrt{m}}\ket{k}_n \ket{0}_{r_{\kappa}}\ket{0}_1 + \ket{\omega_j}
    \end{equation}
    for some state $\ket{\omega_j}$ orthogonal to $\ket{0}_{r_{\kappa}}\ket{0}$ and $\aleph = \frac{\kappa}{m}$. 

    \item Next, we use the phase oracle (see Eq. \ref{eq:phase_oracle}) to get:
    \begin{equation}
        \ket{p}_1\ket{j}_n \frac{1}{\sqrt{\aleph d}}\sum_{k=0}^{N-1} \frac{\kappa_{jk}^p}{\sqrt{m}}\ket{k}_n \ket{0}_{r_{\kappa}}\ket{0}_1 + \ket{\omega_j'}
    \end{equation}

    \item We then employ an additional ancilla qubit and perform the map $\ket{j}_n\ket{k}_n\ket{0}_1 \rightarrow \ket{k}_n\ket{j}_n\ket{1}_1$ if $k<j$ or leave the state $\ket{j}_n\ket{k}_n\ket{0}_1$ invariant otherwise. This can be done by employing an extra flag ancilla qubit. We first perform a quantum comparator on the $\ket{j}_n\ket{k}_n$ register, and store the result in the extra flag qubit. Then, we perform a controlled-$\mathsf{SWAP}$ gate with the flag qubit as the control and the $\ket{j}_n\ket{k}_n$ register as the target. Thus, we have:
    \begin{equation}
    \begin{gathered}
        \ket{p}_1\ket{j}_{n} \frac{1}{\sqrt{\aleph d}}\sum_{k=0}^{N-1}\frac{\kappa_{jk}^p}{\sqrt{m}} \ket{k}_{n}\ket{0}_{r_{\kappa}}\ket{0}_1\ket{0}_1^f +\ket{\omega_j''}\\
        \xrightarrow{\mathsf{QCOMP}} \ket{p}_1\ket{j}_{n} \frac{1}{\sqrt{\aleph d}}\sum_{k=0}^{N-1}\frac{\kappa_{jk}^p}{\sqrt{m}} \ket{k}_{n}\ket{0}_{r_{\kappa}}\ket{0}_1\ket{k<j}_1^f +\ket{\omega_j''}\\
        \xrightarrow{\mathsf{CSWAP}}  \sum_{k<j}\ket{p}_1\frac{\kappa_{jk}^p}{\sqrt{\aleph md}} \ket{k}_{n} \ket{j}_{n}\ket{0}_{r_{\kappa}}\ket{0}_1\ket{1}_1^f + \sum_{k\geq j}\ket{p}_1\frac{\kappa_{jk}^p}{\sqrt{\aleph md}} \ket{j}_{n}\ket{k}_{n}\ket{0}_{r_{\kappa}}\ket{0}_1\ket{0}_1^f +\ket{\omega_j''} \\
    \end{gathered}
    \end{equation}

    \item Then, if we apply $HZ$ on the flag qubit, and consider only the state corresponding to the $\ket{0}_r\ket{0}_1\ket{0}_1$ subspace, we get:
    \begin{equation}
        \ket{p}_1\frac{1}{\sqrt{2\aleph d}} \Bigg(\sum_{k\geq j} \frac{{\kappa}_{jk}^p}{\sqrt{m}}\ket{j}_{n} \ket{k}_{n} - \sum_{k< j} \frac{{\kappa}_{jk}^p}{\sqrt{m}}\ket{k}_{n} \ket{j}_{n}\Bigg)\ket{0}_{r_{\kappa} +2}
    \end{equation}

    \item Finally, we apply a Hadamard on the first qubit $\ket{p}_1$ and project onto the $\ket{0}$ state to get:
    \begin{equation}
                \ket{0}_1\frac{1}{\sqrt{4\aleph d}} \Bigg(\sum_{k\geq j} \frac{{\kappa}_{jk}^p}{\sqrt{m}}\ket{j}_{n} \ket{k}_{n} - \sum_{k< j} \frac{{\kappa}_{jk}^p}{\sqrt{m}}\ket{k}_{n} \ket{j}_{n}\Bigg)\ket{0}_{r_{\kappa} +2}
    \end{equation}
        which the term on the parenthesis corresponds to the action of $\mathbf{B^{\dagger}}$ on a computational basis state $\ket{p}_1\ket{j}_n$.

\end{enumerate}

Hence, we can see that for all $p\in \{0,1\}$ and $j\in [N]$:
\begin{equation}
    \bra{0}_1\otimes\mathds{1}_N\otimes\mathds{1}_N\otimes\bra{0}^{\otimes r_{\kappa} + 2} \mathcal{U}_{\mathbf{B}}^{\dagger}\ket{p}_1\ket{j}_n\ket{0}^{\otimes n}\ket{0}^{\otimes r_{\kappa}+2} \approx \sqrt{\frac{1}{4\aleph d}}\mathbf{B}^{\dagger}\ket{p}_1\ket{j}_n
\end{equation}

Equivalently,
\begin{equation}
    \mathds{1}_2\otimes\mathds{1}_N\otimes\bra{0}_{n}\otimes\bra{0}^{\otimes r_{\kappa} + 2} \mathcal{U}_{\mathbf{B}}\ket{0}_1\otimes \mathds{1}_N\otimes \mathds{1}_N\otimes \ket{0}^{\otimes r_{\kappa}+2} \approx_{\epsilon'} \sqrt{\frac{1}{4\aleph d}}\mathbf{B}
\end{equation}

\emph{To achieve a final error of $\epsilon$, the block encoding should be done to a an error of $\epsilon^{\prime}$ , where $\epsilon^{\prime}=\mathcal{O}(\epsilon/[\tau+\log(1/\epsilon)])$ and $\tau=t\sqrt{\aleph d}$, where $t$ is time, $d$ is the sparsity, and $\aleph=\kappa_{max}/m_{min}$}

Knowing that we need at most an $\epsilon^{\prime}$ error in the block encoding, we can work backwards to determine the number of bits of precision required by the mass and spring constants oracles. For spring constants ($\kappa_{jk}$): we need $r_{\kappa}=\mathcal{O}(\log(1/\epsilon^{\prime}))$ bits, and for masses ($m_j$) we need $r_{m}=\mathcal{O}(\log(m_{\max}/(m_{\min}\epsilon^{\prime})))$ bits.

Once we have that encoding, we can use it to construct the block-encoding for the Hamiltonian $\mathbf{H}$, that was defined in Eq. \eqref{eq:problem_hamiltonian}. Recall that in Sec. \ref{sec:orig_encoding}, we explained how the total information about the kinetic and potential energies can be encoded in $2N + M$ amplitudes of a $2n+2$ qubit system
\begin{equation}
\begin{gathered}
  \frac{1}{\sqrt{m\alpha^2 +4\kappa d\beta^2}}\Bigg(\ket{0}_1\sum_{p,j} \sqrt{m}\dot{u}_{p,j}(0)\ket{p}_1 \ket{j}_{n} \ket{0}_{n} + \\i \ket{1}_1 \sum_{j<k} \sqrt{\kappa_{jk}}\Big[\cos \theta_{jk} \big(u_{0,j}(0) -u_{0,k}(0)\big) +\sin \theta_{jk}  \big(u_{1,j}(0) -u_{1,k}(0)\big)\Big] \ket{0}_1\ket{j}_{n}\ket{k}_{n}\Bigg)
\end{gathered}
\end{equation}

We now need to construct a block encoding of $\mathbf{H}$ that acts nontrivially only on the subspace where we encoded the information of the potential and kinetic energy. Since we padded $\mathbf{B}$ with zeros to make its dimension $2N \times N^2$, the dimension of $\mathbf{H}$ is now $(2N+N^2)\times (2N+N^2)$. However, we will pad additional zeroes to make its dimension $(4N^2)\times (4N^2)$, i.e. $\mathbf{H}$ will act on a space of $2n+2$ qubits. This can be done by transforming the matrices $\mathbf{B},\mathbf{B}^{\dagger}$ to square matrices of size $(2N^2)\times (2N^2)$. The new transformed matrices $\mathbf{B'}, \mathbf{(B')^{\dagger}}$ will act so that they reorder the computational basis states, so that the action of $\mathbf{H}$ satisfies its definition in Eq. \eqref{eq:problem_hamiltonian}. Specifically, we want $\mathbf{B'}$ to act as:

Recall that the target Hamiltonian is of the form:
\begin{equation}
    \mathbf{H} = -\begin{pmatrix} \mathbf{0} & \mathbf{B} \\ \mathbf{B}^\dagger & \mathbf{0} \end{pmatrix},
\label{eq:hamiltonian_block_structure}
\end{equation}
Since $\mathbf{B}$ is of dimension $2N \times N^2$, then $\mathbf{H}$ naturally has a dimension of $(2N + N^2) \times (2N + N^2)$. As shown in~\cite{babbush2023exponential}, we now pad $\mathbf{B}$ and $\mathbf{B}^\dagger$ with zeroes to make them both dimension $2N^2 \times 2N^2$. This expands the full Hamiltonian $\mathbf{H}$ to a dimension of $(4N^2) \times (4N^2)$, meaning it acts on a $2n+2$ qubit space.

The previous padding is equivalent to replacing $\mathbf{B} \ket{j, k} \longrightarrow (\mathbf{B} \ket{j, k}) \otimes \ket{0}^{\otimes n}$ and $\bra{p, j} \mathbf{B} \longrightarrow \bra{0}_1 \otimes (\bra{p, j} \mathbf{B})$.

Because the $D>1$ spatial embedding involves padding along both the rows and columns, we define a system projector $\mathbf{\Pi}_{sys}$ that acts conditionally on the block subspace:
\begin{equation}
    \mathbf{\Pi}_{sys} := \ket{0}\bra{0}_{block} \otimes \mathds{1}_{2N} \otimes \ket{0}\bra{0}^{\otimes n} + \ket{1}\bra{1}_{block} \otimes \ket{0}\bra{0}_1 \otimes \mathds{1}_{N^2}.
\end{equation}

With this small modification, the block encoding implies:
\begin{equation}
    \frac{\mathbf{H}}{\sqrt{4\aleph d}} = \bra{0}^{\otimes r_\kappa+2} \tilde{\mathcal{U}}_{\mathbf{H}} \ket{0}^{\otimes r_\kappa+2},
\end{equation}
where
\begin{equation}
    \tilde{\mathcal{U}}_{\mathbf{H}} := - (\mathbf{\Pi}_{sys} \otimes \mathds{1}^{\otimes r_\kappa+2}) \left( \ket{0}\bra{1}_{block} \otimes \mathcal{U}_{\mathbf{B}} \right) (\mathbf{\Pi}_{sys} \otimes \mathds{1}^{\otimes r_\kappa+2}) + H.c.
\label{eq:non_unitary_H_operator}
\end{equation}
Here, $H.c.$ denotes the conjugate transpose of the first term. This operator is not yet unitary because $\mathbf{\Pi}_{sys}$ is not unitary (it is a projector). However, it is simple to construct a block encoding for a projector as follows. We bring one additional ancilla and implement a conditional unitary operation on the state of the ancilla that is:
\begin{equation}
    \mathcal{U}_{cond} := \ket{+}\bra{+} \otimes (2\mathbf{\Pi}_{sys} - \mathds{1}_{sys}) + \ket{-}\bra{-} \otimes \mathds{1}_{sys}.
\end{equation}

Because $\langle 0 |+\rangle\langle +| 0\rangle = \langle 0 |-\rangle\langle -| 0\rangle = 1/2$, this gives:
\begin{equation}
    \bra{0}_{anc} \left( \ket{+}\bra{+} \otimes (2\mathbf{\Pi}_{sys} - \mathds{1}_{sys}) + \ket{-}\bra{-} \otimes \mathds{1}_{sys} \right) \ket{0}_{anc} = \mathbf{\Pi}_{sys}.
\end{equation}
Applying $\bra{0} \dots \ket{0}$ to this operator gives the block encoding which implements the desired projector. We write $\mathcal{U}_{cond}$ for this unitary when acting on the system of $2n+2$ qubits (the space associated with $\mathbf{H}$) plus $r_\kappa + 3$ ancilla qubits. Hence, our block encoding for $\mathbf{H}$ is:
\begin{equation}
    \frac{1}{\sqrt{4\aleph d}} \mathbf{H} = \bra{0}^{\otimes r_\kappa+3} \mathcal{U}_{\mathbf{H}} \ket{0}^{\otimes r_\kappa+3},
\end{equation}
where
\begin{equation}
    \mathcal{U}_{\mathbf{H}} := -\mathcal{U}_{cond} \left( \ket{0}\bra{1}_{block} \otimes \mathcal{U}_\mathbf{B} \right) \mathcal{U}_{cond} + H.c.
\label{eq:unitary_circuit_H}
\end{equation}

Simulating $\mathcal{U}_{\mathbf{H}}$ with a quantum circuit requires applying a controlled version of $\mathcal{U}_{\mathbf{B}}$ and $\mathcal{U}_{\mathbf{B}}^\dagger$ once, in addition to a simple $X$ gate on the controlled qubit. Simulating $\mathcal{U}_{cond}$ can be done with gate complexity $\mathcal{O}(n)$, since it requires applying a conditional phase on either $\ket{0}^{\otimes n}$ or $\ket{0}_1$ depending on the state of the block flag. 

This block encoding can subsequently be deployed into a Quantum Signal Processing (QSP) protocol \cite{low2017optimal} to approximate the time-evolution operator $e^{-i\mathbf{H}t}$, yielding a total simulation complexity of:
\begin{equation}
    \mathcal{O}\left(t\sqrt{\frac{4\kappa_{max} d}{m_{min}}} + \log(1/\epsilon)\right).
\label{eq:hamiltonian_simulation_complexity}
\end{equation}

\subsection{Measurement}\label{sec:graphene_measurement}

Traditionally, ENMs output a set of low-frequency normal modes that describe the  functional motion of a given molecular assembly. Extracting such normal modes is not, however, possible in our current description of QENM. This necessitates exploration of alternative outputs that can meaningfully explore molecular properties of interest in only $polylog(N)$ complexity. One set of potential quantities that can be measured from currently available encodings are described below. Of course, this short supply of measurable quantities - along with limitations on initial state preparation - means that careful experimental design is required to probe interesting material properties. To this end, we explore two examples of possible QENM applications that are of potential interest. 

The first application, \emph{heat transfer}, is an example of an experiment that requires simulation of long-time dynamics. This is expected to have a super-polynomial quantum advantage over the best classical methods. The second application, \emph{out-of-plane rippling}, requires only short-time dynamics. Here, a dequantized \cite{sakamoto2025quantum} version of the algorithm described in \cite{babbush2023exponential} has been found, meaning that the expected speedup is polynomial. It is however, of high order - see Sec. \ref{sec:discussion} for more details. 

\subsubsection{Observable quantities}
\label{subsec:obs_quants}
In Sec. \ref{sec:orig_encoding}, we described how to extract an estimate of the kinetic energy, $k_V$, of a subset, $V$, of oscillators by employing an oracle, $\mathcal{V}$. Here, we extend \cite{babbush2023exponential} to measure properties of displacement. 

First however, it should be noted that Eq. \ref{eqn:K_epsilon} quotes the error, $\epsilon$, in estimating $K_V(t)/E$. In this case, given a $polylog(N)$ sized subset, an exponential number of calls to $\mathcal{V}$ would be required to estimate $K_V(t)$ with error $\epsilon$. This is due to the fact that, $E$ is exponentially larger than $K_V(t)$ under the assumption of an approximately uniform distribution of energy. Thus, we must restrict measurements to global kinetic and potential energy, or the special case where large proportions of energy are clustered in local regions of the graphene lattice.

Measuring the Mean Squared Displacement (MSD),

\begin{equation}
{M_V(t)} =\frac{1}{|V|} \sum_{i\in V}|{x}_{i}(t)|^2,
\end{equation}
of a subset of nodes is rarely possible using the standard encoding - for example in graphene, none of the atoms are coupled to their rest position. This means that the diagonal entries of the spring constant matrix are all zero, and thus the $N$ entries of the statevector are $\sqrt{\kappa_{j j}} x_j(t)=0$. Instead, one can use the alternate encoding (see Sec. \ref{subsec:alternative_initial_state}), where the displacements are much more accessible. Given an oracle, $\mathcal{V}\subseteq [N]$, which now adds a phase to basis states encoding the displacements of a subset of masses, $V$, the MSD can be calculated using the same technique as the energies. By making $\mathcal{O}(\log (1 / \delta) / \epsilon)$ calls to $\mathcal{V}$, an estimate, $\hat{m}_V(t)$ can be made that is $\epsilon$ close to $\frac{M_V(t)}{F}$

\begin{equation}
\label{eq:msd_err}
|\hat{m}_V(t)-M_V(t) / F| \leq \epsilon .
\end{equation}
Note that in this case, the oracle $\mathcal{V}$ is subtly different to the potential energy case, as we are only interested in adding a phase to basis states corresponding the displacements from rest position. One interesting quantity that follows directly from MSD and can be directly linked to an experimentally obtainable quantity (the Debye-Waller factor) is the thermal $B$-factor:

\begin{equation}
\label{eqn:BFactor}
     B_V = 8\pi^2 \langle M_V\rangle
\end{equation}
where $\langle M_V \rangle$ denotes the time-averaged relative displacement of the subset $V$.

The $B$-factors indicate the relative vibrational motion of different parts of the molecular structure. Specifically, constituents with large $B$-factors indicate that part of the structure is very flexible. In some cases, it is also possible to measure the Mean Squared Velocity (MSV). For example, when the mass of each node is identical, the MSV is trivially related to the kinetic energy. In the case where we have a small set of possible mass values, one can simply iterate through different subsets of nodes corresponding to different mass values, calculating the kinetic energy and thus MSV of each.

There are a variety of application specific quantities that may be required and it is rarely trivial to map the possible outputs of the quantum algorithm described here to these specific quantities. In the next section we provide some example experiments that we are able to efficiently simulate in the case of graphene.

\subsection{Applications}
\subsubsection{Heat transfer}

\begin{figure}[!h]
    \centering
    \begin{minipage}{0.48\textwidth}
        \centering
        \includegraphics[width=\linewidth]{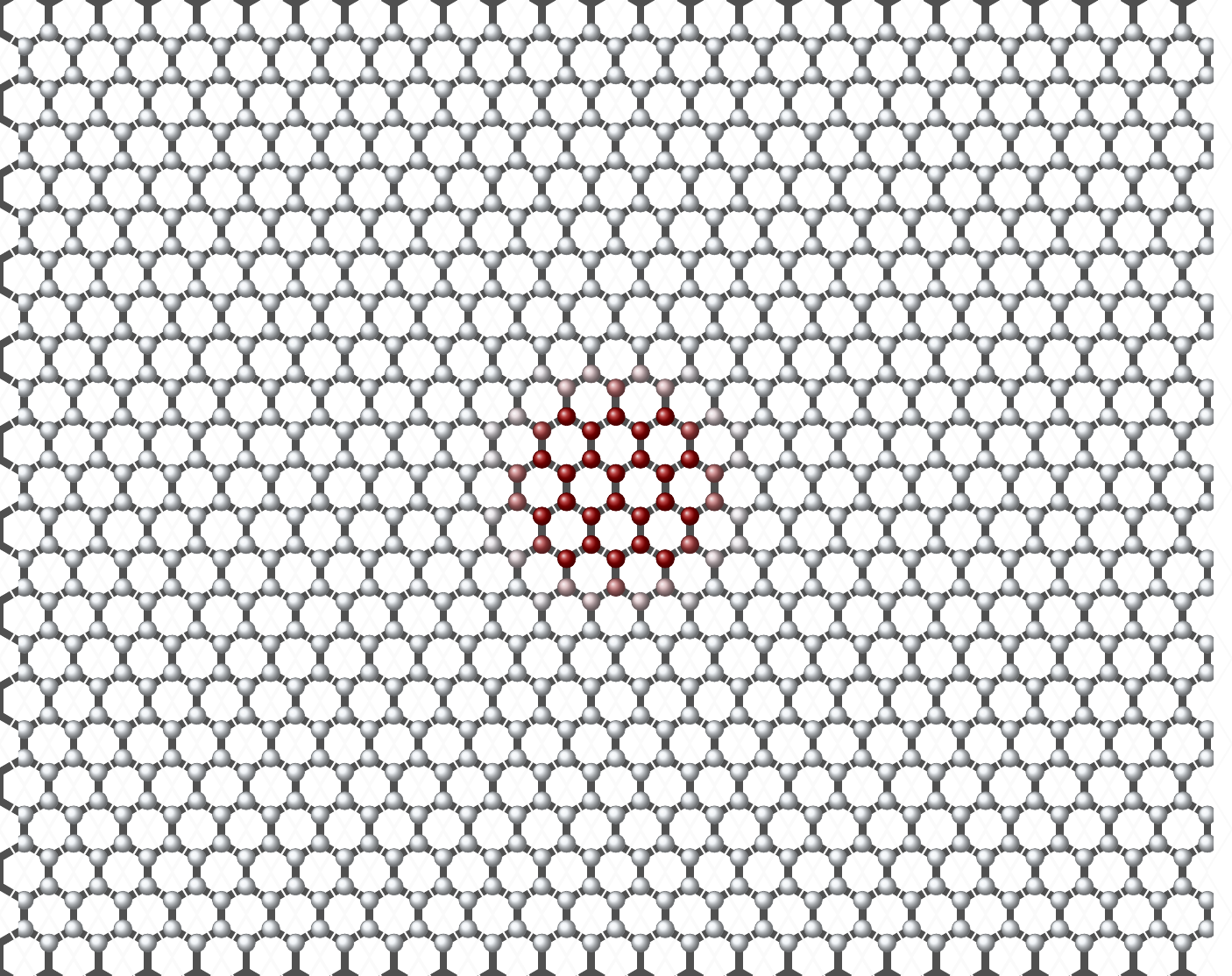}
        
        (a)
    \end{minipage}\hfill
    \begin{minipage}{0.48\textwidth}
        \centering
        \includegraphics[width=\linewidth]{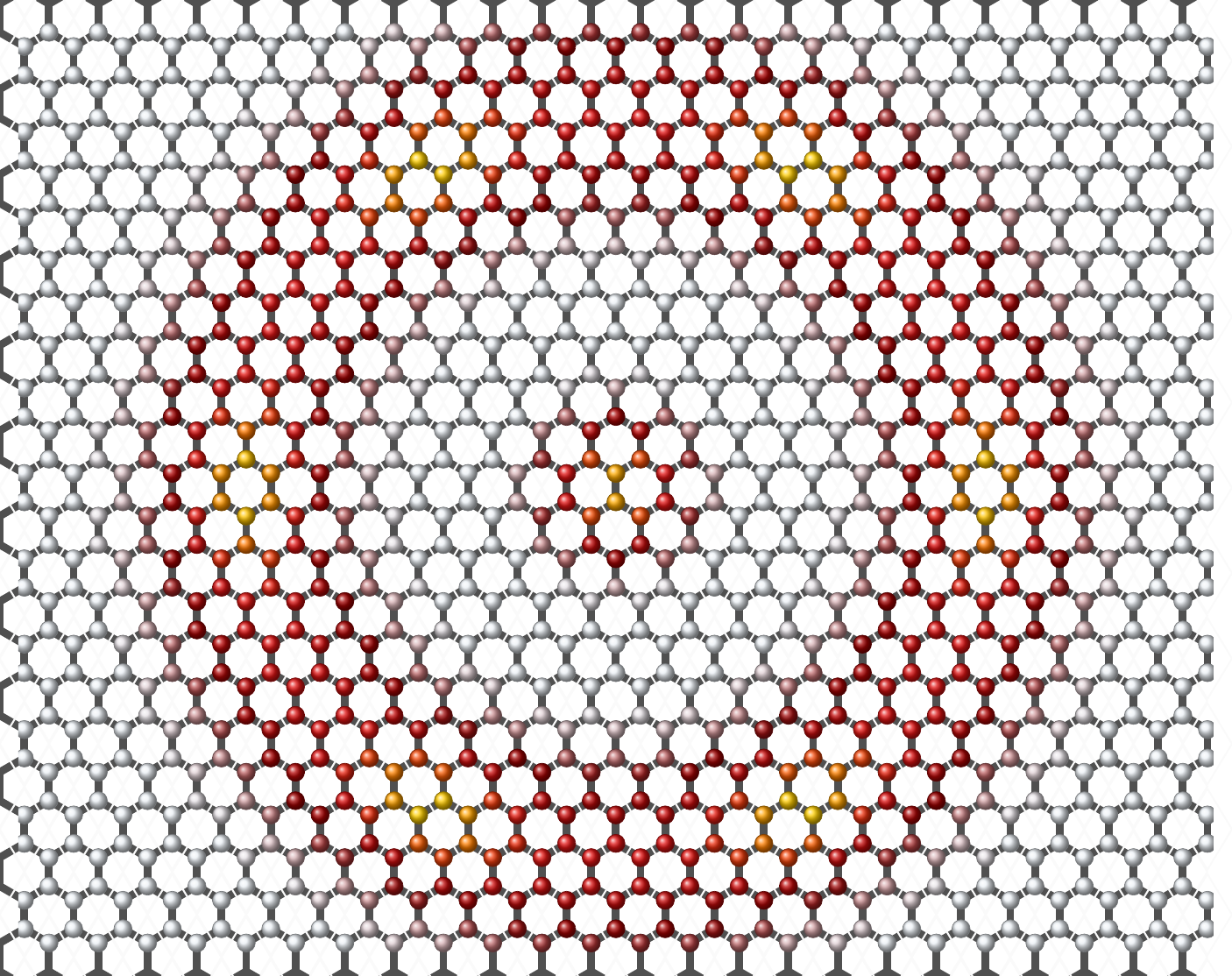}
        
        (b)
    \end{minipage}
    
    \caption{\emph{Illustration of the unsteady ballistic heat transfer experiment with initial circuit temperature profile. (a) Initial state ($t=0$) showing the localized heated boundary nodes prior to propagation. (b) Time-evolved state ($t>0$) showing the thermal wavefront. This is an example of what one would expect to see when tracking the kinetic energy distribution of quasi-particles across the lattice after an initial localized thermal pulse.}}
    \label{fig:heat-transfer}
\end{figure}

In unsteady (as opposed to non-equilibrium steady state) thermal transport we consider the time evolution of some initial temperature distribution. This can be aligned well with experiment, as techniques such as transient thermal grating or simply a laser pulse can create interesting initial temperature distributions in a laboratory environment, before dissipation of heat through a material is tracked \cite{panchenko2022unsteady, johnson2013direct, huberman2019observation}. Before tracking thermal transport (by measuring the kinetic energy of specific carbon atom subsets), we assign an initial temperature distribution and treat those subsets as energetic quasi-particles. This has previously been studied analytically and numerically in the 2D case \cite{panchenko2022unsteady}. Note that in-plane and out-of-plane \cite{kuzkin2019unsteady} dynamics are uncoupled, and are studied separately in the literature. Here, our focus is on in-plane thermal transport. 

In order to track and understand how heat transfers in a planar material such as graphene, we need to be able to measure the kinetic energy of any subset of particles of the system at hand. To do so, we have to use the standard encoding (see Eq. \eqref{eq:initial_state_default_encoding}) that gives direct access to the kinetic (and potential) energy of all atoms.

In this setting, we are not interested in a generic MD simulation; rather, we are interested in understanding how the heat transfers within a graphene sheet, given a initial temperature profile. For example, Panchenko et.al \cite{panchenko2022unsteady} discuss some experimentally realizable temperature profiles in graphene. As such, instead of drawing the initial velocities from the Maxwell-Boltzmann distribution (see Sec. \ref{subsec:loading_intial_velocities}), we can choose the initial velocities to be \cite{panchenko2022unsteady}:
\begin{equation}
    \dot{u}_{p,j} = \sqrt{\frac{k_B}{m}T_0(p,\boldsymbol{r}_j)}
\end{equation}
where $T_0(j)$ is a function that depends on the position vector $\boldsymbol{r}_j$ of the atom $j$ in its reference structure, and on the axis $p\in \{x,y\}$. Thus, the first condition that enables to prepare the initial state with the above initial conditions, is to have to an efficient oracle that acts as:
\begin{equation}
    \ket{p}\ket{j}\ket{0} \xrightarrow{O_T} \ket{p}\ket{j}\ket{T_0(p,\boldsymbol{r}_j)}
\end{equation}
Once we have the circuit $O_T$ that encodes the initial temperature for all masses (and for all axes), we can proceed with the initial state preparation (see Sec. \ref{sec:initial_state_preparation}). Recall that in order for procedure to be efficient, we need to be able to estimate $\alpha$ efficiently, where $\alpha$ is related to the initial kinetic energy of the system. This implies that we need to be able to estimate: 
\begin{equation}
    \alpha = \sum_{p\in\{0,1\}}\sum_{j\in [N]} \dot{u}_{p,j}^2 = \sqrt{\frac{k_B}{m}}\sum_{p\in\{0,1\}}\sum_{j\in [N]} T_0(p,\boldsymbol{r}_j)
\end{equation}
in $\mathcal{O}(\text{poly}(n))$ time.

In \cite{panchenko2022unsteady}, the authors investigated the anisotropy of ballistic heat transport in graphene by choosing two different temperature profiles. The first one is a sinusoidal profile that can be realized in an experimental setting by using two crossed laser pulses \cite{johnson2013direct, huberman2019observation}. The temperature is then described by the function:
\begin{equation}
    T_0 = T_b + \Delta T \sin \frac{2\pi x}{L}
\end{equation}
for a zigzag sinusoidal profile and
\begin{equation}
    T_0 = T_b + \Delta T \sin \frac{2\pi y}{L}
\end{equation}
for an armchair sinusoidal profile, where $T_b$, $\Delta T$ are constants such that $T_b\geq \Delta T$ and $L$ is the wavelength of the sine. Furthermore, the second initial temperature profile that they investigate is a circular profile defined as (and can be visualized in Fig. \ref{fig:heat-transfer}):
\begin{equation}
    T_0(\boldsymbol{r}) = \begin{cases}
        T_1 \;\text{if $\boldsymbol{r}^2 \leq R^2$}\\
        0 \;\;\text{if $\boldsymbol{r}^2> R^2$}
    \end{cases}
\end{equation}
where $R$ is the radius of the circle with non-zero initial temperature.

To study the thermal transport, we track the wavefront by performing a classical binary search on the radius at a given time $t$. We initialize the search by checking the energy within a disk of radius $N_L/4$, where $N_L$ is the lattice dimension. To evaluate the energy sum efficiently, we utilize the amplitude estimation technique described by Babbush \textit{et al.}, which requires performing a phase flip on the nodes corresponding to the subset of atoms within that radius. Depending on whether the energy threshold $\epsilon$ is met, we either bisect the inner region (moving to $N_L/8$) or the outer region (moving to $3N_L/8$). This iterative halving allows us to isolate the leading edge of the ballistic propagation in $O(\text{polylog } N)$ steps.

 There are a few notable limitations to this setup, which we identify as interesting areas for future research. Firstly, although we can measure the kinetic (or potential) energy of an arbitrary subset of carbon atoms, if this energy is exponentially smaller than the total energy, then it cannot be efficiently measured - see Eq. \ref{eqn:K_epsilon}. This presents a problem for our binary search, as we expect to be able to discern an equilibrium region from that of a hotspot, but this is only true if one subset, $V$, has $K_V \approx E$. For an exponential number of atoms, the harmonic approximation would have to be effectively broken to make this true for $T \gg 0$. This means that our method only provides an exponential advantage at very low temperature. In this case, one could trivially perform a polynomial number of measurements of $\ket{\psi(t)}$ in the computational basis, and discern which nodes correspond to kinetic energy basis states of high amplitude.

As described in \cite{babbush2023exponential}, the time complexity of our algorithm is approximately linear in $t$. Thus, given a system with local connectivity and local initialization, the time required for the system to explore the lattice is $t \approx poly(N)$, meaning exponential time in Hamiltonian simulation. Although applications of this type can result in a practical advantage (polynomial in time, exponential in space), it is unlikely to produce an exponential speedup in time complexity \cite{sakamoto2025quantum}. 

Another limitation is the unphysical nature of the harmonic approximation, which excludes fundamental anharmonic terms required for important phenomena like phonon scattering \cite{Gu2019Revisiting}. This is particularly important in the case of thermal transport, as it results in more realistic, non-ballistic behavior, including thermal resistance and other relevant effects.

\subsubsection{Out-of-plane rippling}
\label{sec:out-of-plane_rippling}

\begin{figure}[!h]
    \centering
    \includegraphics[width=0.5\textwidth]{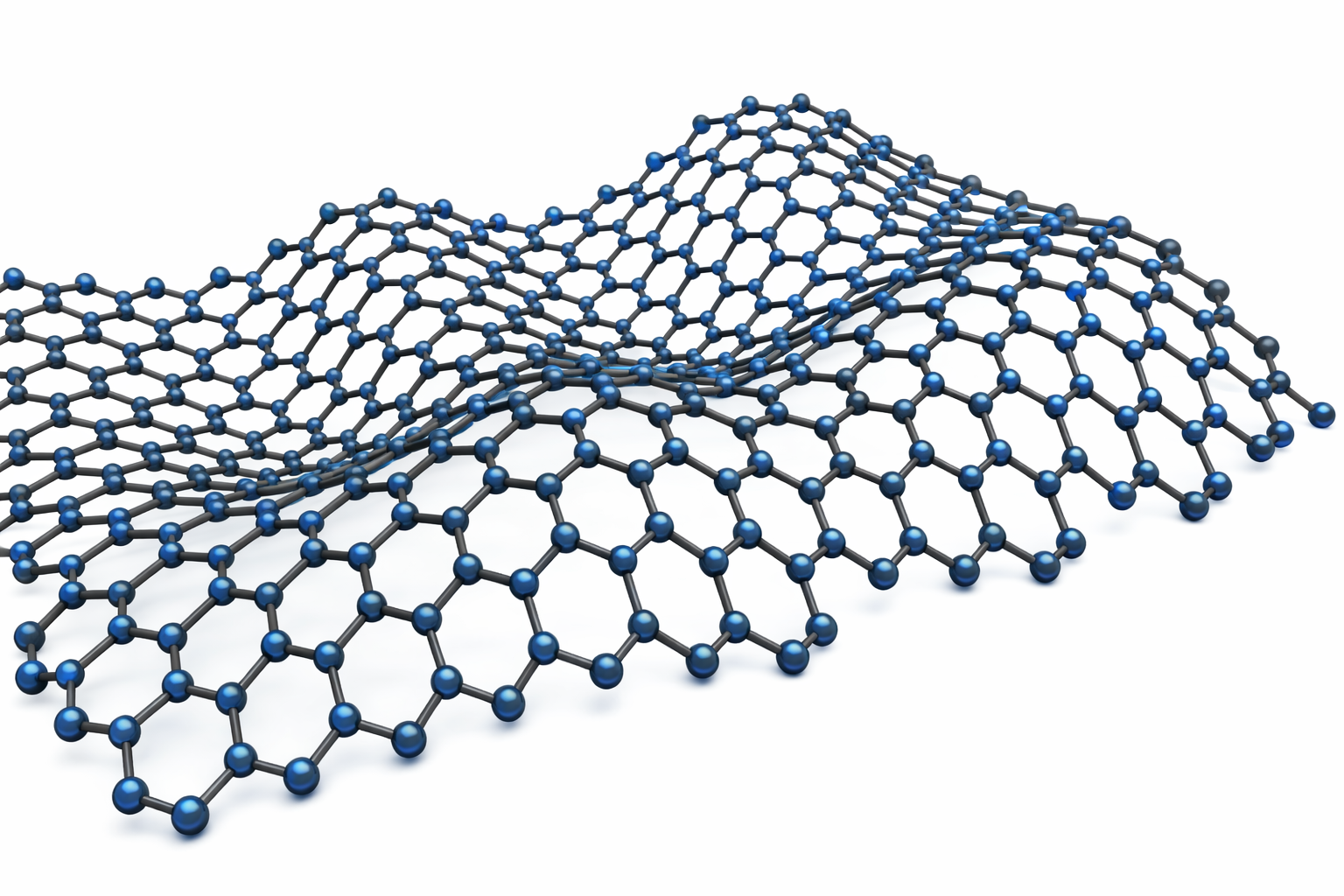}
    \caption{\emph{Illustration of out-of-plane rippling effects of a graphene sheet.}}
    \label{fig:oop-rippling}
\end{figure}

The second application that we consider is the out-of-plane \textit{rippling} effect that is observed in graphene \cite{fasolino2007intrinsic,lopez2022effect},and is visualized in Fig \ref{fig:oop-rippling}. This is directly measurable by scanning tunneling microscopy \cite{xu2014unusual}, however aligning simulation with experiment is often difficult. MSDs measured in simulations do not match with experiment, as the magnitude of the rippling scales with size of the material, thus it becomes necessary to simulate very large lattices \cite{los2009scaling}. Here, we use the out-of-plane (z-axis) MSD calculation to quantify this rippling. This can be related to a host of materials values, 
\begin{equation}
\left\langle M\right\rangle \propto \frac{k_BT A}{\kappa_b},
\end{equation}
where $T$ is the temperature of the sheet, $A$ is the surface area, and $\kappa_b$ is the bending rigidity \cite{fasolino2007intrinsic, wei2013bending}.

In a typical forcefield, one uses harmonic terms to describe the bond interactions. These include not only the bond potential in Eq. \eqref{eq:potential_energy}, which was used to introduce the QENMs, but also harmonic terms that control the geometry between three atoms (see Eq. \eqref{eq:total_harmonic_potentials}). These terms allow for a $z$-axis displacement, as long as the system is initialized with non-zero displacements in the $z$-axis, or if a thermostat is used. Note that one should, however, extend our analysis in Sec. \ref{section:application_to_2d} from 2D to 3D in order to analyze the motion in the 3D space.

In contrast to the heat-transfer application, here we are interested in the MSD of a subset of atoms - explicitly the out-of-plane $z$ displacement of every carbon atom. This means that instead of the standard encoding, which gives direct access to the energies of the system, the alternative encoding in Eq. \eqref{eq:alternative_quantum_state} is more appropriate (see Sec. \ref{sec:initial_state_preparation}).

Starting from the initial state, it is important that the graphene sheet is in equilibrium before any measurements are made. Classical simulation makes use of heat baths to carefully control the temperature; however we employ our Boltzmann velocity initialization (Sec. \ref{subsec:loading_intial_velocities}) to accurately initialize the system very near equilibrium \cite{los2009scaling}. As discussed in \cite{los2009scaling}, starting from a perfectly flat configuration $z=0$ requires significantly long Monte Carlo runs to observe the rippling effect. To ameliorate this effect, the authors suggest a more sophisticated initialization. However, how this can be done efficiently in a quantum computer is out of the scope of this paper and is left for future work. In this work, we propose that a given area of the total graphene sheet is initialized with a non-zero initial displacement $z$.

Finally, we proceed to the measurement part. Measurement is relatively simple in this case. We can find the MSD on the $z$ plane of the entire lattice, with the error given by Eq. \eqref{eq:msd_err}. Averaging this over an appropriate time range ($\approx 1ns$) gives $\left\langle M\right\rangle$.

There are certain limitations to this application, which are discussed below. One possible limitation lies in the thermalisation. Typical simulations of this sort employ a thermostat (for example, Nosé–Hoover \cite{Hoover1985Canonical}), which is currently beyond our current framework. Next, in order to have even more accurate simulations, and to observe the temperature dependence of the ripples correctly, we need terms that penalize this deviation from planarity. This is possible if we consider additional anharmonic terms in the potential, such as a dihedral potential (see discussion on future research directions in Sec. \ref{sec:discussion}).

Finally, although this experiment is efficient in both time and space, it is an example of short-time dynamics, which has recently been dequantized. This means that an exponential quantum speedup in this case is unlikely; however high-order polynomial time advantage is expected - see our discussion in the next section.

\section{Discussion: Road to realistic molecular dynamics simulations}
\label{sec:discussion}

This paper introduces a foundational framework to perform molecular dynamics simulations on a quantum computer, by assuming that atoms interact through harmonic forces. This opens up a vast number of research directions in order to make our model more realistic, and by adapting known classical molecular dynamics (MD) simulation techniques in a quantum setting.

First and foremost, real atomic interactions are not harmonic. Therefore, a natural first step to improve our model is by incorporating, in our algorithm, the appropriate \emph{force fields}. When aiming for a high-precision molecular dynamics simulation, certain extra interactions must be taken into account. Specifically, the total interaction potential for standard MD simulations can be decomposed into a sum of potentials as:
\begin{equation}
    V_{\text{total}} = \sum_{i=1}^{N_{\text{bond}}} V_{\text{bond}} + \sum_{i=1}^{N_{\text{angle}}} V_{\text{angle}} + \sum_{i=1}^{N_{\text{dihed}}} V_{\text{dihed}} + \sum_{i=1}^{N_{\text{nb}}} V_{\text{nb}}
\label{eq:molecular_potential}
\end{equation}
The first two terms in Eq. \eqref{eq:molecular_potential} are described by a harmonic potential as:
\begin{equation}
    \begin{gathered}
    V_{\text{bond}} = \frac{1}{2}k_b(r-r_0)^2 \\
    V_{\text{angle}} = \frac{1}{2}k_a(\theta-\theta_0)^2
    \end{gathered}
\label{eq:total_harmonic_potentials}
\end{equation}
Specifically, $V_{\text{bond}}$ penalizes atoms that move away from their equilibrium, while $V_{\text{angle}}$ maintains the angle between three atoms. On top of that, the latter two terms do not correspond to harmonic potentials. $V_{\text{dihed}}$ describes the dihedral potential that involves four atoms, while $V_{\text{nb}}$ is the non-bonded potential that includes Van der Waals forces (modeled by the Lennard-Jones potential) and Coulomb forces. $V_{\text{dihed}}$ is frequently modeled using Fourier series, but in the case of graphene a harmonic potential applied to improper dihedral angles could be used to describe out-of-plane vibrations of the sheet (see Sec. \ref{sec:out-of-plane_rippling}).

Harmonic potentials cannot describe bond breaking and formation; therefore, more sophisticated force fields are required to accurately model the dihedral and non-bonded terms. However, incorporating non-linear interactions into our quantum algorithm is highly non-trivial. A very promising first step was recently made by \cite{muraleedharan2025simulating}, where the authors introduced a technique named ``Nonlinear Schrödingerization''  and showed that the algorithm in \cite{babbush2023exponential} can be modified to include certain weak non-linear interactions while preserving the advantage over classical methods. On top of that, \cite{jennings2025quantum} also showed how Carleman embeddings can be used to linearize non-linear interactions by introducing auxiliary variables. As such, it would be very interesting to examine whether these techniques can be adapted to extend our model to include important non-linear interactions. This would also allow for more accurate bond potentials \cite{Tersoff1988Empirical, Tersoff1988New, Abell1985Empirical, brenner1990Empirical, brenner2002second, stuart2000reactive}. It should be noted, that although assuming harmonic potentials is an approximation, it is known to be accurate for short time scales and for small displacements. This means that under certain conditions such simulations can be accurate \cite{kuzkin2020ballistic, kuzkin2020equilibration, gendelman2021kapitza, korznikova2020equilibration}. In fact, this has been well quantified in the case of graphene by the so-called Ginzburg criterion \cite{fasolino2007intrinsic}.

Our current work assumes an isolated graphene sheet. Coupling the system to a heat bath would enable simulations in the canonical ensemble at temperatures matching experimental conditions. This could be pursued, for example, by extending the framework to incorporate Nosé-Hoover chains \cite{martyna1992nose}. However, these models are typically constructed by incorporating auxiliary masses coupled to every atom in the system, thereby breaking the sparsity condition of the algorithm. A way around this problem might be to use techniques such as the one proposed in \cite{villanyi2025exponential}, where the system of interest will be coupled to an external bath (defined as a system of oscillators), while also maintaining a low sparsity.

Our framework also allows for the material structure to subtly altered. \textit{Doping} is one way that graphene can be ``customized'' for different practical application, where carbon atoms are replaced by a different element. Important physical properties of the resulting material are disrupted, including both thermal and electrical conduction. This is often a more interesting case to simulate, and current research in this direction already relies heavily on classical MD simulations \cite{joucken2015charge,  hao2011mechanical, goharshadi2015thermal}. Additionally, in reality, perfect graphene sheets do not exist. For example, when made using chemical vapor deposition, they usually contain \emph{defects}. In certain cases, carbon atoms are not present (vacancies) or the connectivity between two $\pi$-bonded carbon atoms changes (Stone-Wales defects). In these scenarios, a more accurate model should include such defects, by replacing certain carbon atoms, thus leading to more accurate simulations for lab-grown materials. In this paper, we made a first-step on how to incorporate defects (through doping) in our model (see the end of Sec. \ref{sec:connectivity_oracle}).

The applications discussed in Sec. \ref{sec:graphene_measurement} are carefully chosen to represent two distinct situations, with different potentials for quantum speedup. \emph{Heat transfer} represents long-time dynamics, while \emph{out-of-plane rippling effects} is an example of short-time dynamics. Long-time dynamics, (those for which $t=\mathcal{O}(\text{poly}N)$), is currently known to have a super-polynomial advantage. By this, we mean an exponential advantage in space complexity and polynomial advantage in time complexity over classical approaches \cite{sakamoto2025quantum}. The exponential space saving makes this approach incredibly attractive. 
According to Lemma 8 of \cite{babbush2023exponential}, the qubit cost of the block encoding scales as $\sim 2 \log_2 N + r + 2$ where $r$ denotes the number of bits of precision in the mass and spring constant oracles.
For a 1 cm$^2$ graphene sheet, node indexing requires $2 \log_2 N \approx 104$ qubits, and even with the high precision of $r \approx 50$ qubits needed to suppress errors in long-time dynamics, the total requirement is only $\sim 160$ logical qubits compared to more than 180 petabytes of classical memory. Although this estimate excludes specific overheads for subroutines, such as state preparation, we do not expect these to substantially increase the logical qubit count. However, since long-time dynamics imply large circuit depths requiring fault-tolerant execution, the error correction overhead must be quantified to determine the true extent of the physical resource savings. 
Addressing this overhead poses an important open question for establishing a practical advantage over classical limits.

As proven in \cite{sakamoto2025quantum}, the previous exponential advantage for short-time dynamics (where $t=\mathcal{O}(\text{poly}\log N)$) for locally-connected systems can be ``dequantized'', meaning that there is a classical polynomial-time probabilistic algorithm to simulate the system, provided that certain assumptions hold. However, the quantum algorithm achieves a significant polynomial advantage, since the algorithm of \cite{sakamoto2025quantum} runs in $\tilde{\mathcal{O}}(t^{3+2D}\log(1/\delta)/\epsilon^2)$ time when the task is to estimate the kinetic or potential energy of a $D$ dimensional system after time $t=\mathcal{O}(\text{poly}\log N)$. Although there is no hope for an exponential advantage here, for $D=2$, one can expect the classical algorithm to run in $\tilde{\mathcal{O}}(t^{7}\log(1/\delta)/\epsilon^2)$, compared to $\tilde{\mathcal{O}}(t\log(1/\delta)/\epsilon)$ for the quantum algorithm \cite{sakamoto2025quantum}. This is a large polynomial advantage which, for example, could theoretically reduce the runtime to simulate centimeter scale sheets of graphene ($N \approx 3.8\times10^{15}$) for $t = log(N)$ by a factor of up to $10^{10}$. 

From these two applications, it is clear that although exponential advantage is in practice unlikely, high order polynomial and super-polynomial advantages still have vast potential for utility in fault-tolerant quantum computing. Additionally, when the physical system under study becomes more complex, for example, by doping, defects, or more complex potentials, it is not immediately clear how the quantum or dequantized algorithm will perform, and further research is required to establish if one could expect a greater, similar, or smaller quantum advantage.

\section{Conclusion}
\label{sec:conclusion}
In this paper, we introduced a quantum algorithm that brings material and molecular dynamics simulations a step closer to practical implementations on a quantum computer. Our algorithm was based on the novel algorithm designed by Babbush et al. \cite{babbush2023exponential}, in which the authors proved that the dynamics of a system of coupled classical oscillators can be simulated exponentially faster on a quantum computer.

However, the question of whether this algorithm can be used for real-world applications remained open. The main reason is that their framework is constrained by several assumptions. Specifically, they assume that certain operations (oracles) can be constructed (and implemented) efficiently, that the system is sparsely connected, and finally that the number of arbitrary non-zero initial conditions is bounded by $\mathcal{O}(\text{poly}\log N)$, where $N$ is the system size.

The main inspiration for our work has been elastic network models (ENMs) which are widely used for simulating the slow dynamics (and conformational changes) of large molecules. In this framework, the complex interactions between atoms of molecular systems are replaced by harmonic interactions, provided that the distance of two atoms is within a threshold $R_c$. As such, ENMs provide a coarse-grained description of the potential energy surface, simplifying the molecular topology into a network of nodes and springs

In this paper, we propose the quantum elastic network model, which allows to map the classical ENM system onto a quantum mechanical system whose dynamics are governed by the Schrödinger equation. We show that all the building blocks (quantum oracles) can be constructed efficiently for these models, provided that the molecules exhibit an inherent well-defined structure. Moreover, we provide a methodology to initialize the system, similar to a standard molecular dynamics simulation. To be precise, we discuss the initial displacements that can be chosen for such a simulation and then prove how the user can efficiently load $2^n$ initial velocities that are sampled from the Maxwell-Boltzmann distribution onto a quantum state in $\text{poly}(n)$ time.

As an application for our model, we choose to simulate the dynamics of a 2D graphene sheet; a well-studied material that exhibits a hexagonal lattice structure. We first provide a methodology that allows an intuitive way of extending the algorithm of \cite{babbush2023exponential} in a 2D system. Then, we thoroughly analyze the exact steps and resources needed for the graphene molecule to prepare alternative initial states (which are solutions to the Schrödinger equation) and give rise to different observables at the end of the simulation. Moreover, we explain how the system is then simulated for a given time $t$. 

Finally, we stress that the computational advantage (in both time and space) is inextricably linked not only to the physical system to be simulated, but also to the computational experiment that a practitioner wishes to investigate. The first application that we introduce, \emph{out-of-plane rippling effects}, is an example of short-time dynamics, which has been successfully dequantized \cite{sakamoto2025quantum} however there remains ample room for a high order polynomial advantage. On the other hand, \emph{heat transfer} is an example of long-time dynamics, which exhibits a super-polynomial advantage, with a particularly interesting exponential advantage in space complexity. The effects that defects, doping and more complex interaction potentials have on simulability, classical or quantum, is a fascinating area of research to look forward to. This forms the next natural step in bridging the gap towards practical computational science with quantum computers.

\subsection*{Acknowledgements}
We would like to thank Niall Moroney, Mario Herrero-González, Elham Kashefi, Qisheng Wang and Jinge Bao, for valuable discussions. P.W. acknowledges funding from EPSRC grants EP/X026167/1 and EP/Z53318X/1. S.F. was supported by EPSRC DTP
studentship grant EP/W524311/1, S.T.  by EPSRC DTP studentship grant
EP/T517884/1. A.S. was supported by the Engineering and Physical Sciences Research Council (grant number EP/W524384/1).

\printbibliography

@article{babbush2023exponential,
  title={Exponential quantum speedup in simulating coupled classical oscillators},
  author={Babbush, Ryan and Berry, Dominic W and Kothari, Robin and Somma, Rolando D and Wiebe, Nathan},
  journal={Physical Review X},
  volume={13},
  number={4},
  pages={041041},
  year={2023},
  publisher={APS}
}

@article{lezon2009elastic,
  title={Elastic network models for biomolecular dynamics: theory and application to membrane proteins and viruses},
  author={Lezon, Timothy R and Shrivastava, Indira H and Yang, Zheng and Bahar, Ivet},
  journal={Handbook on biological networks},
  pages={129--58},
  year={2009},
  publisher={World Scientific Hackensack, NJ}
}

@article{cuccaro2004new,
  title={A new quantum ripple-carry addition circuit},
  author={Cuccaro, Steven A and Draper, Thomas G and Kutin, Samuel A and Moulton, David Petrie},
  journal={arXiv preprint quant-ph/0410184},
  year={2004}
}

@article{togashi2018coarse,
  title={Coarse-grained protein dynamics studies using elastic network models},
  author={Togashi, Yuichi and Flechsig, Holger},
  journal={International journal of molecular sciences},
  volume={19},
  number={12},
  pages={3899},
  year={2018},
  publisher={MDPI}
}

@article{durrant2011molecular,
  title={Molecular dynamics simulations and drug discovery},
  author={Durrant, Jacob D and McCammon, J Andrew},
  journal={BMC biology},
  volume={9},
  pages={1--9},
  year={2011},
  publisher={Springer}
}

@article{tirion1996large,
  title={Large amplitude elastic motions in proteins from a single-parameter, atomic analysis},
  author={Tirion, Monique M},
  journal={Physical review letters},
  volume={77},
  number={9},
  pages={1905},
  year={1996},
  publisher={APS}
}

@article{bahar1997direct,
  title={Direct evaluation of thermal fluctuations in proteins using a single-parameter harmonic potential},
  author={Bahar, Ivet and Atilgan, Ali Rana and Erman, Burak},
  journal={Folding and Design},
  volume={2},
  number={3},
  pages={173--181},
  year={1997},
  publisher={Elsevier}
}

@article{haliloglu1997gaussian,
  title={Gaussian dynamics of folded proteins},
  author={Haliloglu, Turkan and Bahar, Ivet and Erman, Burak},
  journal={Physical review letters},
  volume={79},
  number={16},
  pages={3090},
  year={1997},
  publisher={APS}
}

@article{hayward2008normal,
  title={Normal modes and essential dynamics},
  author={Hayward, Steven and De Groot, Bert L},
  journal={Molecular Modeling of Proteins},
  pages={89--106},
  year={2008},
  publisher={Springer}
}

@article{bacstuug2012molecular,
  title={Molecular dynamics simulations of membrane proteins},
  author={Ba{\c{s}}tu{\u{g}}, Turgut and Kuyucak, Serdar},
  journal={Biophysical reviews},
  volume={4},
  pages={271--282},
  year={2012},
  publisher={Springer}
}

@article{kim2014vibrational,
  title={Vibrational characteristics of graphene sheets elucidated using an elastic network model},
  author={Kim, Min Hyeok and Kim, Daejoong and Choi, Jae Boong and Kim, Moon Ki},
  journal={Physical Chemistry Chemical Physics},
  volume={16},
  number={29},
  pages={15263--15271},
  year={2014},
  publisher={Royal Society of Chemistry}
}

@article{zhang2016optimized,
  title={An optimized cross-linked network model to simulate the linear elastic material response of a smart polymer},
  author={Zhang, Jinjun and Koo, Bonsung and Subramanian, Nithya and Liu, Yingtao and Chattopadhyay, Aditi},
  journal={Journal of Intelligent Material Systems and Structures},
  volume={27},
  number={11},
  pages={1461--1475},
  year={2016},
  publisher={SAGE Publications Sage UK: London, England}
}

@article{low2017optimal,
  title={Optimal Hamiltonian simulation by quantum signal processing},
  author={Low, Guang Hao and Chuang, Isaac L},
  journal={Physical review letters},
  volume={118},
  number={1},
  pages={010501},
  year={2017},
  publisher={APS}
}

@article{low2019hamiltonian,
  title={Hamiltonian simulation by qubitization},
  author={Low, Guang Hao and Chuang, Isaac L},
  journal={Quantum},
  volume={3},
  pages={163},
  year={2019},
  publisher={Verein zur F{\"o}rderung des Open Access Publizierens in den Quantenwissenschaften}
}

@article{gur2013global,
  title={Global transitions of proteins explored by a multiscale hybrid methodology: application to adenylate kinase},
  author={Gur, Mert and Madura, Jeffry D and Bahar, Ivet},
  journal={Biophysical journal},
  volume={105},
  number={7},
  pages={1643--1652},
  year={2013},
  publisher={Elsevier}
}

@article{costa2015exploring,
  title={Exploring free energy landscapes of large conformational changes: molecular dynamics with excited normal modes},
  author={Costa, Mauricio GS and Batista, Paulo R and Bisch, Paulo M and Perahia, David},
  journal={Journal of chemical theory and computation},
  volume={11},
  number={6},
  pages={2755--2767},
  year={2015},
  publisher={ACS Publications}
}

@article{knill2007optimal,
  title={Optimal quantum measurements of expectation values of observables},
  author={Knill, Emanuel and Ortiz, Gerardo and Somma, Rolando D},
  journal={Physical Review A—Atomic, Molecular, and Optical Physics},
  volume={75},
  number={1},
  pages={012328},
  year={2007},
  publisher={APS}
}

@article{muraleedharan2025simulating,
  title={Simulating Time Dependent and Nonlinear Classical Oscillators through Nonlinear Schr$\backslash$" odingerization},
  author={Muraleedharan, Abhinav and Wiebe, Nathan},
  journal={arXiv preprint arXiv:2505.17170},
  year={2025}
}

@article{luongo2024measurement,
  title={Measurement-based uncomputation of quantum circuits for modular arithmetic},
  author={Luongo, Alessandro and Miti, Antonio Michele and Narasimhachar, Varun and Sireesh, Adithya},
  journal={arXiv preprint arXiv:2407.20167},
  year={2024}
}

@article{sanders2019black,
  title={Black-box quantum state preparation without arithmetic},
  author={Sanders, Yuval R and Low, Guang Hao and Scherer, Artur and Berry, Dominic W},
  journal={Physical review letters},
  volume={122},
  number={2},
  pages={020502},
  year={2019},
  publisher={APS}
}

@article{berry2007efficient,
  title={Efficient quantum algorithms for simulating sparse Hamiltonians},
  author={Berry, Dominic W and Ahokas, Graeme and Cleve, Richard and Sanders, Barry C},
  journal={Communications in Mathematical Physics},
  volume={270},
  number={2},
  pages={359--371},
  year={2007},
  publisher={Springer}
}

@inproceedings{berry2014exponential,
  title={Exponential improvement in precision for simulating sparse Hamiltonians},
  author={Berry, Dominic W and Childs, Andrew M and Cleve, Richard and Kothari, Robin and Somma, Rolando D},
  booktitle={Proceedings of the forty-sixth annual ACM symposium on Theory of computing},
  pages={283--292},
  year={2014}
}

@article{costa2019quantum,
  title={Quantum algorithm for simulating the wave equation},
  author={Costa, Pedro CS and Jordan, Stephen and Ostrander, Aaron},
  journal={Physical Review A},
  volume={99},
  number={1},
  pages={012323},
  year={2019},
  publisher={APS}
}

@book{rapaport2004art,
  title={The art of molecular dynamics simulation},
  author={Rapaport, Dennis C},
  year={2004},
  publisher={Cambridge university press}
}

@book{frenkel2023understanding,
  title={Understanding molecular simulation: from algorithms to applications},
  author={Frenkel, Daan and Smit, Berend},
  year={2023},
  publisher={Elsevier}
}

@book{monticelli2013biomolecular,
  title={Biomolecular simulations: methods and protocols},
  author={Monticelli, Luca and Salonen, Emppu},
  volume={924},
  year={2013},
  publisher={Springer}
}

@article{yang2009protein,
  title={Protein elastic network models and the ranges of cooperativity},
  author={Yang, Lei and Song, Guang and Jernigan, Robert L},
  journal={Proceedings of the National Academy of Sciences},
  volume={106},
  number={30},
  pages={12347--12352},
  year={2009},
  publisher={National Academy of Sciences}
}

@article{panchenko2022unsteady,
  title={Unsteady ballistic heat transport in two-dimensional harmonic graphene lattice},
  author={Panchenko, A Yu and Kuzkin, VA and Berinskii, IE},
  journal={Journal of Physics: Condensed Matter},
  volume={34},
  number={16},
  pages={165402},
  year={2022},
  publisher={IOP Publishing}
}

@article{huberman2019observation,
  title={Observation of second sound in graphite at temperatures above 100 K},
  author={Huberman, Samuel and Duncan, Ryan A and Chen, Ke and Song, Bai and Chiloyan, Vazrik and Ding, Zhiwei and Maznev, Alexei A and Chen, Gang and Nelson, Keith A},
  journal={Science},
  volume={364},
  number={6438},
  pages={375--379},
  year={2019},
  publisher={American Association for the Advancement of Science}
}

@article{johnson2013direct,
  title={Direct Measurement of Room-Temperature Nondiffusive Thermal Transport Over Micron Distances in a Silicon Membrane},
  author={Johnson, Jeremy A and Maznev, AA and Cuffe, John and Eliason, Jeffrey K and Minnich, Austin J and Kehoe, Timothy and Torres, Clivia M Sotomayor and Chen, Gang and Nelson, Keith A},
  journal={Physical review letters},
  volume={110},
  number={2},
  pages={025901},
  year={2013},
  publisher={APS}
}

@article{kuzkin2019unsteady,
  title={Unsteady ballistic heat transport in harmonic crystals with polyatomic unit cell},
  author={Kuzkin, Vitaly A},
  journal={Continuum Mechanics and Thermodynamics},
  volume={31},
  number={6},
  pages={1573--1599},
  year={2019},
  publisher={Springer}
}

@article{kuzkin2020equilibration,
  title={Equilibration of kinetic temperatures in face-centered cubic lattices},
  author={Kuzkin, Vitaly A and Liazhkov, Sergei D},
  journal={Physical Review E},
  volume={102},
  number={4},
  pages={042219},
  year={2020},
  publisher={APS}
}

@article{gendelman2021kapitza,
  title={Kapitza thermal resistance in linear and nonlinear chain models: Isotopic defect},
  author={Gendelman, OV and Paul, Jithu},
  journal={Physical Review E},
  volume={103},
  number={5},
  pages={052113},
  year={2021},
  publisher={APS}
}

@article{korznikova2020equilibration,
  title={Equilibration of sinusoidal modulation of temperature in linear and nonlinear chains},
  author={Korznikova, Elena A and Kuzkin, Vitaly A and Krivtsov, Anton M and Xiong, Daxing and Gani, Vakhid A and Kudreyko, Aleksey A and Dmitriev, Sergey V},
  journal={Physical Review E},
  volume={102},
  number={6},
  pages={062148},
  year={2020},
  publisher={APS}
}

@article{kuzkin2020ballistic,
  title={Ballistic resonance and thermalization in the Fermi-Pasta-Ulam-Tsingou chain at finite temperature},
  author={Kuzkin, Vitaly A and Krivtsov, Anton M},
  journal={Physical Review E},
  volume={101},
  number={4},
  pages={042209},
  year={2020},
  publisher={APS}
}

@article{xu2014unusual,
  title={Unusual ultra-low-frequency fluctuations in freestanding graphene},
  author={Xu, P and Neek-Amal, M and Barber, SD and Schoelz, JK and Ackerman, ML and Thibado, PM and Sadeghi, A and Peeters, FM},
  journal={Nature communications},
  volume={5},
  number={1},
  pages={3720},
  year={2014},
  publisher={Nature Publishing Group UK London}
}

@article{los2009scaling,
  title={Scaling properties of flexible membranes from atomistic simulations: application to graphene},
  author={Los, JH and Katsnelson, Mikhail I and Yazyev, OV and Zakharchenko, KV and Fasolino, Annalisa},
  journal={Physical Review B—Condensed Matter and Materials Physics},
  volume={80},
  number={12},
  pages={121405},
  year={2009},
  publisher={APS}
}

@book{chung1997spectral,
  title={Spectral graph theory},
  author={Chung, Fan RK},
  volume={92},
  year={1997},
  publisher={American Mathematical Soc.}
}

@article{Tersoff1988New,
  title = {New empirical approach for the structure and energy of covalent systems},
  author = {Tersoff, J.},
  journal = {Phys. Rev. B},
  volume = {37},
  issue = {12},
  pages = {6991--7000},
  numpages = {0},
  year = {1988},
  month = {4},
  publisher = {American Physical Society},
  doi = {10.1103/PhysRevB.37.6991},
  url = {https://link.aps.org/doi/10.1103/PhysRevB.37.6991}
}

@article{Tersoff1988Empirical,
  title = {Empirical Interatomic Potential for Carbon, with Applications to Amorphous Carbon},
  author = {Tersoff, J.},
  journal = {Phys. Rev. Lett.},
  volume = {61},
  issue = {25},
  pages = {2879--2882},
  numpages = {0},
  year = {1988},
  month = {12},
  publisher = {American Physical Society},
  doi = {10.1103/PhysRevLett.61.2879},
  url = {https://link.aps.org/doi/10.1103/PhysRevLett.61.2879}
}

@article{Abell1985Empirical,
  title = {Empirical chemical pseudopotential theory of molecular and metallic bonding},
  author = {Abell, G. C.},
  journal = {Phys. Rev. B},
  volume = {31},
  issue = {10},
  pages = {6184--6196},
  numpages = {0},
  year = {1985},
  month = {5},
  publisher = {American Physical Society},
  doi = {10.1103/PhysRevB.31.6184},
  url = {https://link.aps.org/doi/10.1103/PhysRevB.31.6184}
}

@article{brenner1990Empirical,
  title = {Empirical potential for hydrocarbons for use in simulating the chemical vapor deposition of diamond films},
  author = {Brenner, Donald W.},
  journal = {Phys. Rev. B},
  volume = {42},
  issue = {15},
  pages = {9458--9471},
  numpages = {0},
  year = {1990},
  month = {11},
  publisher = {American Physical Society},
  doi = {10.1103/PhysRevB.42.9458},
  url = {https://link.aps.org/doi/10.1103/PhysRevB.42.9458}
}

@article{brenner2002second,
  title={A second-generation reactive empirical bondorder (REBO) potential energy expression for hydrocarbons},
  author={Brenner, Donald W and Shenderova, Olga A and Harrison, Judith A and Stuart, Steven J and Ni, Boris and Sinnott, Susan B},
  journal={Journal of Physics: Condensed Matter},
  volume={14},
  number={4},
  pages={783},
  year={2002},
  publisher={IOP Publishing}
}

@article{stuart2000reactive,
  title={A reactive potential for hydrocarbons with intermolecular interactions},
  author={Stuart, Steven J and Tutein, Alan B and Harrison, Judith A},
  journal={The Journal of chemical physics},
  volume={112},
  number={14},
  pages={6472--6486},
  year={2000},
  publisher={American Institute of Physics}
}

@article{childs2017quantum,
  title={Quantum algorithm for systems of linear equations with exponentially improved dependence on precision},
  author={Childs, Andrew M and Kothari, Robin and Somma, Rolando D},
  journal={SIAM Journal on Computing},
  volume={46},
  number={6},
  pages={1920--1950},
  year={2017},
  publisher={SIAM}
}

@article{berinskii2020equilibration,
  title={Equilibration of energies in a two-dimensional harmonic graphene lattice},
  author={Berinskii, I and Kuzkin, VA},
  journal={Philosophical Transactions of the Royal Society A},
  volume={378},
  number={2162},
  pages={20190114},
  year={2020},
  publisher={The Royal Society Publishing}
}

@article{Gu2019Revisiting,
  title = {Revisiting phonon-phonon scattering in single-layer graphene},
  author = {Gu, Xiaokun and Fan, Zheyong and Bao, Hua and Zhao, C. Y.},
  journal = {Phys. Rev. B},
  volume = {100},
  issue = {6},
  pages = {064306},
  numpages = {14},
  year = {2019},
  month = {8},
  publisher = {American Physical Society},
  doi = {10.1103/PhysRevB.100.064306},
  url = {https://link.aps.org/doi/10.1103/PhysRevB.100.064306}
}

@article{hollingsworth2018molecular,
  title={Molecular dynamics simulation for all},
  author={Hollingsworth, Scott A and Dror, Ron O},
  journal={Neuron},
  volume={99},
  number={6},
  pages={1129--1143},
  year={2018},
  publisher={Elsevier}
}

@article{google2025quantum,
  title={Quantum error correction below the surface code threshold},
  author = {Google Quantum AI and Collaborators},
  journal={Nature},
  volume={638},
  number={8052},
  pages={920--926},
  year={2025},
  publisher={Nature Publishing Group UK London}
}

@article{robledo2025chemistry,
  title={Chemistry beyond the scale of exact diagonalization on a quantum-centric supercomputer},
  author={Robledo-Moreno, Javier and Motta, Mario and Haas, Holger and Javadi-Abhari, Ali and Jurcevic, Petar and Kirby, William and Martiel, Simon and Sharma, Kunal and Sharma, Sandeep and Shirakawa, Tomonori and others},
  journal={Science Advances},
  volume={11},
  number={25},
  pages={eadu9991},
  year={2025},
  publisher={American Association for the Advancement of Science}
}

@article{alam2025fermionic,
  title={Fermionic dynamics on a trapped-ion quantum computer beyond exact classical simulation},
  author={Alam, Faisal and Bosse, Jan Lukas and {\v{C}}epait{\.e}, Ieva and Chapman, Adrian and Clinton, Laura and Crichigno, Marcos and Crosson, Elizabeth and Cubitt, Toby and Derby, Charles and Dowinton, Oliver and others},
  journal={arXiv preprint arXiv:2510.26300},
  year={2025}
}

@article{babbush2025grand,
  title={The grand challenge of quantum applications},
  author={Babbush, Ryan and King, Robbie and Boixo, Sergio and Huggins, William and Khattar, Tanuj and Low, Guang Hao and McClean, Jarrod R and O'Brien, Thomas and Rubin, Nicholas C},
  journal={arXiv preprint arXiv:2511.09124},
  year={2025}
}

@article{vcepaite2025quantum,
  title={Quantum-Enhanced Optimization by Warm Starts},
  author={{\v{C}}epait{\.e}, Ieva and Vaishnav, Niam and Zhou, Leo and Montanaro, Ashley},
  journal={arXiv preprint arXiv:2508.16309},
  year={2025}
}

@article{cerezo2021variational,
  title={Variational quantum algorithms},
  author={Cerezo, Marco and Arrasmith, Andrew and Babbush, Ryan and Benjamin, Simon C and Endo, Suguru and Fujii, Keisuke and McClean, Jarrod R and Mitarai, Kosuke and Yuan, Xiao and Cincio, Lukasz and others},
  journal={Nature Reviews Physics},
  volume={3},
  number={9},
  pages={625--644},
  year={2021},
  publisher={Nature Publishing Group UK London}
}

@article{kolotouros2025accelerating,
  title={Accelerating quantum imaginary-time evolution with random measurements},
  author={Kolotouros, Ioannis and Joseph, David and Narayanan, Anand Kumar},
  journal={Physical Review A},
  volume={111},
  number={1},
  pages={012424},
  year={2025},
  publisher={APS}
}

@article{mcardle2019variational,
  title={Variational ansatz-based quantum simulation of imaginary time evolution},
  author={McArdle, Sam and Jones, Tyson and Endo, Suguru and Li, Ying and Benjamin, Simon C and Yuan, Xiao},
  journal={npj Quantum Information},
  volume={5},
  number={1},
  pages={75},
  year={2019},
  publisher={Nature Publishing Group UK London}
}

@article{layden2023quantum,
  title={Quantum-enhanced markov chain monte carlo},
  author={Layden, David and Mazzola, Guglielmo and Mishmash, Ryan V and Motta, Mario and Wocjan, Pawel and Kim, Jin-Sung and Sheldon, Sarah},
  journal={Nature},
  volume={619},
  number={7969},
  pages={282--287},
  year={2023},
  publisher={Nature Publishing Group UK London}
}

@article{ferguson2025quantum,
  title={Quantum-enhanced Markov chain Monte Carlo for systems larger than a quantum computer},
  author={Ferguson, Stuart and Wallden, Petros},
  journal={Physical Review Research},
  volume={7},
  number={1},
  pages={013231},
  year={2025},
  publisher={APS}
}

@article{harrow2009quantum,
  title={Quantum algorithm for linear systems of equations},
  author={Harrow, Aram W and Hassidim, Avinatan and Lloyd, Seth},
  journal={Physical review letters},
  volume={103},
  number={15},
  pages={150502},
  year={2009},
  publisher={APS}
}

@inproceedings{grover1996fast,
  title={A fast quantum mechanical algorithm for database search},
  author={Grover, Lov K},
  booktitle={Proceedings of the twenty-eighth annual ACM symposium on Theory of computing},
  pages={212--219},
  year={1996}
}

@article{shor1999polynomial,
  title={Polynomial-time algorithms for prime factorization and discrete logarithms on a quantum computer},
  author={Shor, Peter W},
  journal={SIAM review},
  volume={41},
  number={2},
  pages={303--332},
  year={1999},
  publisher={SIAM}
}

@article{simon2024improved,
  title={Improved precision scaling for simulating coupled quantum-classical dynamics},
  author={Simon, Sophia and Santagati, Raffaele and Degroote, Matthias and Moll, Nikolaj and Streif, Michael and Wiebe, Nathan},
  journal={PRX Quantum},
  volume={5},
  number={1},
  pages={010343},
  year={2024},
  publisher={APS}
}

@article{ollitrault2020nonadiabatic,
  title={Nonadiabatic molecular quantum dynamics with quantum computers},
  author={Ollitrault, Pauline J and Mazzola, Guglielmo and Tavernelli, Ivano},
  journal={Physical Review Letters},
  volume={125},
  number={26},
  pages={260511},
  year={2020},
  publisher={APS}
}

@article{fedorov2021ab,
  title={Ab initio molecular dynamics on quantum computers},
  author={Fedorov, Dmitry A and Otten, Matthew J and Gray, Stephen K and Alexeev, Yuri},
  journal={The Journal of Chemical Physics},
  volume={154},
  number={16},
  year={2021},
  publisher={AIP Publishing}
}

@article{o2022efficient,
  title={Efficient quantum computation of molecular forces and other energy gradients},
  author={O'Brien, Thomas E and Streif, Michael and Rubin, Nicholas C and Santagati, Raffaele and Su, Yuan and Huggins, William J and Goings, Joshua J and Moll, Nikolaj and Kyoseva, Elica and Degroote, Matthias and others},
  journal={Physical Review Research},
  volume={4},
  number={4},
  pages={043210},
  year={2022},
  publisher={APS}
}

@article{eyal2006anisotropic,
  title={Anisotropic network model: systematic evaluation and a new web interface},
  author={Eyal, Eran and Yang, Lee-Wei and Bahar, Ivet},
  journal={Bioinformatics},
  volume={22},
  number={21},
  pages={2619--2627},
  year={2006},
  publisher={Oxford University Press}
}

@article{fasolino2007intrinsic,
  title={Intrinsic ripples in graphene},
  author={Fasolino, Annalisa and Los, JH and Katsnelson, Mikhail I},
  journal={Nature materials},
  volume={6},
  number={11},
  pages={858--861},
  year={2007},
  publisher={Nature Publishing Group UK London}
}

@article{wei2013bending,
  title={Bending rigidity and Gaussian bending stiffness of single-layered graphene},
  author={Wei, Yujie and Wang, Baoling and Wu, Jiangtao and Yang, Ronggui and Dunn, Martin L},
  journal={Nano letters},
  volume={13},
  number={1},
  pages={26--30},
  year={2013},
  publisher={ACS Publications}
}

@article{Hoover1985Canonical,
  title = {Canonical dynamics: Equilibrium phase-space distributions},
  author = {Hoover, William G.},
  journal = {Phys. Rev. A},
  volume = {31},
  issue = {3},
  pages = {1695--1697},
  numpages = {0},
  year = {1985},
  month = {3},
  publisher = {American Physical Society},
  doi = {10.1103/PhysRevA.31.1695},
  url = {https://link.aps.org/doi/10.1103/PhysRevA.31.1695}
}

@article{martyna1992nose,
  title={Nos{\'e}--Hoover chains: The canonical ensemble via continuous dynamics},
  author={Martyna, Glenn J and Klein, Michael L and Tuckerman, Mark},
  journal={The Journal of chemical physics},
  volume={97},
  number={4},
  pages={2635--2643},
  year={1992},
  publisher={American Institute of Physics}
}

@article{villanyi2025exponential,
  title={Exponential Quantum Advantage for Simulating Open Classical Systems},
  author={Villanyi, Agi and Yanay, Yariv and Mizel, Ari},
  journal={arXiv preprint arXiv:2503.11483},
  year={2025}
}

@article{sakamoto2025quantum,
  title={On the quantum computational complexity of classical linear dynamics with geometrically local interactions: Dequantization and universality},
  author={Sakamoto, Kazuki and Fujii, Keisuke},
  journal={arXiv preprint arXiv:2505.10445},
  year={2025}
}

@article{joucken2015charge,
  title={Charge transfer and electronic doping in nitrogen-doped graphene},
  author={Joucken, Fr{\'e}d{\'e}ric and Tison, Yann and Le F{\`e}vre, Patrick and Tejeda, Antonio and Taleb-Ibrahimi, Amina and Conrad, Edward and Repain, Vincent and Chacon, Cyril and Bellec, Amandine and Girard, Yann and others},
  journal={Scientific reports},
  volume={5},
  number={1},
  pages={14564},
  year={2015},
  publisher={Nature Publishing Group UK London}
}

@article{hao2011mechanical,
  title={Mechanical and thermal transport properties of graphene with defects},
  author={Hao, Feng and Fang, Daining and Xu, Zhiping},
  journal={Applied physics letters},
  volume={99},
  number={4},
  year={2011},
  publisher={AIP Publishing}
}

@article{goharshadi2015thermal,
  title={Thermal conductivity and heat transport properties of nitrogen-doped graphene},
  author={Goharshadi, Elaheh K and Mahdizadeh, Sayyed Jalil},
  journal={Journal of Molecular Graphics and Modelling},
  volume={62},
  pages={74--80},
  year={2015},
  publisher={Elsevier}
}

@article{khade2020characterizing,
  title={Characterizing and predicting protein hinges for mechanistic insight},
  author={Khade, Pranav M and Kumar, Ambuj and Jernigan, Robert L},
  journal={Journal of molecular biology},
  volume={432},
  number={2},
  pages={508--522},
  year={2020},
  publisher={Elsevier}
}

@article{feher2014computational,
  title={Computational approaches to mapping allosteric pathways},
  author={Feher, Victoria A and Durrant, Jacob D and Van Wart, Adam T and Amaro, Rommie E},
  journal={Current opinion in structural biology},
  volume={25},
  pages={98--103},
  year={2014},
  publisher={Elsevier}
}

@article{lopez2022effect,
  title={The effect of rippling on the mechanical properties of graphene},
  author={Lopez-Polin, Guillermo and Gomez-Navarro, Cristina and Gomez-Herrero, Julio},
  journal={Nano Materials Science},
  volume={4},
  number={1},
  pages={18--26},
  year={2022},
  publisher={Elsevier}
}

@article{jennings2025quantum,
  title={Quantum algorithms for general nonlinear dynamics based on the Carleman embedding},
  author={Jennings, David and Korzekwa, Kamil and Lostaglio, Matteo and Sornborger, Andrew T and Subasi, Yigit and Wang, Guoming},
  journal={arXiv preprint arXiv:2509.07155},
  year={2025}
}

\appendix

\section{Loading non uniform velocity distributions}
In the main text, we discussed a method to load a set of velocities sampled from a discretized uniform distribution. We now detail the general case scenario here for a non uniform distribution. This is required to prepare the initial state for the QENM simulation. Formally, we construct a map $f: [N] \to [k]$ assigning each node to a bucket, such that the preimage sizes $\lvert f^{-1}(b) \rvert$, for $b\in [k]$, follow a discretized distribution of ones choice. This reduces the problem to loading a bucket-index state coherently, which can be done with $\mathcal{O}(\log N)$ resources. For this construction, one needs to use an appropriate family of Pseudorandom Functions (PRF). A PRF is a deterministic function that, when initialized with a random key, produces outputs that are computationally indistinguishable from those of a truly random function.

\noindent \textbf{Step 1: Compute PRF.}
First, we sample a function $\operatorname{PRF}_\theta(i)$ from a family of $polylog(N)$ sized PRFs, parametrized by $\theta$. We then coherently apply this function to a quantum state:
\begin{equation}
\ket{i}\ket{0} \rightarrow \ket{i}\ket{\operatorname{PRF}_\theta(i)}
\end{equation}

\noindent \textbf{Step 2: Inverse CDF Sampling.}
Next, we use the PRF output as a proxy for a random number to determine the velocity bucket index. We define a discrete cumulative distribution function (CDF) $C(j)$ for the desired Maxwell-Boltzmann distribution. The bucket index $f_\theta(i)$ for each node $i$ is determined by finding the smallest index $j$ for which the PRF output is less than the CDF value, effectively performing an inverse CDF sampling. This step is represented as:
\begin{equation}
f_\theta(i) = \min\{j | \operatorname{PRF}_\theta(i) < C(j)\}
\end{equation}
This operation is described as follows:
\begin{equation}
    \ket{i}\ket{\operatorname{PRF}_\theta(i)} \rightarrow \ket{i}\ket{f_\theta(i)}
\end{equation}

\noindent \textbf{Step 3: Map Bucket to Velocity.}
Finally, we assign a specific discretized velocity value $v_{f_\theta(i)}$ to each bucket index $f_\theta(i)$. This can be done via a lookup table or inline logic (e.g. using a single multiplication). After this step, the state of the system is a superposition of all nodes, each coherently assigned its appropriate initial velocity. The final state is:
\begin{equation}
\ket{i}\ket{f_\theta(i)} \rightarrow \ket{i}\ket{v_{f_\theta(i)}}
\end{equation}
The result is a coherent quantum state that holds the sampled Boltzmann distribution in its basis states. 
\begin{equation} 
\frac{1}{\sqrt{N}}\sum_{i=0}^{N-1} |i\rangle|v_{f_\theta(i)}\rangle
\end{equation}
This state can then be used to prepare a final amplitude-encoded state, where the amplitudes themselves correspond to the desired distribution, using a method such as \emph{inequality testing}.

We start by preparing a seperate register on an equal superposition of $r$ qubits with $r$ such that $\max{v_{f_\theta(i)}} \leq 2^r$:
\begin{equation}
    \ket{0}_r \xrightarrow{H^{\otimes r}} \frac{1}{\sqrt{2^r}}\sum_{x=0}^{2^r-1}\ket{x}_r
\end{equation}
The overall state is thus:
\begin{equation}
    \frac{1}{\sqrt{N}}\frac{1}{\sqrt{2^r}}\sum_{i=0}^{N-1}\ket{i}\ket{v_{f_\theta(i)}}\sum_{x=0}^{2^r-1}\ket{x}_r
\end{equation}
At this stage, we need to apply a quantum comparator $Q_{\mathsf{COMP}}$ \cite{luongo2024measurement, sanders2019black}, defined to act as:
\begin{equation}
    Q_{\mathsf{COMP}} \ket{a}\ket{b}\ket{0} = \begin{cases}
        \ket{a}\ket{b}\ket{0} \; \text{if $a<b$} \\
        \ket{a}\ket{b}\ket{1} \; \text{if $a\geq b$} 
    \end{cases}
\end{equation}
Thus, introducing an extra ancilla qubit and applying a comparator between the second, third and ancilla registers gives:
\begin{equation}
    \frac{1}{\sqrt{N2^r}}\sum_{i=0}^{N-1}\ket{i}\ket{v_{f_\theta(i)}}\Big(\sum_{x=0}^{v_{f_\theta(i)}} \ket{x}_r\ket{0}_1 + \sum_{v_{f_\theta(i)}+1}^{2^r-1}\ket{x}_r \ket{1}_1\Big)
\end{equation}
Applying another series of Hadamard rotations on the third register:
\begin{equation}
        \frac{1}{\sqrt{N2^r}}\sum_{i=0}^{N-1}\ket{i}\ket{v_{f_\theta(i)}}(v_{f_\theta(i)} \ket{0}_r\ket{0}_1 + \ket{\text{other}})
\end{equation}
projecting onto the $\ket{0}_{r+1}$, and then uncomputing the second register will result in the $\thicksim \sum_{i=0}^{N-1} v_{f_\theta(i)}\ket{i}$ state, which corresponds to the quantum state that encodes the initial velocities that are sampled from the appropriate distribution.

\end{document}